%% file: ROB_ROC_2022-06-22.tex
\documentclass[12pt]{article} 
\usepackage{graphicx}
\usepackage{color}
\usepackage{rotating}
\usepackage{amsmath}
\usepackage{amssymb}
\usepackage{amsfonts}
\usepackage{amsthm}
\usepackage{float}
\usepackage{url}

\newtheorem{theorem}{Theorem} 
\theoremstyle{plain}

\newtheorem{corollary}{Corollary} 
\newtheorem{remark}{Remark} 

\newtheorem{proposition}{Proposition}
\newtheorem{lemma}{Lemma} 

\usepackage{array,arydshln}

\usepackage{epstopdf}

\usepackage{enumerate}
\usepackage{enumitem}
\include{definiciones}

\usepackage[utf8]{inputenc}
\usepackage{xr}

\begin{document}
	
	\title{A robust approach for ROC curves with covariates}
	\author{Ana M. Bianco$^1$, Graciela Boente$^1$ and  
Wenceslao Gonz\'alez--Manteiga$^2$\\
$^1$ Universidad de Buenos Aires and CONICET\\
$^2$ Universidad de Santiago de Compostela
}
	\date{}
	\maketitle
	
	\small 
	\begin{abstract}
		The Receiver Operating Characteristic (ROC) curve is a useful tool that measures the discriminating power of a continuous variable or the accuracy of a pharmaceutical or medical test to distinguish between two conditions or classes.
		In certain situations, the practitioner may be able to measure some covariates related to the diagnostic variable which can increase the discriminating power of the ROC curve. To protect against the existence of atypical data among the observations, a procedure to obtain   robust estimators for the ROC curve in presence of covariates is introduced. The considered proposal focusses on a semiparametric approach which fits a location-scale regression model to the diagnostic variable and considers empirical estimators of the regression residuals distributions. Robust parametric estimators are combined with adaptive weighted empirical distribution estimators to down-weight the influence of outliers.  The uniform consistency of the proposal is derived under mild assumptions.
		A Monte Carlo study is carried out to compare the performance of the robust proposed estimators  with the classical ones  both, in clean and contaminated samples.  A real data set is also analysed.
		
	\end{abstract}


	\normalsize
	
	\section{Introduction}{\label{intro}}

	The Receiver Operating Characteristic (ROC) curve is a useful tool to size up the capability of a continuous variable or the accuracy of a pharmaceutical or medical test to distinguish between two conditions.
	ROC curves are a very well known technique in medical studies where a continuous variable or marker (biomarker) is used to diagnose a disease or to evaluate  the progression of  a disease.
	The use  of ROC curves has become more and more popular in medicine from the early 60's (see Goncalves \textsl{et al.}, 2014, for a historical note and Krzanowsk  and Hand, 2009 for further details).
	
	ROC curves can also be extended to other general statistical situations such as classification or discrimination, where  we typically have a set of individuals or items assigned to one of two  classes on the basis of disposable information of that individual. A ROC curve is essentially a plot that represents the diagnostic skill of a binary classifier as the discriminating threshold varies. Assignations are not perfect and may lead to classification errors. In fact, during the assignment procedure some errors may occur,  in the sense that an individual or object may be allocated into a wrong class. At this point, ROC curves become an interesting strategy either to evaluate the quality of a given assignment rule or to compare two available procedures.
	
	To be more precise, assume that we deal with two populations, henceforth, identified as diseased (\textsl{D}) and healthy (\textsl{H}) and  that a continuous score usually called \textsl{biomarker} or \textsl{diagnostic variable}, $Y$, is considered for the assignment purpose and whose rule is based on a cut--off value $c$. Thus, according to this assignment rule, an individual is classified as diseased if $Y \ge c$ and as healthy when $Y < c$. Let $F_{D}$ be the distribution of the marker on the diseased population and $F_{H}$ the distribution of $Y$ in the healthy one. From now on, for practical reasons, we denote  as $Y_D \sim F_D$ the marker in the diseased population and $Y_{H} \sim F_{H}$ the score in the healthy one.  Without loss of generality, we will assume that $Y_D$ is stochastically greater than  $Y_{H}$, that is,  $\prob(Y_{D} \le c) \le \prob(Y_{H} \le c)$ for all  $c$.
	It is clear that the classification errors depend on the threshold $c$. Therefore, it becomes of interest to study the triplets  $\{(c,1-F_{H}(c),1-F_{D}(c)),\; c \in \real \}$,
	which describes a geometrical object called ROC curve, that reflects the discriminatory capability of the marker.  This suggests a different parametrization of this curve in terms of the false positive rate, $1-F_{H}(c)$, leading to  $\{(p, 1-F_{D}(F_{H}^{-1}(1-p))), \; p \in (0,1) \}$
	and therefore, to $\ROC(p)= 1-F_{D}(F_{H}^{-1}(1-p))), \quad p \in (0,1) $.
	In this manner, the ROC curve is a complete picture of the performance of the assignment procedure over all the possible threshold values.   
	
	In practical situations, the   discriminatory effectiveness of the biomarker   may be improved by several factors. Thus, when for each individual there is additional information contained in measured covariates, it is sensible to include them in the ROC analysis. Through examples Pepe (2003)  illustrates how the discriminatory capability of a test is improved by the presence of covariates. For an overview on this topic, we refer to Pardo-Fern\'andez \textsl{et al}. (2014). 
	In brief, we may say that the information registered all along the covariates may impact the discrimination capability of the ROC curve. In this situation,   in order to have a deeper comprehension of the effect of the covariates, it would be advisable to incorporate this additional  covariates information to the ROC analysis instead of considering a \textsl{joint} ROC curve, that may lead to  oversimplification. This issue can be accomplished in different ways. In the direct methodology, the ROC curve is directly regressed onto the covariates by means of a generalized linear model. Among others,  Alonzo and Pepe (2002), Pepe (2003) and  Cai (2004)  follow this approach.
	In contrast, in the induced methodology, the markers distribution in each population is modelled separately in terms of the covariates and just after, the induced ROC curve is computed. The papers by Pepe (1998), Faraggi (2003), Gonz\'alez-Manteiga \textsl{et al.} (2011) and  Rodr\'{\i}guez-\'Alvarez \textsl{et al.} (2011a)   go in this direction.
	Besides, In\'acio de Carvalho \textsl{et al.} (2013) follow a Bayesian nonparametric approach to fit covariate--dependent ROC curves using probability models in each population, while Rodr\'{\i}guez-\'Alvarez \textsl{et al.} (2011b) perform a comparative study of the direct and induced methodologies. From now on,  we denote as $\bX_D$ and $\bX_H$ the covariates for the disease and healthy populations and we  assume  that they have the same dimension. In such case, for any $\bx$ in the common support of 	$\bX_D$ and $\bX_H$, the conditional ROC curve is defined as 
	\begin{equation}
		{\ROC}_{\bx}(p) =  1-F_{D}(F_{H}^{-1}(1-p|\bx)|\bx) \,,
		\label{eq:ROCx}
	\end{equation}
	where $F_{j}(\cdot|\bx)$ stands for conditional distribution of $Y_j|\bX_j=\bx$, $j=H, D$.
	In this paper, we focus on the latter approach through a general  regression model.
	
	The general methodology to estimate the conditional ROC curve consists in a plug--in procedure where estimators of the regression and of the variance functions together with  empirical distribution  and quantile function estimators based on the residuals are plugged into the general expression of the conditional ROC curve. Pepe (1997, 1998, 2003), Faraggi (2003), Gonz\'alez-Manteiga \textsl{et al.}  (2011) propose estimators that implement these ideas.  
	Since most of these estimators are based on classical least squares procedures or local averages, they may be very sensitive to anomalous data or small deviations from the model assumptions.  The bi--normal model, in which both populations are assumed to be normal,  is a very popular choice to fit a ROC curve and one justification for its broad use is its  robustness. The term robustness may have different interpretations; in fact, Gon\c{c}ales \textsl{et al.} (2014) discuss the scope of the so--called robustness in the ROC curve scenario.  Walsh (1997)  performs a simulation study that shows that the bi--normal estimator is sensitive to model misspecifications and to the location of the decision thresholds. 
	
	In this paper, we focus on robustness, that is,  resistance to deviations from the underlying model plus efficiency when this central model holds.  During the last decades, robust statistics has pursued the aim of developing  procedures that enable reliable inference results, even if small deviations from the model assumptions occur or in the presence of a moderate percentage of outliers. Even when these efforts have been sustained over time  across different statistical areas, up to our knowledge, ROC curves have received little attention from this robustness point of view. When no covariates are available,  robust estimators of the area under the ROC curve were given in Greco and Ventura (2011) assuming that the distribution functions are known up to a finite--dimensional parameter (see also Farcomeni and Ventura, 2012). In this sense, when covariates are recorded to improve the discrimination power of the biomarker, the main contribution of our paper is to bridge the gap between ROC curves and robustness. We achieve this goal by  fitting a location-scale regression model to the diagnostic variable and considering adaptive empirical estimators of the regression residuals distributions. In this respect, our proposal is  semiparametric since the errors distribution is not assumed to be known, for example, as  in the bi--normal model.
	
\begin{figure}[ht!]
	\begin{center}
		\footnotesize
\renewcommand{\arraystretch}{0.1}
		\begin{tabular}{ccc}
	 {(a)}  & (b) & (c)\\
			\includegraphics[scale=0.3]{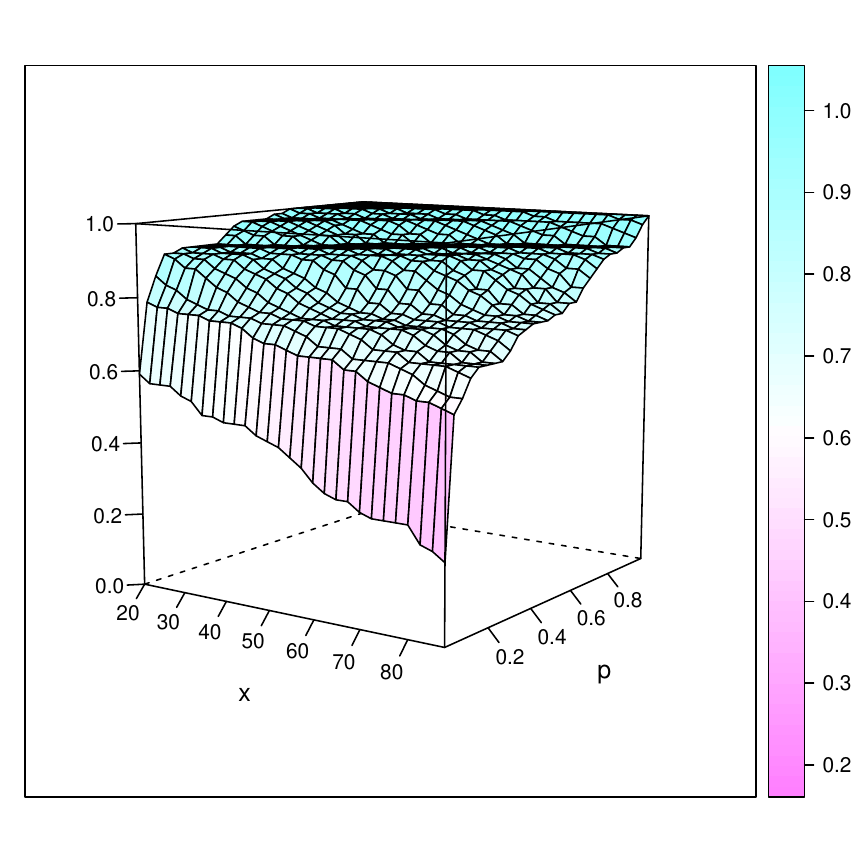} & 
			\includegraphics[scale=0.3]{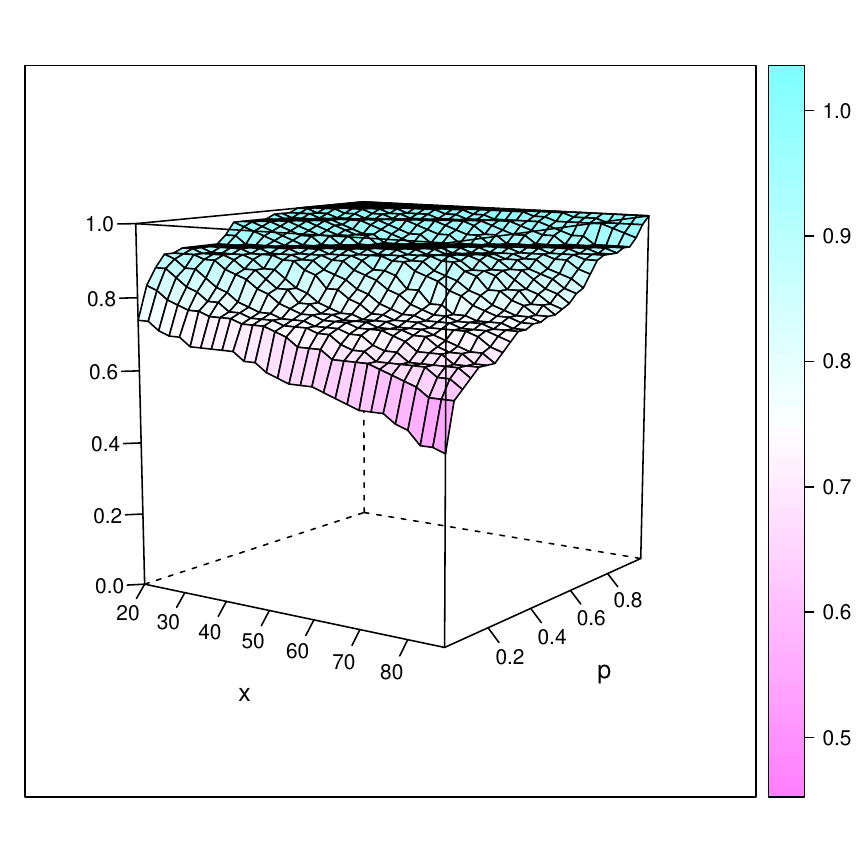} &
			\includegraphics[scale=0.3]{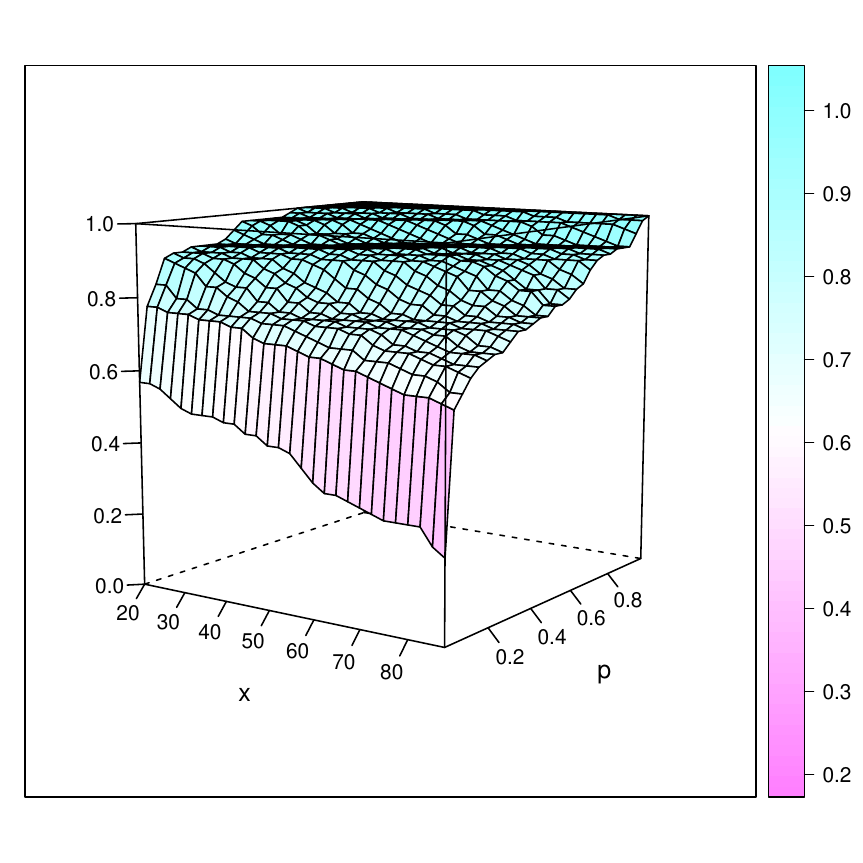} 
		\end{tabular}
		\vskip-0.1in \caption{\label{fig:ROC_diabetes} \small Estimated ROC surfaces for the Diabetes Data using Age ($X$) as covariate: (a) Classical estimator, (b) Classical estimator without the detected outliers and (c) Naive estimator.}
	\end{center} 
\end{figure}

	Our motivating example consists of the real dataset of a marker for diabetes previously  analysed in Faraggi (2003) and  Pardo--Fern\'andez \textsl{et al}. (2014), in which we add to their analysis a robust perspective focussing on the potential effect of influential data. The observations, that come from a population-based pilot survey of diabetes mellitus in Cairo, Egypt, consist of postprandial blood glucose measurements ($Y$) from a fingerstick in 286 subjects who were divided into healthy (198) and diseased (88) groups according to  gold standard criteria of the World Health Organization (1985). It is believed that the aging process may be associated with  resistance or relative insulin deficiency among healthy people, therefore postprandial fingerstick glucose levels would be expected to be higher for older persons who do not have diabetes. According to this belief, Smith and Thompson (1996) adjust the ROC curve analysis for covariate information using age ($X$). The obtained ROC curve of the transformed biomarker is given in Figure \ref{fig:ROC_diabetes}  together with the ROC curve obtained after removing the  6 outliers detected in the healthy sample through a robust regression fit. Figure \ref{fig:ROC_diabetes} also displays the ROC curve built using the naive approach of using robust regression estimators combined with the  usual  empirical distribution  and quantile function estimators based on the residuals. These plots illustrate that the use of robust regression and variance estimators are not enough to protect the estimation of the ROC curve from the influence of atypical data. This effect may be explained by the fact that large residuals are still present when empirical distribution estimators are computed. This motivates the need of defining appropriate robust estimators of the ROC curve.

\normalsize

In the rest of the paper we will introduce a robust proposal and we will study some of its properties.
The paper is organized as follows. Section \ref{sec:prelim} reviews some general concepts regarding the conditional ROC curve, while Section \ref{sec:propuesta} introduces  the robust proposal  to estimate the ROC curve focussing in the special situation of a parametric regression model. Section \ref{sec:consist} presents some consistency results of the proposed procedure. Finally, in Section \ref{sec:monte},  a numerical study is conducted   to examine the small sample properties of the proposed procedures  under a  linear and a nonlinear regression model, while the advantages of the proposed methodology are illustrated in Section \ref{sec:realdata} on a real data set. All proofs are relegated to the Appendices.


\section{Preliminaries}{\label{sec:prelim}}
In this section, we recall the approach considered to model the induced ROC curve when covariates are measured. 
For that purpose, denote  as  $Y_D$ and $\bX_{D}$ the biomarker and the covariates measured in the diseased population and as $Y_H$ and $\bX_{H}$ the corresponding ones in the healthy individuals. For the sake of simplicity, we will assume that the covariates of interest are the same in both populations.

  A general way to include covariates is through a general location--scale regression model which, for simplicity of presentation, we assume homoscedastic, that is,
\begin{eqnarray}
Y_{D} &=& \mu_{0,D}(\bX_{D})+ \sigma_{0,D} \; \epsilon_{D}\;, \label{modeloD}\\
Y_{H} &=& \mu_{0,H}(\bX_{H})+ \sigma_{0,H}  \;\epsilon_{H}\;, \label{modeloH} 
\end{eqnarray}
where, for $j=D,H$, $\mu_{0,j}$ is the true regression function and   $\sigma_{0, j}$ corresponds to the model dispersion, respectively. It is also assumed that  the errors $\epsilon_j \sim G_{j}$ are  independent of $\bX_j$, for $j=D,H$ and have scale $1$ to properly identify $\sigma_{0,j}$.  Furthermore, to identify the regression function  in the classical framework it is assumed that $\esp \epsilon_j=0$, for $j=D,H$. Instead, in the robust setting it is usual to avoid the existence of moments. For that reason,  to ensure  consistency of the robust estimators to the target regression function $\mu_{0,j}$, it is standard to assume that $G_j$ has a symmetric distribution. Otherwise, Fisher--consistency of the related regression functionals should be required.  Denote as ${\cal S}$ the common support of $\bX_{D}$ and $\bX_{H}$.
It is worth noticing that since the errors and the covariates are independent, for a given $\bx \in \itS$, we have that 
\begin{eqnarray*}
F_{Y_{D}}(y|\bx)&=& F_{Y_{D}|X_{D}}(y|\bx)=  \prob(Y_{D}\le y|\bX_{D}=\bx) = \prob( \mu_{0,D}(\bX_{D})+ \sigma_{0,D} \;\epsilon_{D} \le y|\bX_{D}=\bx)\\
&=&  G_{D}\left(\dfrac{y-\mu_{0,D}(\bx)}{\sigma_{0,D}}\right) \;.
\end{eqnarray*}
Analogously,   we get that the conditional distribution in the healthy distribution satisfies
$$F_{Y_{H}}(y|\bx)=F_{Y_{H}|X_{H}}(y|\bx)= G_{H}\left(\dfrac{y-\mu_{0,H}(\bx)}{\sigma_{0,H}}\right) \, .$$
As a consequence, the quantiles of the conditional distributions are related to those of the errors through 
$F_{j}^{-1}(p|\bx)= \sigma_{0,j}\,  G_{j}^{-1}(p) +\mu_{0,j}(\bx)$, for $ j=D,H $,
where $G_{j}^{-1}(\cdot)$ denotes the quantile function of the errors $\epsilon_j$.
Thus, the conditional ROC curve given $\bx \in \itS$ defined in \eqref{eq:ROCx}  can be computed as
\begin{eqnarray}
{\ROC}_{\bx}(p)     
&=& 1-G_{D}\left(\frac{\mu_{0,H}(\bx) -\mu_{0,D}(\bx)}{\sigma_{0,D}} + \frac{\sigma_{0,H}}{\sigma_{0,D} } \, G_{H}^{-1}(1-p)\right)\,. \label{rocx}
\end{eqnarray}
One advantage of this approach is that it  enables a very general modelling of the regression  functions $\mu_{0,j}$, for $ j=D, H$, since this task  can be accomplished from different perspectives. This means that according to the information about the relationship between the biomarker and the covariates and the user's  preferences,  the regression  functions may be modelled  parametrically or either nonparametrically or partly parametrically, even when these last two approaches will be subject of future work.

As mentioned in the Introduction, expression \eqref{rocx} of the conditional ROC curve suggests a natural   estimation procedure. First, compute estimators of the regression   function and the dispersion parameter which allow to obtain the corresponding residuals. Then,  estimate $G_{D}$ and $G_{H}^{-1}$ by  empirical distribution  and quantile function estimators based on the residuals, respectively. Finally, using these estimators in \eqref{rocx} and  plugging there--in the obtained estimators of the regression functions and variance parameters, we obtain an estimator of the conditional ROC.  Our goal is to introduce a procedure to get reliable and stable $\ROC_{\bx}$ estimators, even when  a moderate percentage of outliers arise in one sample or in both of them.  

Different summary measures of the ROC curve are useful to sum up particular features of the curve. One of the most popular indices is the conditional \textsl{area under the curve} (AUC$_{\bx}$), which is computed as $\AUC_{\bx}= \int_0^1 \ROC_{\bx}(p) dp$.  

\section{Proposal}{\label{sec:propuesta}}
\subsection{The general procedure}{\label{sec:general}}
Suppose that  we have  a  sample from the diseased population, $(y_{D,i},\bx_{D,i})$, $1\le i\le n_D$,  that verifies model \eqref{modeloD} and one from the healthy population, $(y_{H,i},\bx_{H,i})$, $1\le i\le n_H$, verifying model \eqref{modeloH}. Furthermore, assume that the samples are  independent from each other.

As mentioned above,  since the conditional ROC curve is given in equation \eqref{rocx}, an estimation procedure can be obtained following the next steps:  i) compute estimators of the regression functions and variance parameters, ii) calculate the corresponding residuals and replace the distribution and quantile functions, $G_D$ and $G_H^{-1}$, by  suitable estimators and iii) plug--in estimators of the regression functions and variance parameters in \eqref{rocx}. 

In order to obtain a final robust estimator of the ROC and AUC curves, it is necessary to consider robust estimators not only in the first step of the described procedure, but also in the second one. In fact, if robust estimators are only considered for the estimation of the regression and variance functions, large residuals would influence the classical empirical distribution and quantile function estimators wasting the efforts  made in the first step to get robustness. Taking these ideas into account, we propose the following stepwise procedure: 

\begin{enumerate}

\item[\textbf{Step 1.}] Estimate $\mu_{0,H}(\bx)$, $\sigma_{0,H}$, $\mu_{0,D}(\bx)$, $\sigma_{0,D}$ in a robust fashion from the samples \linebreak  $(y_{H,1},\bx_{H,1}),\dots,$ $(y_{H, n_H},\bx_{H, n_H})$ and $(y_{D,1},\bx_{D,1}),\dots,(y_{D, n_D},\bx_{D, n_D})$, respectively. Denote the resulting estimators by $\wmu_{H}(\bx)$, $\wsigma_{H}$, $\wmu_{D}(\bx)$ and $\wsigma_{D}$.

\item[\textbf{Step 2.}] Compute for each sample the standardized regression residuals  
$$r_{H,i}= \dfrac{y_{H,i}-\wmu_{H}(\bx_{H,i})}{\wsigma_{H}}\quad \mbox{and} \quad r_{D,i}= \dfrac{y_{D,i}-\wmu_{D}(\bx_{D,i})}{\wsigma_{D}}\,.$$
From these residuals, evaluate  robust estimators of the    distribution and quantile functions, denoted, $\wG_{D}$ and $\wG_{H}^{-1}$, respectively.

\item[\textbf{Step 3.}] Plug--in the robust estimators computed in the first two steps into equation \eqref{rocx} to obtain
\begin{eqnarray*}
\widehat{\ROC}_{\bx}(p) = 1-\wG_{D}\left(\frac{\wmu_{H}(\bx) -\wmu_{D}(\bx)}{\wsigma_{D}} + \frac{\wsigma_{H} }{\wsigma_{D}  } \, \wG_{H}^{-1}(1-p)\right) \, . \label{rocxhat}
\end{eqnarray*}
\end{enumerate}

A key point of the above procedure is to provide robust and consistent estimators in the first and second steps. Regarding \textbf{Step 1}, the considered regression models \eqref{modeloD} and \eqref{modeloH} may be either parametric, nonparametric or semiparametric.  In each case, suitable robust estimators must be used. In particular, in the parametric case, linear or nonlinear models may be adequate. For instance, when the conditional model is a  linear model, the $MM-$estimators introduced in Yohai (1987) are a recommended option, while under a  nonlinear one the weighted $MM-$estimators presented in Bianco and Spano (2019) may be used.  

Beyond the robust estimation of the regression  functions and the scales $\sigma_j$, it is necessary to detect outliers in order to obtain a robust version of the empirical distribution and quantile function  estimators. Unlike the classical empirical estimators, where all the observations have the same weight, downweighting in the second step atypical points, i.e., those values that lie far away from the bulk of the data,   may result in a more resistant procedure.

 \subsection{Regarding the estimation of the residual's distribution}
As in Gervini and Yohai (2002), we consider adaptive  weights computed from the empirical distribution of the residuals obtained from a
robust fit. To describe the extension of their proposal and to fix ideas, let us consider a general homoscedastic nonlinear regression model. Similar arguments can be consider when the model is fully nonparametric, semiparametric or even heteroscedastic. 

Assume that we have a random sample $(y_{1},\bx_{1}),\dots,(y_{ n},\bx_{n})$, where $\bx_i$ is a vector of  $p$ explanatory variables and $y_i$ is a response variable that satisfies
\begin{equation}
y_i= \mu(\bx_i) + u_i= f(\bx_i, \bbe_0) +\sigma_0 \epsilon_i\,, \qquad i=1 \dots n \, , \label{general}
\end{equation} 
with $\bbe \in \real^q$, $\sigma_0$   the scale parameter and $f$ a known function. Note that the dimension of the regression parameter $\bbe$ may be equal  or not to that of the covariates. The errors $\epsilon_i$ are independent and identically distributes (i.i.d.) with unknown distribution $G_0 $  and independent of the covariates $\bx_i$. We will assume that $G_0$ is symmetric around 0.

Consider robust estimators of regression and scale, let us say  $\wmu(\cdot)$ and $\wsigma$, and compute standardized residuals
$r_{i}= ({y_{i}-\wmu(\bx_{i})})/{\wsigma} $.
In particular, under the nonlinear regression model \eqref{general},  $\wmu(\bx_{i})=f(\bx_{i}, \wbbe)$, where for instance, $\wbbe$ is an $S-$ or an $MM-$estimator. 
On the basis of these residuals, the classical empirical distribution at point $t$ can be computed as
$\wG_{n,\mbox{\sc \tiny emp}}(t)= (1/n) \sum_{i=1}^n \indica_{r_i \le t}$.
Large values of $|r_i|$ suggest that the corresponding   pairs $(y_{i},\bx_{i})$ may be outliers. In that case, under a  normal error model, it seems wise to consider as atypical those points whose residuals are larger than a certain cut--off value $t^{\star}$, that is,  such that $|r_i|>t^{\star}$. Typically, $t^{\star}$ is chosen as 2.5by  taking the standard normal distribution as a benchmark. To take into account these considerations, weighting may be a useful alternative in the computation of the empirical distribution estimator. However, in order to make the cut--off criterion more flexible and more data--driven, adaptive cut--off values could be considered in this process.

We compute the adaptive weighted empirical distribution at point $t$ as:
\begin{equation}
\wG_n(t)= \frac{1}{\sum_{\ell=1}^n w_{\ell}}\sum_{i=1}^n w_i \indica_{r_i \le t} \, , \label{empiricalw}
\end{equation}
where the weights $w_i \ge 0$  are based on a weight function $w: \real \to [0,1]$ non-increasing, even, right continuous,
continuous in a neighbourhood of  $0$, $w(0)=1$, $w(u)>0$ for $0 < u <1$ and $w(u)=0$ for $   u \ge 1$. The fact that  $w(u)=0$ for $   u \ge 1$ ensures that $w_i = 0$ when $|r_i|$ is larger than the selected cut--of value, so, as mentioned in Gervini and Yohai (2002), observations with large
absolute residuals are completely eliminated in the weighted estimators.    It is worth noting  that beneath this criterion to downweight large residuals  lays the idea that   the errors distribution $G_j$ is symmetric, since the  weights will remove an equal amount of   large positive and negative residuals.  If the practitioner suspects that a skewed distribution underlies,     another kind of weights, such as asymmetric ones, may be preferable. 

To define the adaptive cut--off values,  consider the empirical  distribution function of the absolute
standardized residuals $r_{i}$ given by
$${G}^{+}_n(t)= \frac 1n \sum_{i=1}^n  \indica_{|r_i| \le t}\, .$$
and let ${G}^{+}_0(t)$ be the distribution of the absolute errors when $\epsilon_i \sim G_0$. As noted in 
Gervini and Yohai (2002), if  for a large $t$ it happens that ${G}^{+}_n(t) < {G}^{+}_0(t)$, we have that the sample proportion of absolute residuals that exceeds $t$ is  greater than  the theoretical proportion   suggesting that outliers are present among the data.
 
Since in practice the actual distribution of $\epsilon_i$ is unknown, an hypothetical distribution $G$, such as the standardized normal distribution, is assumed.  
Gervini and Yohai(2002) consider as a measure of the percentage of atypical data 
$$d_n= \sup_{t \ge 0\eta} \{G^{+}(t)-G^{+}_n(t)\}^+=\sup_{t \ge 0} \{\max\left(G^{+}(t)-G^{+}_n(t),0\right)\} \, ,$$
where $\{\cdot\}^+$ denotes the positive part, $G^{+}$ is the distribution of the random variable $|V|$ when $V \sim G$  and $\eta$ is some large quantile of $G^{+}$, that is,  $\eta=(G^{+})^{-1}(p)$ for some $p$ close to 1.
 Let $|r|_{(1)}\le |r|_{(2)} \le \dots\le |r|_{(n_D)} $ denote the order statistics of the standardized residuals. As those authors note
$$d_{n }= \max_{i_0 \le i \le n } \left\{ \max\left(G ^{+}(|r|_{(i)})-\dfrac{(i-1)}{n },0\right)\right\}  \, ,$$
where $i_0=\max\{i: |r|_{(i)} <\eta\}$. Therefore, a possible cut--off value may be
\begin{equation}
t_{n }= |r|_{i_{n }}= \min\{t: {G}^{+}_n(t)\ge 1- d_{n }\}\,,
\label{eq:tn}
\end{equation}
 where $i_{n }=n -[n \, d_{n }]$.

With this adaptive cut--off value, by means of the weight function $w: \real \to [0,1]$, we define 
\begin{equation}
w_i= w\left(\dfrac{ r_i }{t_n}\right)\,,
\label{eq:pesos}
\end{equation}
and the adaptive weighted empirical distribution as in \eqref{empiricalw}, which allows to define also the weighted quantile function.
Appendix \ref{sec:consistG} provides some   uniform consistency results for the adaptive weighted empirical distribution $\wG_n$ defined through \eqref{empiricalw} and \eqref{eq:pesos}, under mild conditions.

\section{Consistency results}{\label{sec:consist}}
The results in this section are based on those concerning the uniform consistency of the weighted distribution function defined in \eqref{empiricalw} which are given in Appendix \ref{sec:consistG}. We will consider a general nonlinear regression model, extensions to other settings, such as nonparametric regression models, can be obtained similarly.
Henceforth,    $(y_{j,i},\bx_{j,i})$, $1\le i\le n_{\ell}$, for $j=D, H$, stand for independent  random samples from the diseased and healthy populations with the same distribution as $(Y_D, \bX_D)\in \real^{p+1}$ and $(Y_H, \bX_H)\in \real^{p+1}$, respectively, where $(Y_D, \bX_D)$ satisfy \eqref{modeloD} and $(Y_{H},\bX_{H})$ fulfils \eqref{modeloH}. The errors $\epsilon_j \sim G_{j}$ are  independent of $\bX_j$, for $j=D,H$. 
In this situation, using \eqref{rocx}, we get that 
$${\ROC}_{\bx}(p) = 1-G_{D}\left(\frac{\mu_{0,H}(\bx) -\mu_{0,D}(\bx)}{\sigma_{0,D}} + \frac{\sigma_{0,H}}{\sigma_{0,D}} \, G_{H}^{-1}(1-p)\right)\,. $$
To avoid burden notation, for   $j=D, H$, we will denote as $\wG_{j}=\wG_{j,n_j}$ the weighted empirical distribution function defined in \eqref{empiricalw} using the sample $\left(y_{j,i},
\bx_{j,i}\right)$, $1\le i \le n_j$  and robust consistent estimators $\wmu_j$ and $\wsigma_j$ of $\mu_{0,j}$ and $\sigma^2_{0,j}$, respectively. Then, the estimator of the ROC curve whose uniform consistency we will study is given by
\begin{eqnarray*}
\widehat{\ROC}_{\bx}(p) = 1-\wG_{D}\left(\frac{\wmu_{H}(\bx) -\wmu_{D}(\bx)}{\wsigma_{D} } + \frac{\wsigma_{H} }{\wsigma_{D}  } \, \wG_{H}^{-1}(1-p)\right) \, . \label{rocxhatreg}
\end{eqnarray*}

We will need the following assumptions on the errors distributions and on their estimates:

\begin{enumerate}[label = \textbf{A\arabic*}]
\item\label{ass:A1}  $G_H : \real\to  (0, 1)$ has an associated density $g_H$ such that $g_H(y)>0$, for all $y\in \real$.
\item\label{ass:A2} $G_D : \real\to  (0, 1)$ is continuous.
\item\label{ass:A3} $\|\wG_{j}-G_j\|_{\infty}\convpp 0$, $j=D,H$.
\item\label{ass:A61}  For each fixed $\bx$,  $ |\wmu_{j}(\bx)-\mu_{0,j}(\bx)\|\convpp 0$, $j=D,H$. 
\item\label{ass:A6}  For any compact set $\itK \subset {\cal S}$,  $\sup_{\bx\in \itK}|\wmu_{j}(\bx)-\mu_{0,j}(\bx)\|\convpp 0$, $j=D,H$. 
\item\label{ass:A7} The regression functions $\mu_{0,j}$ are such that, for any compact set $\itK$   $\sup_{\bx\in \itK} |\mu_{0,j}(\bx)|=A_j<\infty$.
\end{enumerate}

\begin{remark}\label{remark:casonolineal}
If we are dealing with a parametric regression model, i.e., when $\mu_{0,D}(\bx)= f_{D}(\bx, \bbe_{0,D})$ and $\mu_{0,H}(\bx)=f_{H}(\bx, \bbe_{0,H})$ and   $\wbbe_j$ and $\wsigma_j$ stand for  robust consistent estimators of $\bbe_{0,j}$ and $\sigma^2_{0,j}$, respectively, the estimator of the ROC curve equals
$$
\widehat{\ROC}_{\bx}(p) = 1-\wG_{D}\left(\frac{f_{H}(\bx, \wbbe_H) -f_{D}(\bx, \wbbe_D)}{\wsigma_{D} } + \frac{\wsigma_{H} }{\wsigma_{D}  } \, \wG_{H}^{-1}(1-p)\right) \, . $$
In this framework, conditions under which \ref{ass:A3} holds for the linear model $f_{j}(\bx, \bbe_{0,j})=\bx\trasp \bbe_{0,j}$ or more generally, for a nonlinear model  are given in the Appendix \ref{sec:consistG}. The derivation of conditions that guarantee the validity \ref{ass:A3}  under   nonparametric or  semiparametric models are beyond the scope of this paper.

On the other hand, \ref{ass:A61} to \ref{ass:A7}  hold if the  non--linear regression functions are such that
\begin{enumerate}[label = \textbf{A\arabic*}]
\setcounter{enumi}{6}
\item\label{ass:A4} For each fixed $\bx$, the  regression functions $f_j(\bx, \bb)$ are  continuous in $\bb$.
\item\label{ass:A5} The  functions $f_j$ are such that, for any compact set $\itK$ and any sequence $\bbe_n\to \bbe_{0,j}$, we have
$\sup_{\bx \in \itK} |f_j(\bx, \bbe_n)- f_j(\bx, \bbe_{0,j})|\to 0$. Further, $\sup_{\bx\in \itK} |f_j(\bx, \bbe_j)|=A_j<\infty$.
\end{enumerate}
In particular, these  assumptions hold if the regression model is a linear one.
\end{remark}

\begin{theorem}  \label{theo:consist.2} Let  $\left(y_{j,i},
\bx_{j,i}\right)$, $1\le i \le n_j$, $j=D, H$, be independent observations satisfying \eqref{modeloD} and \eqref{modeloH}, respectively and assume that  $\wmu_j$ and $\wsigma_j$ are strongly consistent estimators of $\mu_{0,j}$ and $\sigma_{0,j}$, respectively. Then, under \ref{ass:A1} to \ref{ass:A3} and \ref{ass:A61},
\begin{enumerate}
\item[(i)] $\sup_{0<p<1} |\widehat{\ROC}_{\bx}(p)- {\ROC}_{\bx}(p)|\convpp 0$.
\item[(ii)] If, in addition,  \ref{ass:A6} holds, $G_D $ has a  bounded density $g_D$ and the regression functions $\mu_{0,j}$ satisfy \ref{ass:A7}, then, for any $\delta>0$ 
$\sup_{\delta<p<1-\delta} \sup_{\bx\in \itK} |\widehat{\ROC}_{\bx}(p)- {\ROC}_{\bx}(p)|\convpp 0$.
\item[(iii)] Furthermore, assume that   $G_D $ has a  bounded density $g_D$, the regression functions $\mu_{0,j}$ satisfy \ref{ass:A7} and the conditional ROC function is such that, for any $\epsilon>0$, there exists $0<\eta<1$   such that, for any $\bx\in \itK$, ${\ROC}_{\bx}(\eta)<\epsilon$ and $1-{\ROC}_{\bx}(1-\eta)<\epsilon$,  then 
$\sup_{0<p<1 } \sup_{\bx\in \itK} |\widehat{\ROC}_{\bx}(p)- {\ROC}_{\bx}(p)|\convpp 0$.
\end{enumerate}
\end{theorem}

As a consequence of Theorem \ref{theo:consist.2}, we immediately get the following result.

\begin{corollary}  \label{theo:consist.1} Let  $\left(y_{j,i},
\bx_{j,i}\right)\sim (Y_j, \bX_j)$, $1\le i \le n_j$, $j=D, H$, be independent observations satisfying 
\begin{eqnarray*}
Y_{D} = f_{D}(\bX_{D}, \bbe_{0,D})+ \sigma_{0,D}  \epsilon_{D} \qquad \qquad 
Y_{H}= f_{H}(\bX_{H}, \bbe_{0,H})+ \sigma_{0,H}  \epsilon_{H} \label{modeloregDH}\;,
\end{eqnarray*}
where, for $j=D,H$, the errors $\epsilon_j \sim G_{j}$ are  independent of $\bX_j$, for $j=D,H$. Assume that  $\wbbe_j$ and $\wsigma_j$ are strongly consistent estimators of $\bbe_{0,j}$ and $\sigma_{0,j}$, respectively. Then, under \ref{ass:A1} to \ref{ass:A4},
\begin{enumerate}
\item[(i)]
$\sup_{0<p<1} |\widehat{\ROC}_{\bx}(p)- {\ROC}_{\bx}(p)|\convpp 0$.
\item[(ii)] If, in addition,  $G_D $ has a  bounded density $g_D$ and the regression functions $f_j$ satisfy \ref{ass:A5}, then  $\sup_{\bx \in \itK} |\widehat{\ROC}_{\bx}(p) - {\ROC}_{\bx}(p)| \convpp 0$. 
\item[] Moreover, for any $\delta>0$ 
$\sup_{\delta<p<1-\delta} \sup_{\bx\in \itK} |\widehat{\ROC}_{\bx}(p)- {\ROC}_{\bx}(p)|\convpp 0$.
\item[(iii)] Furthermore, assume that   $G_D $ has a  bounded density $g_D$, the regression functions $f_j$ satisfy \ref{ass:A5} and the conditional ROC function is such that, for any $\epsilon>0$, there exists $0<\eta<1$   such that, for any $\bx\in \itK$, ${\ROC}_{\bx}(\eta)<\epsilon$ and $1-{\ROC}_{\bx}(1-\eta)<\epsilon$,  then 
$\sup_{0<p<1 } \sup_{\bx\in \itK} |\widehat{\ROC}_{\bx}(p)- {\ROC}_{\bx}(p)|\convpp 0$.
\end{enumerate}
\end{corollary}

It is worth noticing that the requirement  $\sup_{\bx\in \itK}{\ROC}_{\bx}(\eta)<\epsilon$ and $\sup_{\bx\in \itK}1-{\ROC}_{\bx}(1-\eta)<\epsilon$ in (iii)
is satisfied when  \ref{ass:A5}  holds and $G_H$ has support on the whole line as stated in  \ref{ass:A1}.

\section{Monte Carlo study}{\label{sec:monte}}

In this section, we summarize the results of a simulation study conducted to study the small sample performance of the proposal given in Section \ref{sec:propuesta}.    The goal of this numerical experiment is two--fold. On the one hand, we want to illustrate the sensitivity of the classical methods to deviations from the central model. On the other hand, we want to evaluate the performance of our robust proposal under different contamination schemes and to compare it with the classical one. For that purpose, we considered  different scenarios and contaminations schemes. In all cases, we generate $Nrep=1000$ datasets  of size $n_{D}=n_{H}=n=100$  and $n_{D}=n_{H}=n=200$. To evaluate if the advantages to be observed in the robust procedure depend on linearity, we considered two regression models, a linear and a nonlinear one. Besides, different contaminating schemes are analysed either contaminating one or both populations.

To summarize the discrepancy between the estimator and the true ROC surface, we consider two grids of points: $\itG_p=\{p_j\}_{j=1}^{N_p}$ corresponding to   equidistant values between $0.01$ and $0.99$ with step $0.01$ and $\itG_x=\{x_i\}_{i=1}^{N_x}$  where the net has step $0.05$ within the interval $[a,b]$ with $a=-1$ and $b=1$ for the linear model, while $a=-0.5$ and $b=0.5$ for the nonlinear one.  The estimators performance is then evaluated using the mean over replications of   
\begin{itemize}
\item the  Mean Squared Error ($ MSE $) given by 
$$MSE=\frac{1}{N_x N_p} \sum_{i=1}^{N_x}\sum_{j=1}^{N_p} \left(\widehat{\ROC}_{x_i}(p_j) -{\ROC}_{x_i}(p_j)\right)^2 \, ,$$

\item a measure inspired on the Kolmogorov distance ($ KS $) calculated as
$$KS=\sup_{1\le i N_x} \sup_{1\le j \le N_p} \left| \widehat{\ROC}_{x_i}(p_j) - {\ROC}_{x_i}(p_j) \right| \, ,$$
\end{itemize}
that give a global summary of the mismatch  between the estimated $\ROC$ curves and the true ones.

\subsection{Numerical study under a linear model}{\label{sec:linearmodel}}

In the first scenario, we consider different homoscedastic linear-mean regression
models for the two populations. We considered the same conditions as  in In\'acio  de Carvalho \textsl{et al.} (2013), that is,   following the  linear regression models 
\begin{eqnarray}
y_{D,i} &=& 2+ 4 x_{D,i} + \sigma_{D}  \;  \epsilon_{D,i} \label{trued}\\
y_{H,i} &=&  0.5+   x_{H,i} + \sigma_{H} \; \epsilon_{H,i} \label{trueh}\; ,
\end{eqnarray}
for all $i=1,\dots,n$ $\epsilon_{j,i} \sim N(0,1)$ are independent and independent from   $x_{j,i} \sim U(-1,1)$, for $j=D,H$, $ \sigma_D=2 $ and $ \sigma_{H}=1.5$. Besides, the sample from one population was generated independently from the other one.

\begin{center}
\begin{figure}[ht!]
\hspace{3cm} \includegraphics[scale=0.38]{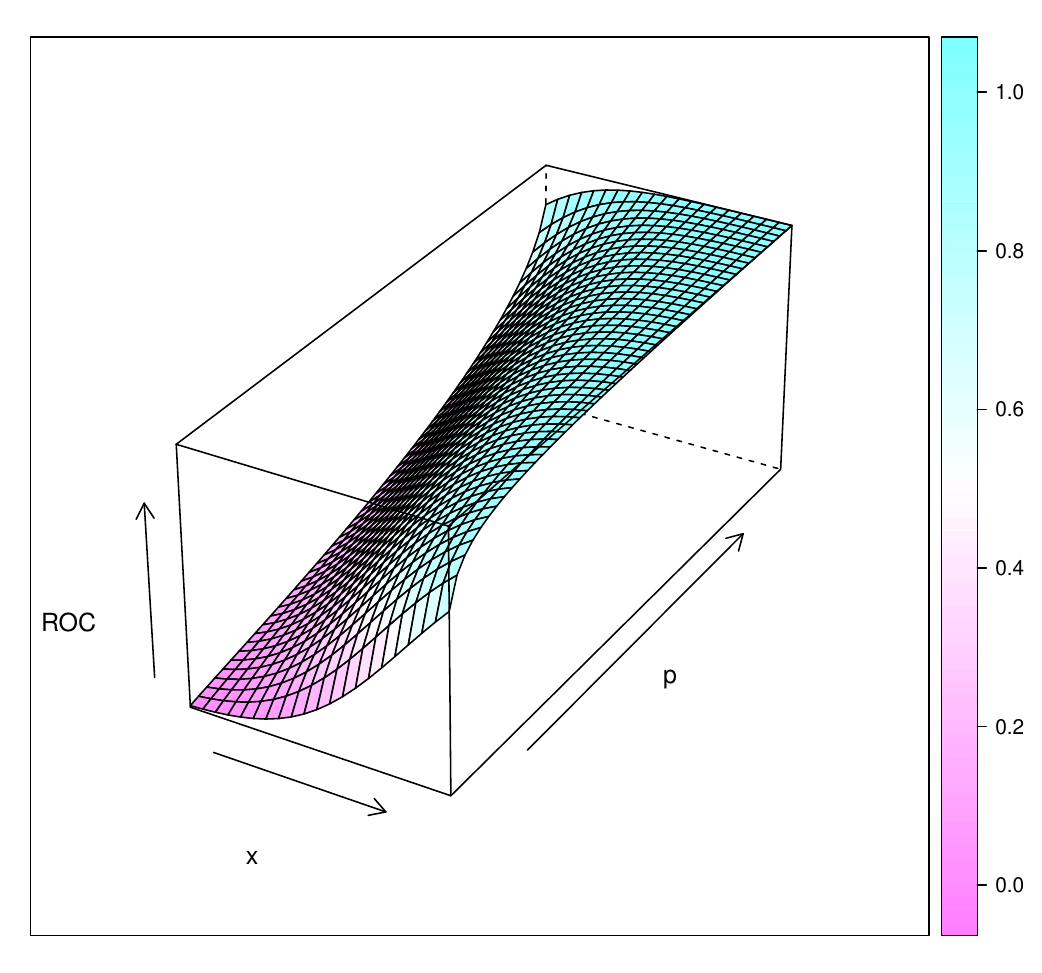}
\vskip-0.1in\caption{\label{fig:surface_true} True ROC surface under the central model given by equations \eqref{trued} and \eqref{trueh} under the   linear model.}
\end{figure}
\end{center}

Figure \ref{fig:surface_true}  displays the surface corresponding to the true ROC curves generated under the central model given by equations \eqref{trued} and \eqref{trueh}.

To evaluate the sensitivity of the classical conditional ROC curve and the robust proposal given in Section \ref{sec:propuesta}, we consider different contamination schemes by varying the sample where we introduce atypical points, the percentage of anomalous data  and the size of the  outliers.

\begin{itemize}
\item $C^{H}_{\delta}$: is a contamination in the healthy sample introduced so as to affect the estimation of quantiles of the healthy population.
In order to introduce atypical observations, we generate \textsl{shift outliers} as follows. The first $m=n \delta$  observations in the healthy dataset were replaced by observations following the model  
$y_{H,i} = 0.5+   x_{H,i} + S \,\sigma_{H} + \sigma_{H} \epsilon_{H,i} $,
where the shift $S\in \itS=\{ 2.5, 5, 7.5, 10, 12.5, 15, 17.5, 20\}$. 
\item $C_{\delta}^D$: corresponds to contaminating  the diseased population  introduced so as to affect the estimation of the empirical distribution of the diseased population. The   atypical observations are introduced in the same fashion as in $C^{H}_{\delta}$, that is, the first $m=n \delta$  observations in the diseased dataset were replaced by observations following the model  
$y_{D,i} = 0.5+   x_{D,i} + S \,\sigma_{D} + \sigma_{D} \epsilon_{D,i} $, 
where the shift  $S\in \itS$. 
\item $C_{\delta}$: we generate now \textsl{shift outliers} in both samples simultaneously. For this end, the first  $m=n \delta$ observations in each dataset were replaced by observations generated as follows
\begin{eqnarray}
y_{D,i} &=& 2+ 4\, x_{D,i} + 20\, \sigma_{D} + \sigma_{D} \epsilon_{D,i} \label{contaminad}\\
y_{H,i} &=& 0.5+   x_{H,i} + 15\, \sigma_{H} + \sigma_{H} \epsilon_{H,i}  \label{contaminah}\; ,
\end{eqnarray}
\end{itemize}
We choose two possible contaminating percentages $\delta=0.05$ and $0.10$, that is, a 5\% or a 10\% of observations are modified, respectively.  To avoid burden notation, in all Figures and Tables, $C_0$ stands for the situation of clean samples.

\begin{figure}[ht!]
 \begin{center}
 \footnotesize
 \renewcommand{\arraystretch}{0.4}
 \newcolumntype{M}{>{\centering\arraybackslash}m{\dimexpr.1\linewidth-1\tabcolsep}}
   \newcolumntype{G}{>{\centering\arraybackslash}m{\dimexpr.4\linewidth-1\tabcolsep}}
\begin{tabular}{M GG}
 & Classical Estimators & Robust Estimators \\
$C_0$ & 
\includegraphics[scale=0.35]{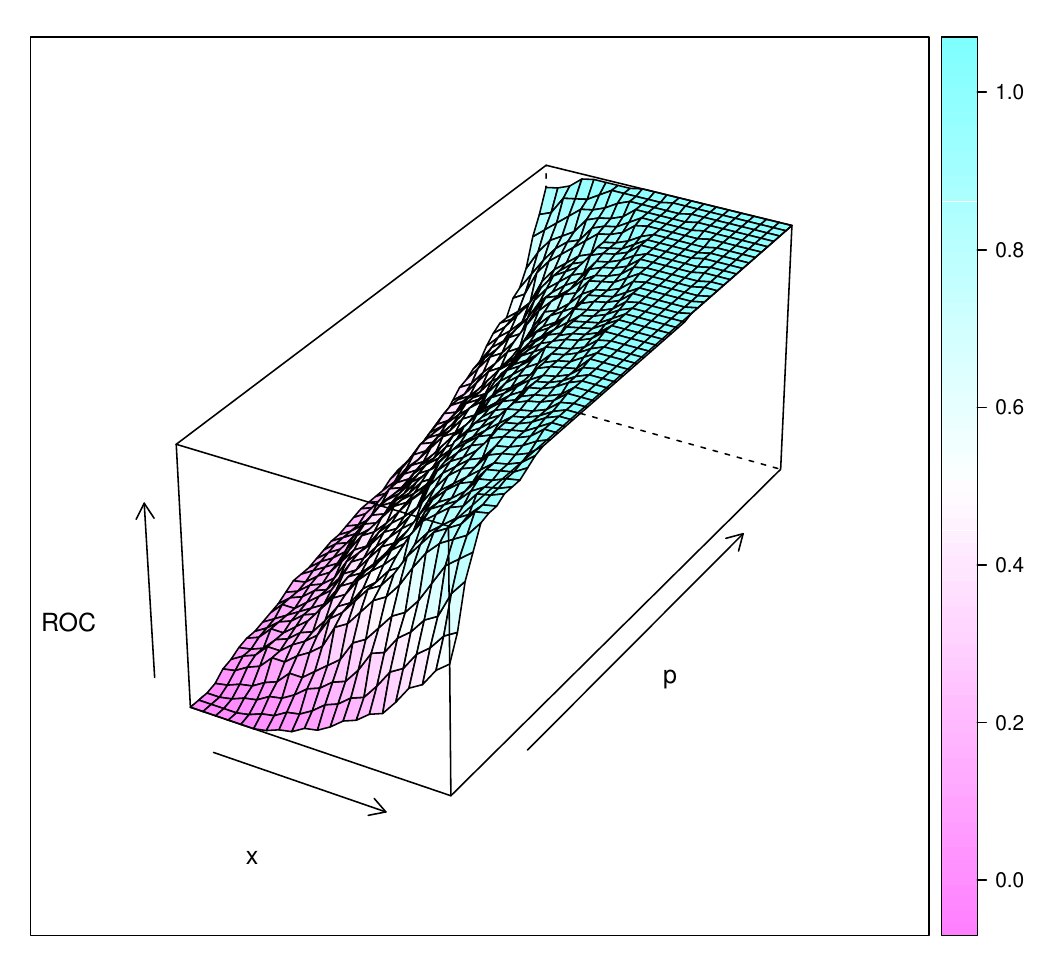} & 
\includegraphics[scale=0.35]{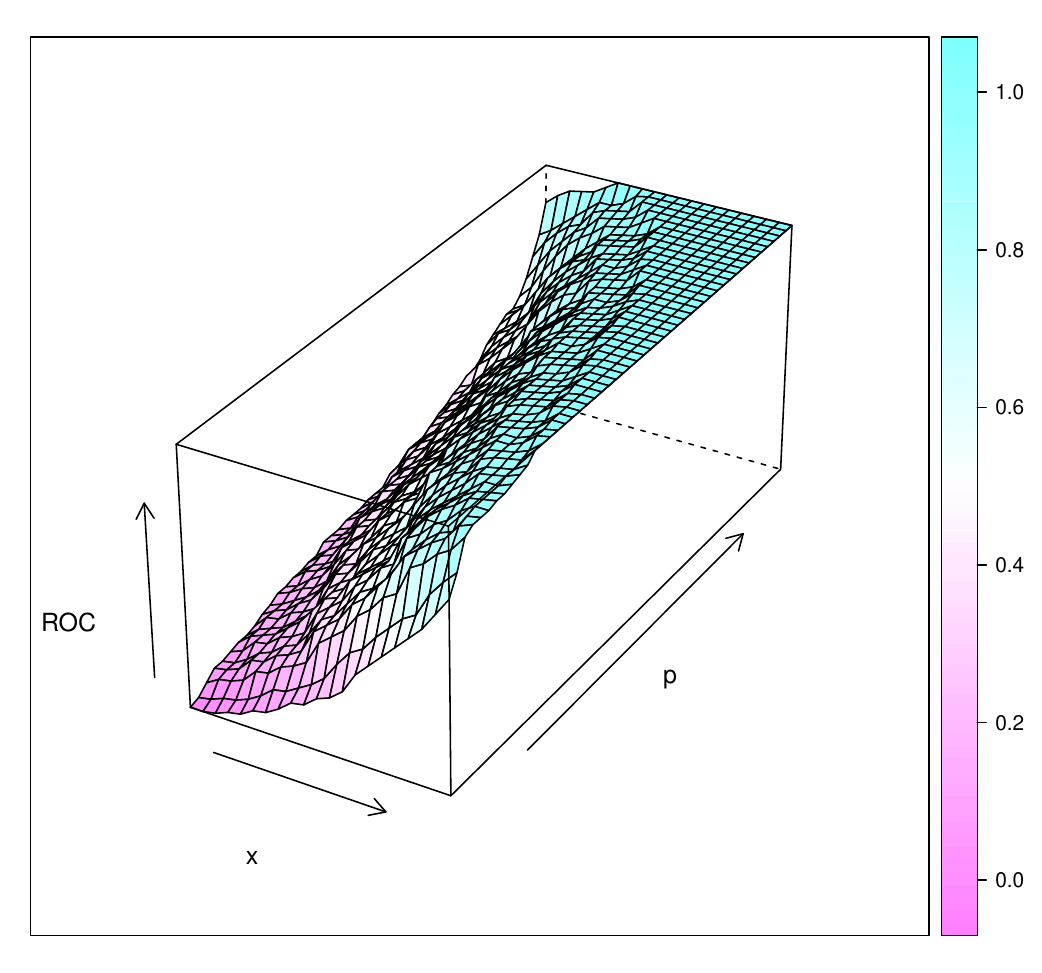} \\
$C_{0.05}$ &  
\includegraphics[scale=0.35]{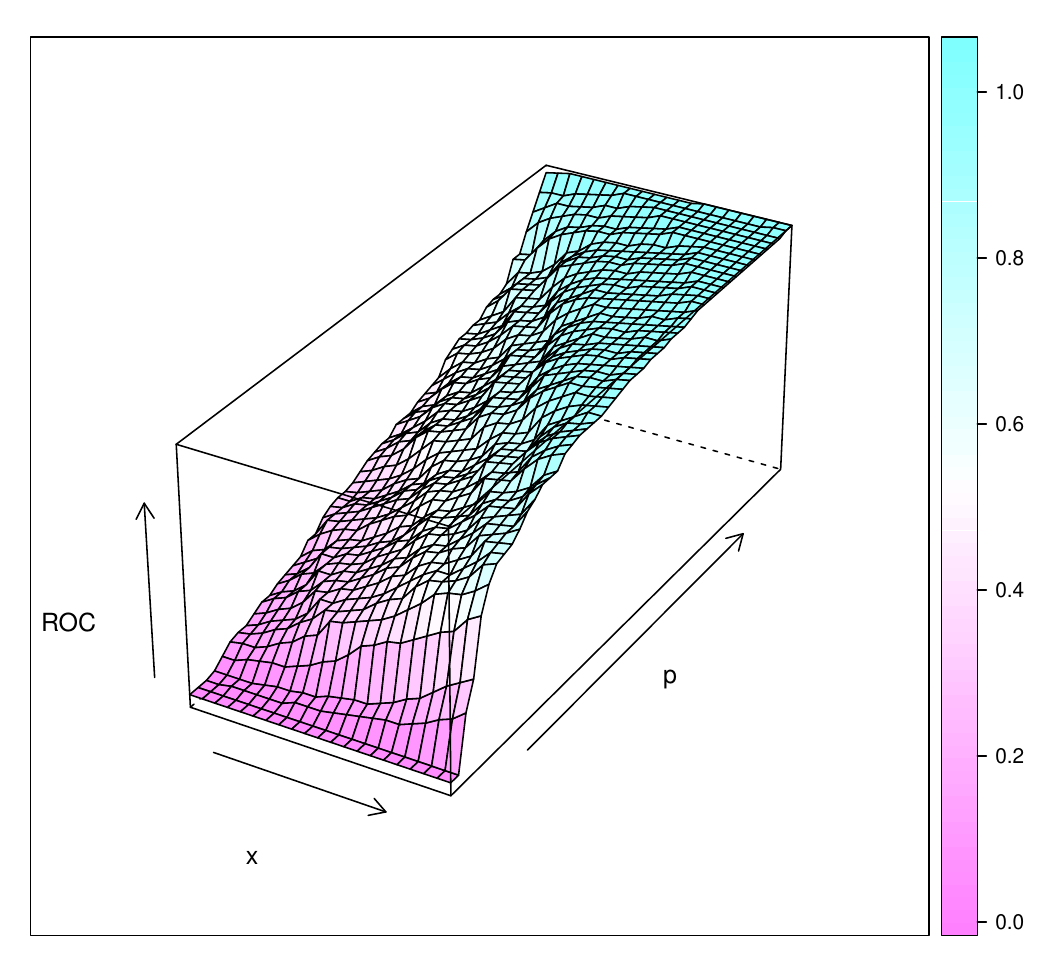} & 
\includegraphics[scale=0.35]{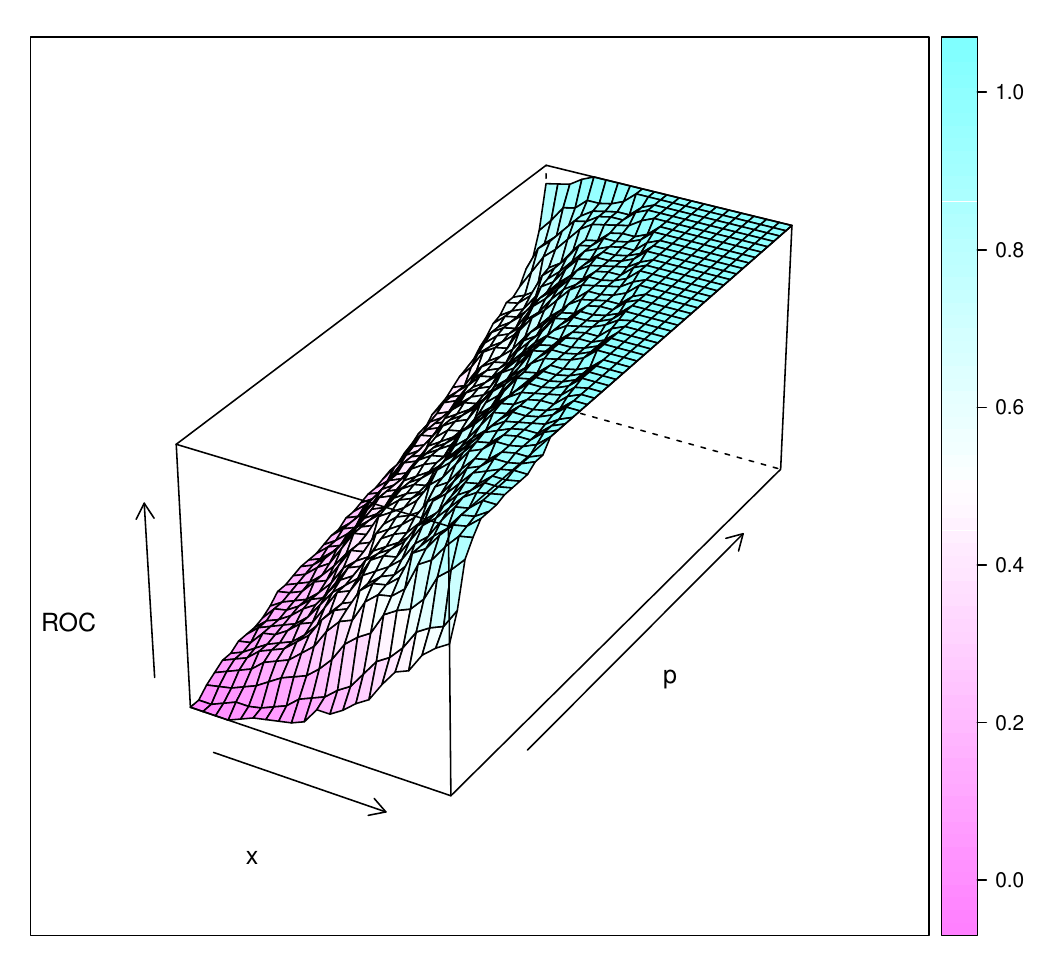}\\
$C_{0.10}$ &  
\includegraphics[scale=0.35]{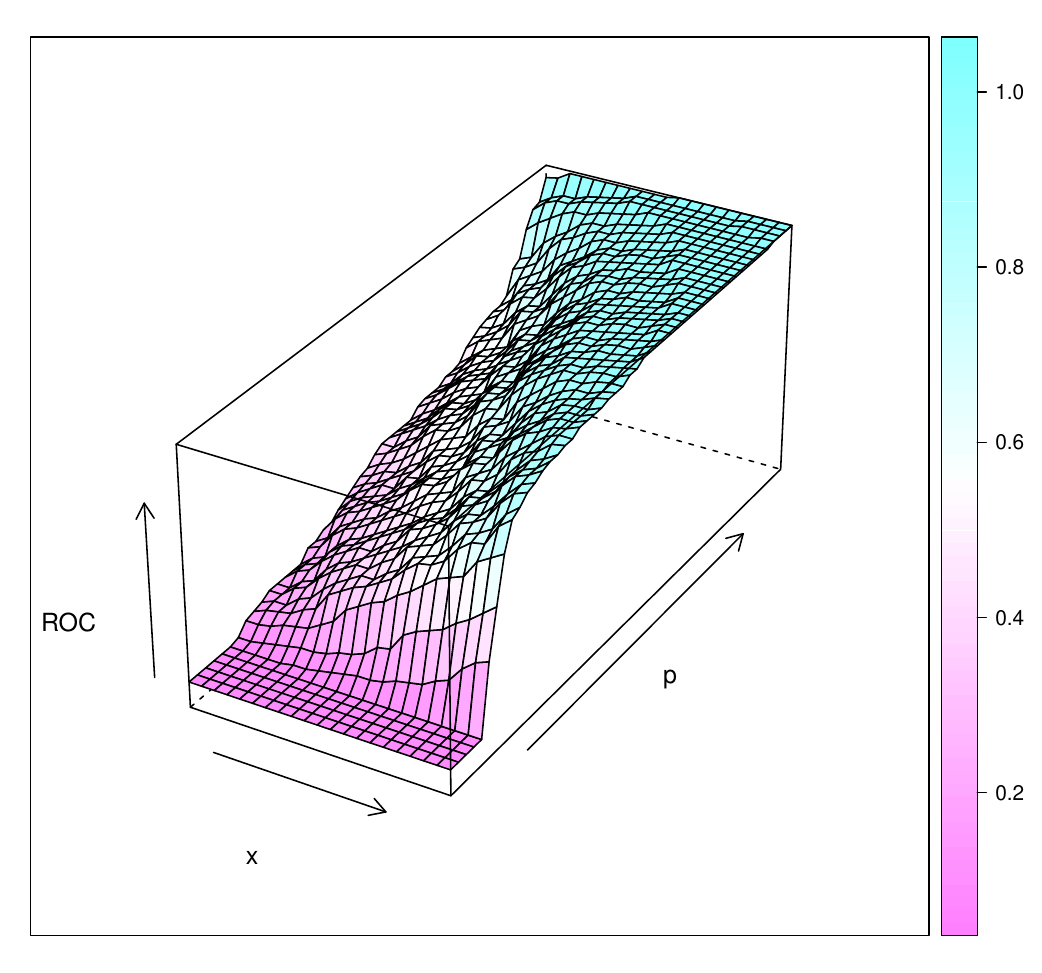} &  
\includegraphics[scale=0.35]{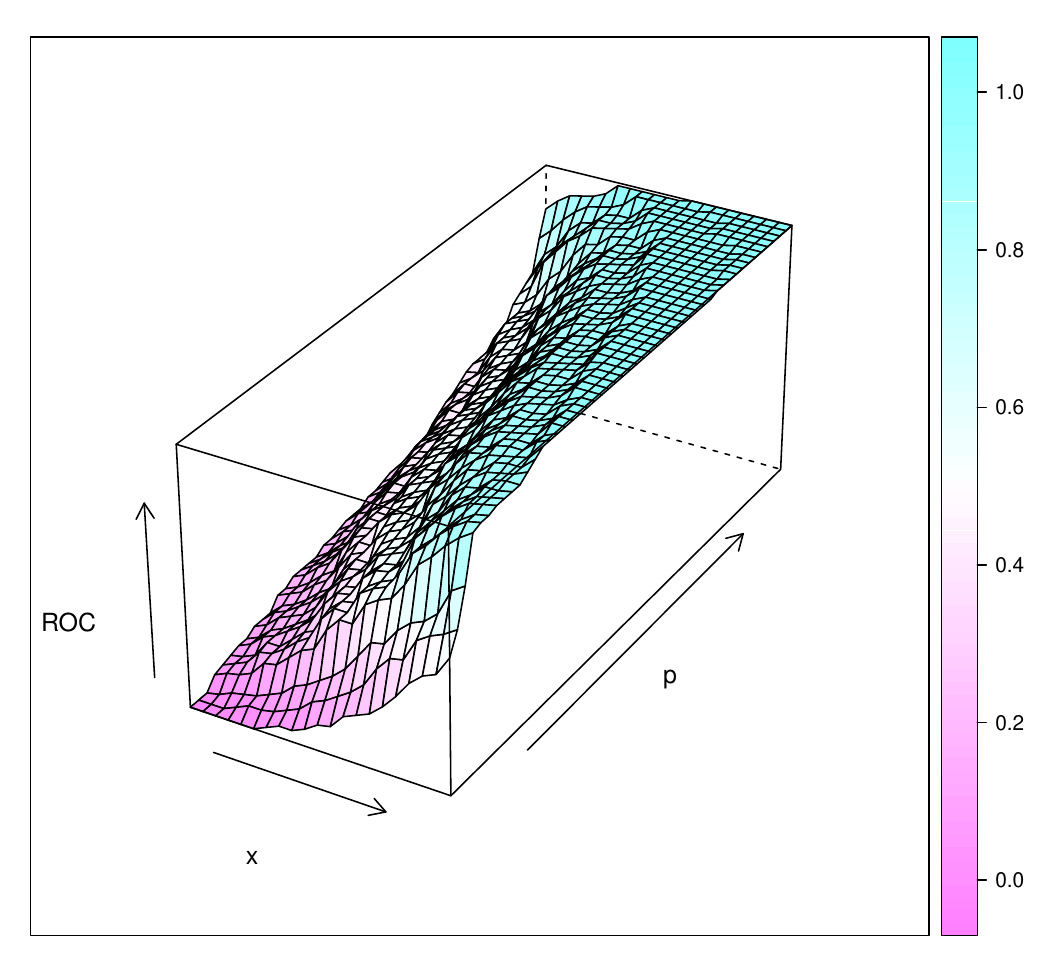}\\
\end{tabular}
\vskip-0.1in  \caption{\label{fig:surface} Estimated surfaces  for $n_{D}=n_{H}=100$ under the   linear model \eqref{trued} and \eqref{trueh} for a clean and contaminated sample.}
\end{center} 
\end{figure}

To illustrate the behaviour of the ROC curves for clean and contaminated samples, Figure \ref{fig:surface}    shows the estimated surfaces obtained with the classical and robust estimators from one of the clean samples generated when  $n=100$ and when the same sample is corrupted with the shifted outliers generated as in equations \eqref{contaminad} and \eqref{contaminah}. The estimators of the conditional $\ROC$ curves were computed on the net of points $\itG_x$ and quantiles $\itG_p$, described above.  The right panel in Figure \ref{fig:surface}  illustrates the stability of the proposed method, since the three figures on the right  panel  are quite similar. On the other hand, the classical estimators are distorted in the presence of outliers, the surface being shifted towards 1 in the central region and flatten towards $0$ specially under $C_{0.10}$.

\normalsize

\begin{figure}[ht!]
 \begin{center}
 \footnotesize
 \renewcommand{\arraystretch}{0.1}
 \newcolumntype{M}{>{\centering\arraybackslash}m{\dimexpr.1\linewidth-1\tabcolsep}}
   \newcolumntype{G}{>{\centering\arraybackslash}m{\dimexpr.4\linewidth-1\tabcolsep}}
\begin{tabular}{M GG}
 & $n=100$ & $n=200$ \\[-2ex]
$C_{0.05}^H$ &  
\includegraphics[scale=0.3]{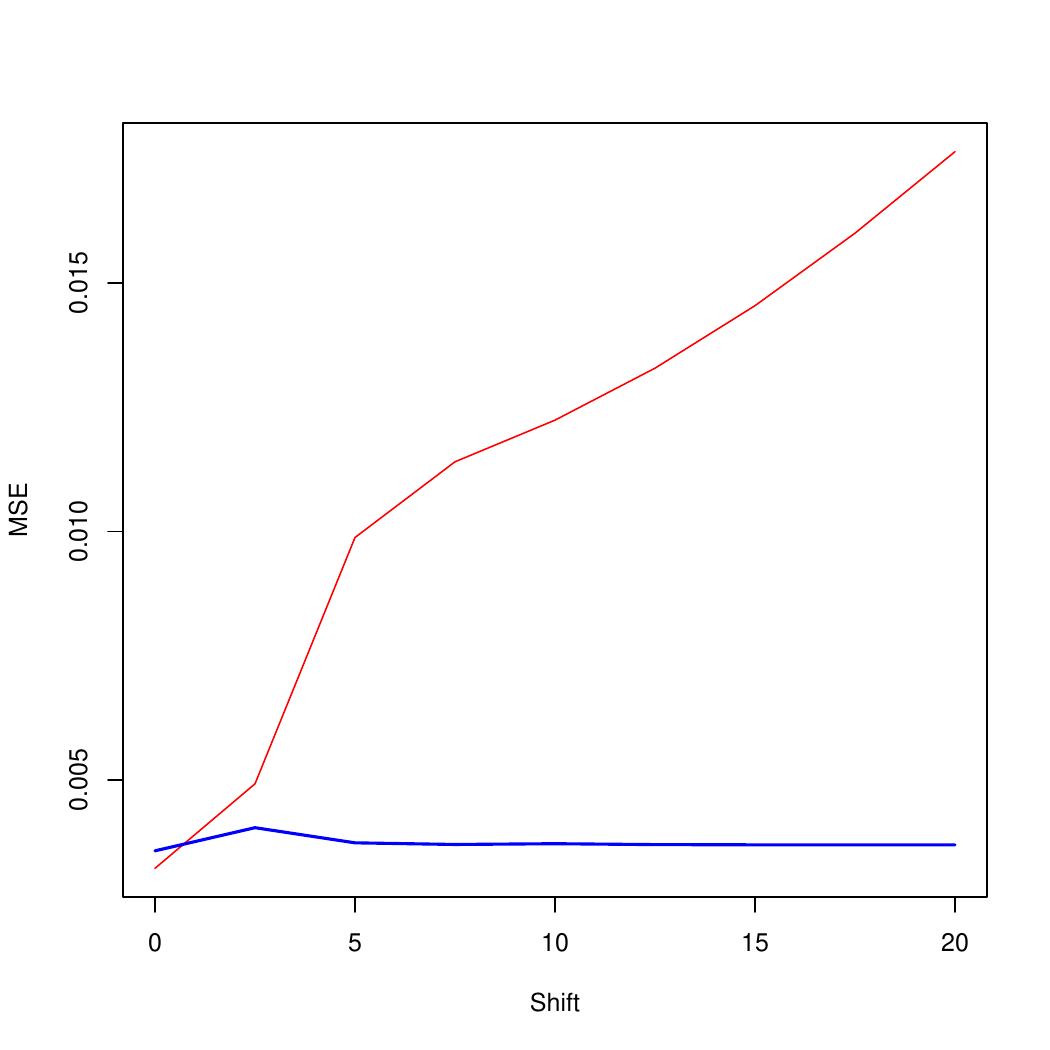} & 
\includegraphics[scale=0.3]{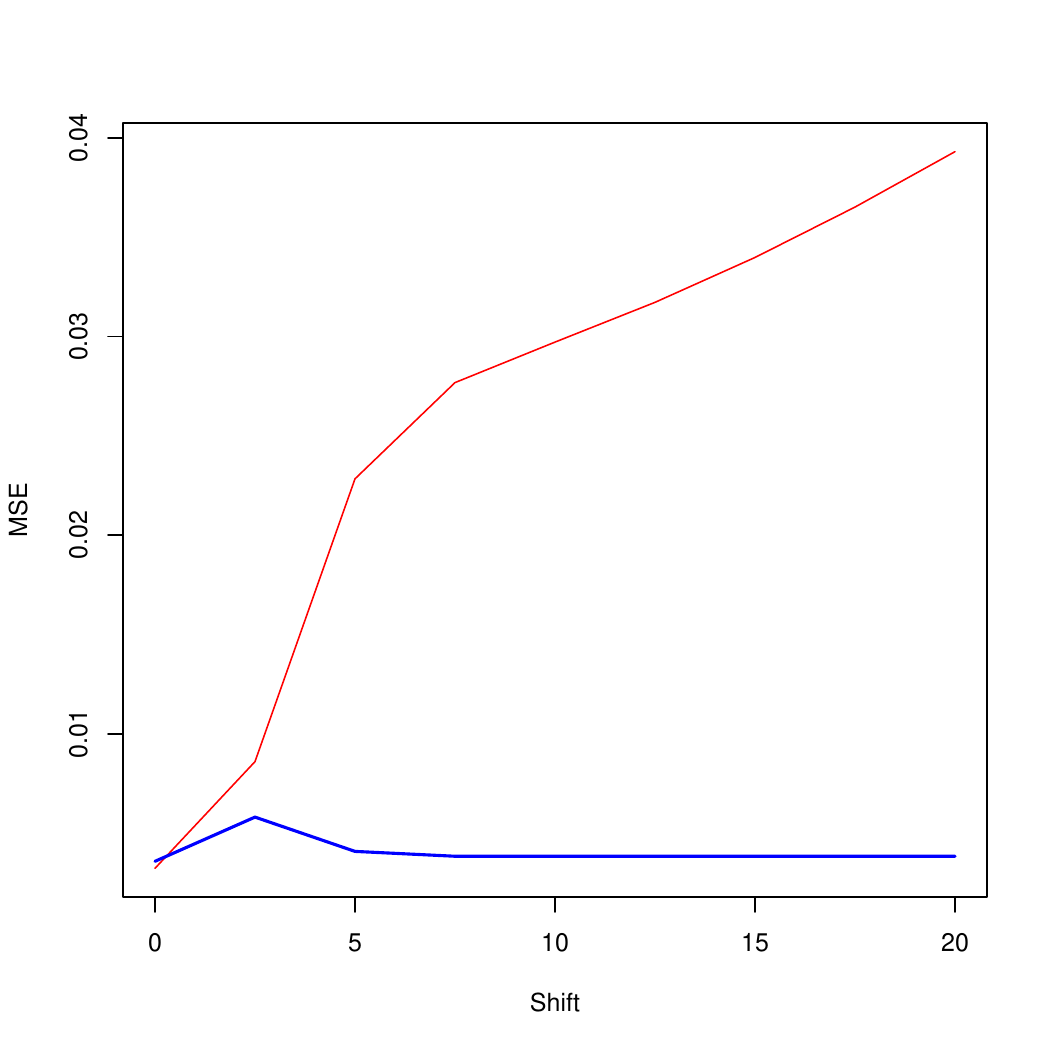}\\[-5ex]
$C_{0.10}^H$ &  
\includegraphics[scale=0.3]{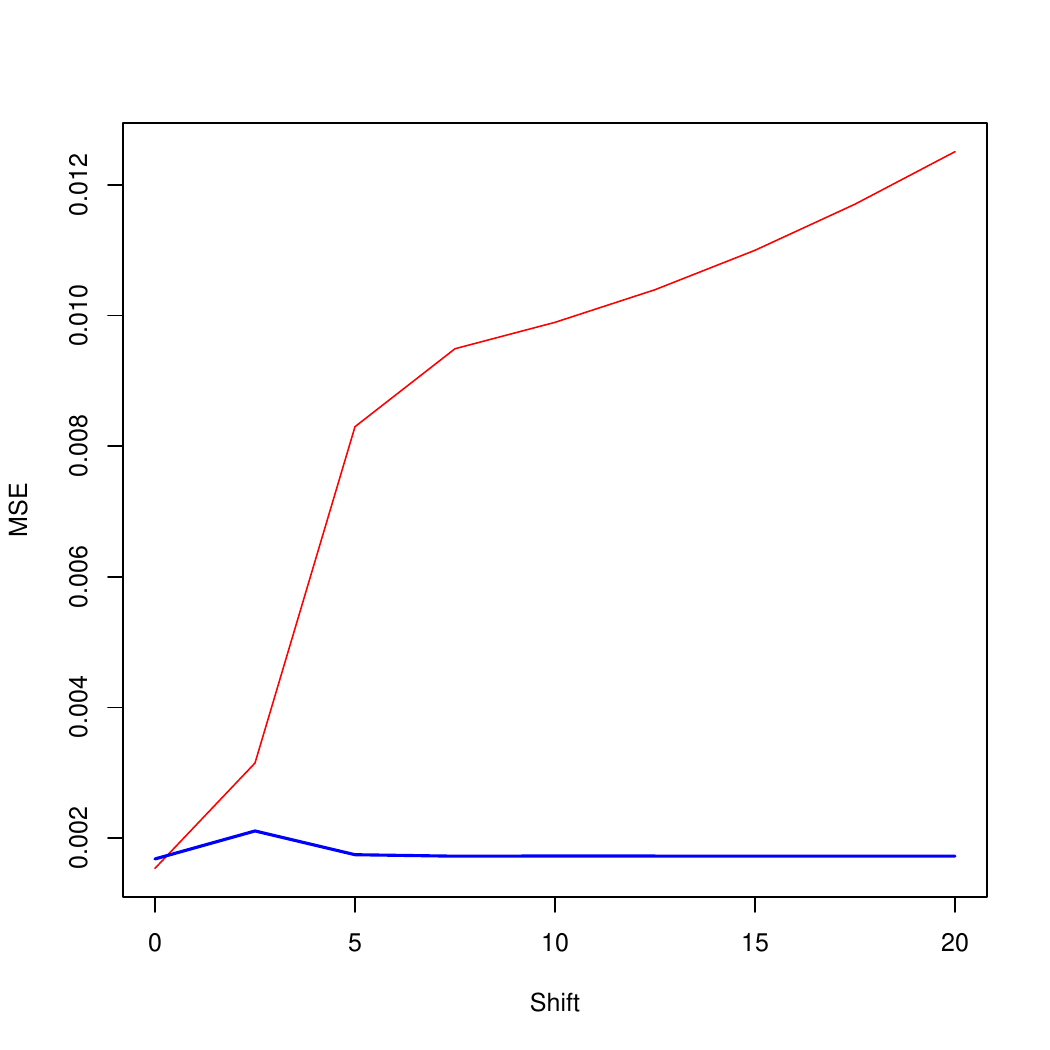} &  
\includegraphics[scale=0.3]{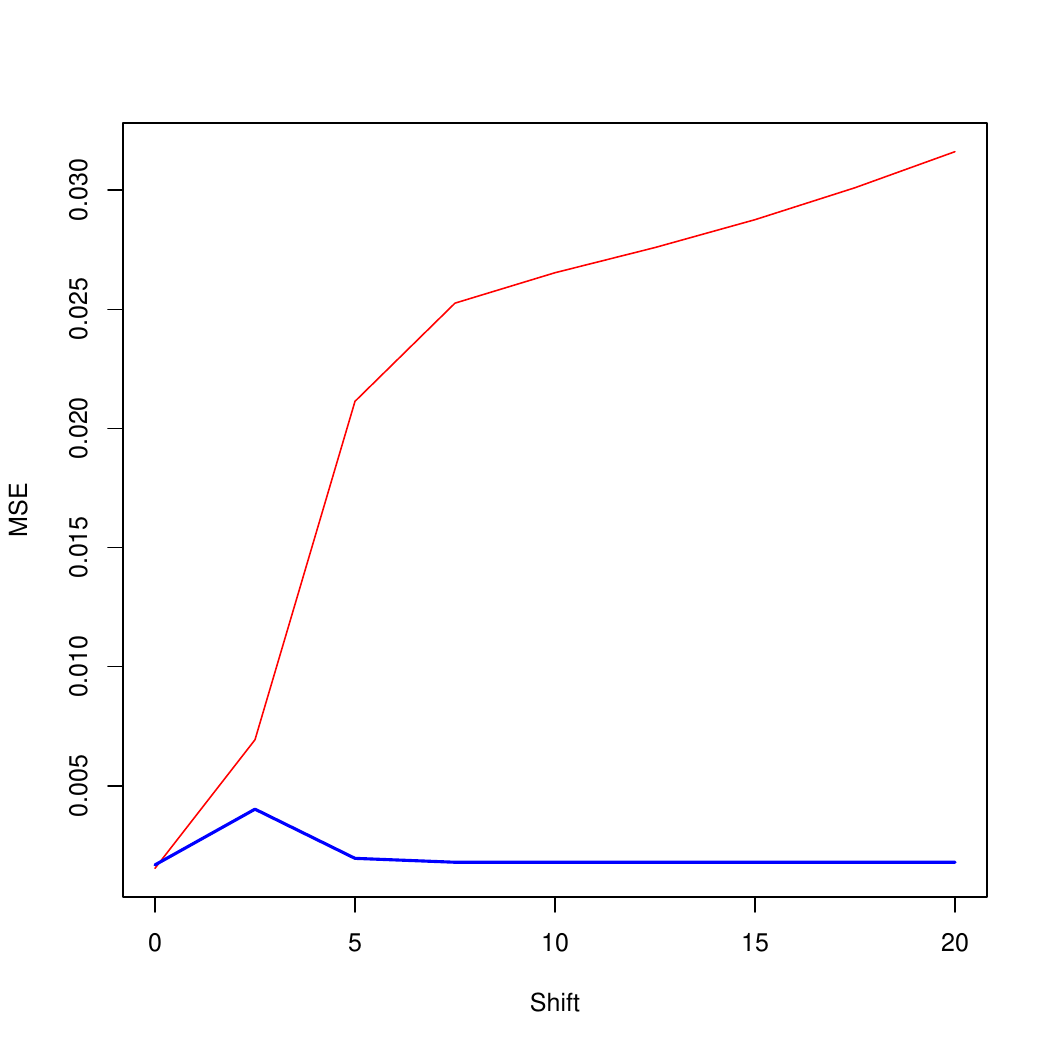} 
\end{tabular}
\vskip-0.1in  \caption{\label{fig:MSE_CH} \footnotesize Sensitivity to shift size $S$: $MSE$ under $C_{\delta}^H$ when $n_{D}=n_{H}=100$ and $n_{D}=n_{H}=200$ for $ \delta=0.05, 0.10 $. The red line corresponds to the classical procedure, while the blue one the robust one.}
\end{center} 
\end{figure}

\begin{figure}[ht!]
 \begin{center}
 \footnotesize
 \renewcommand{\arraystretch}{0.2}
 \newcolumntype{M}{>{\centering\arraybackslash}m{\dimexpr.1\linewidth-1\tabcolsep}}
   \newcolumntype{G}{>{\centering\arraybackslash}m{\dimexpr.4\linewidth-1\tabcolsep}}
\begin{tabular}{M GG}
 & $n=100$ & $n=200$ \\[-2ex]
$C_{0.05}^D$ &  
\includegraphics[scale=0.3]{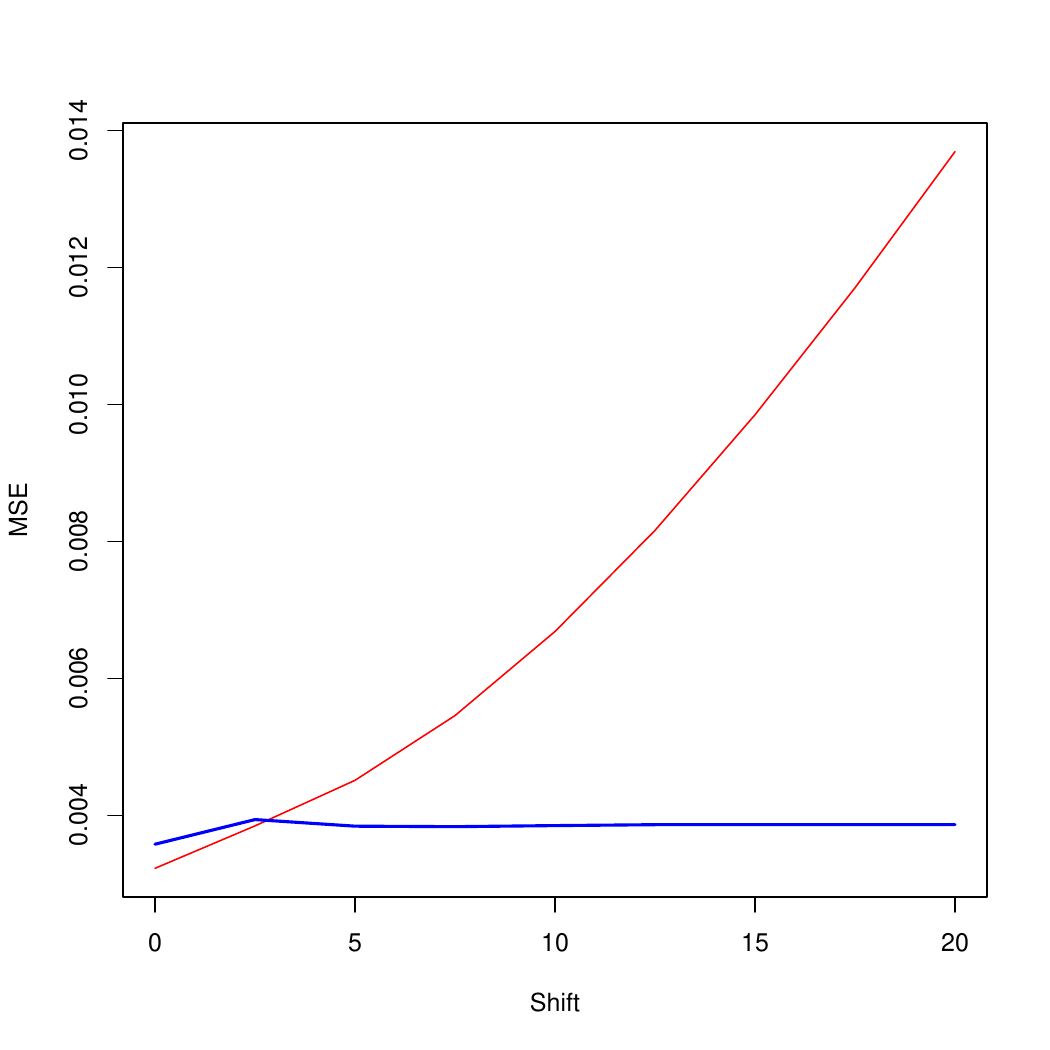} & 
\includegraphics[scale=0.3]{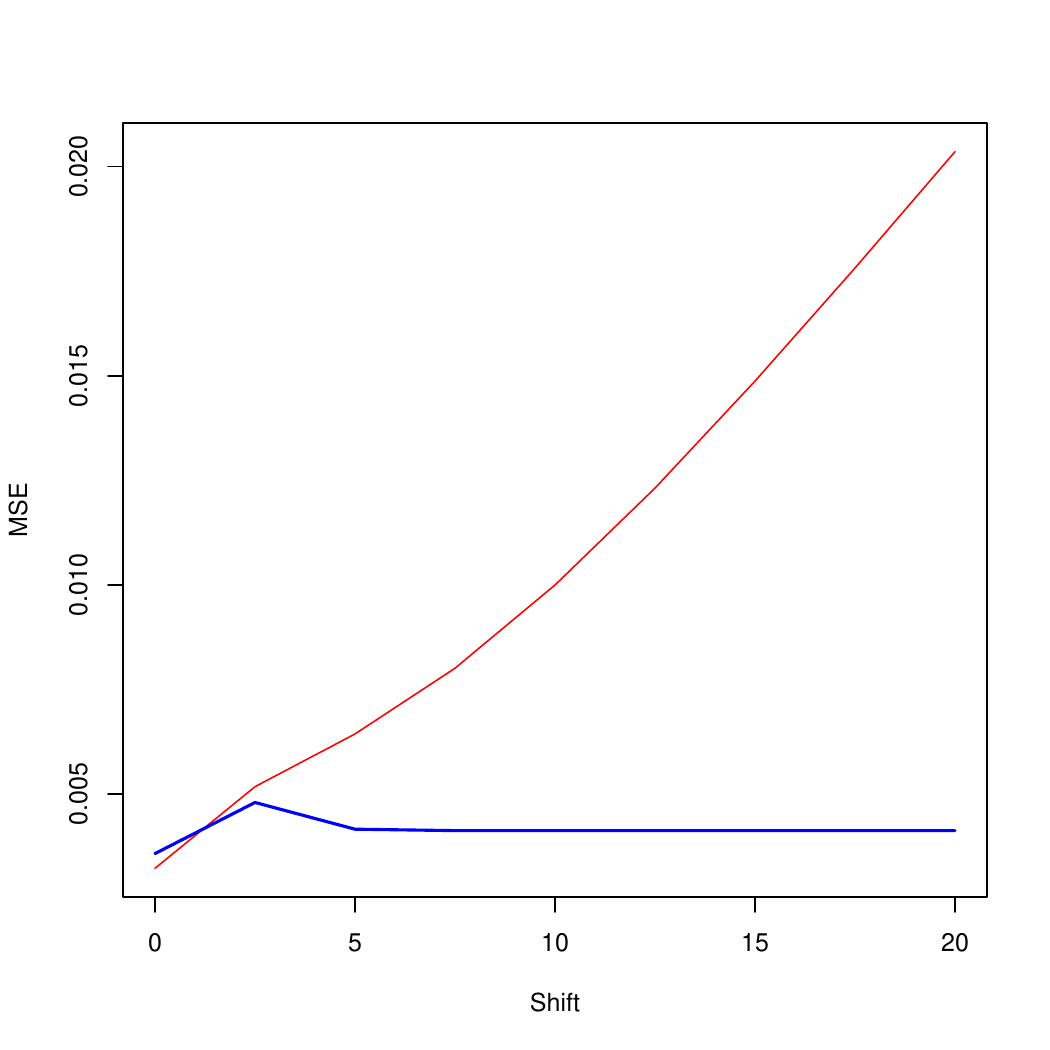}\\[-5ex]
$C_{0.10}^D$ &  
\includegraphics[scale=0.3]{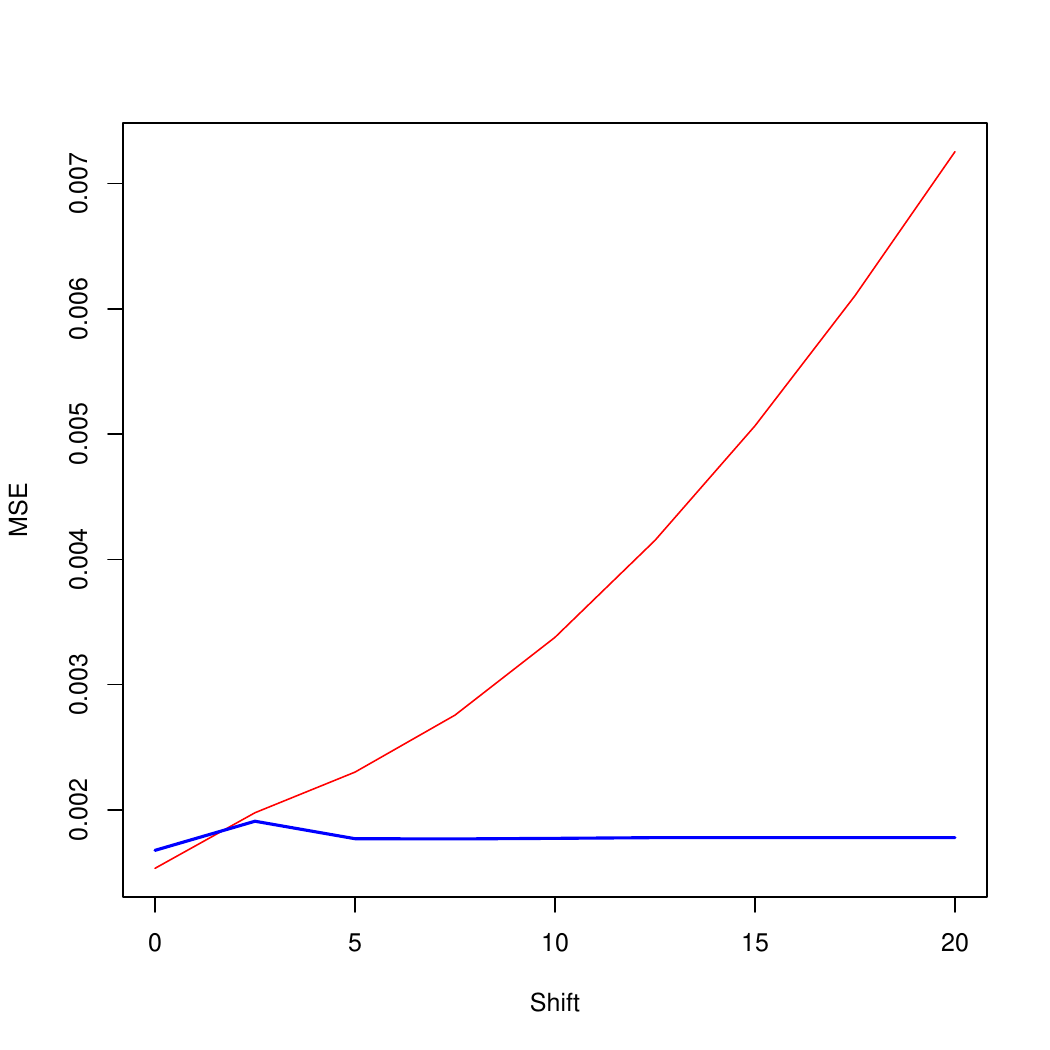} &  
\includegraphics[scale=0.3]{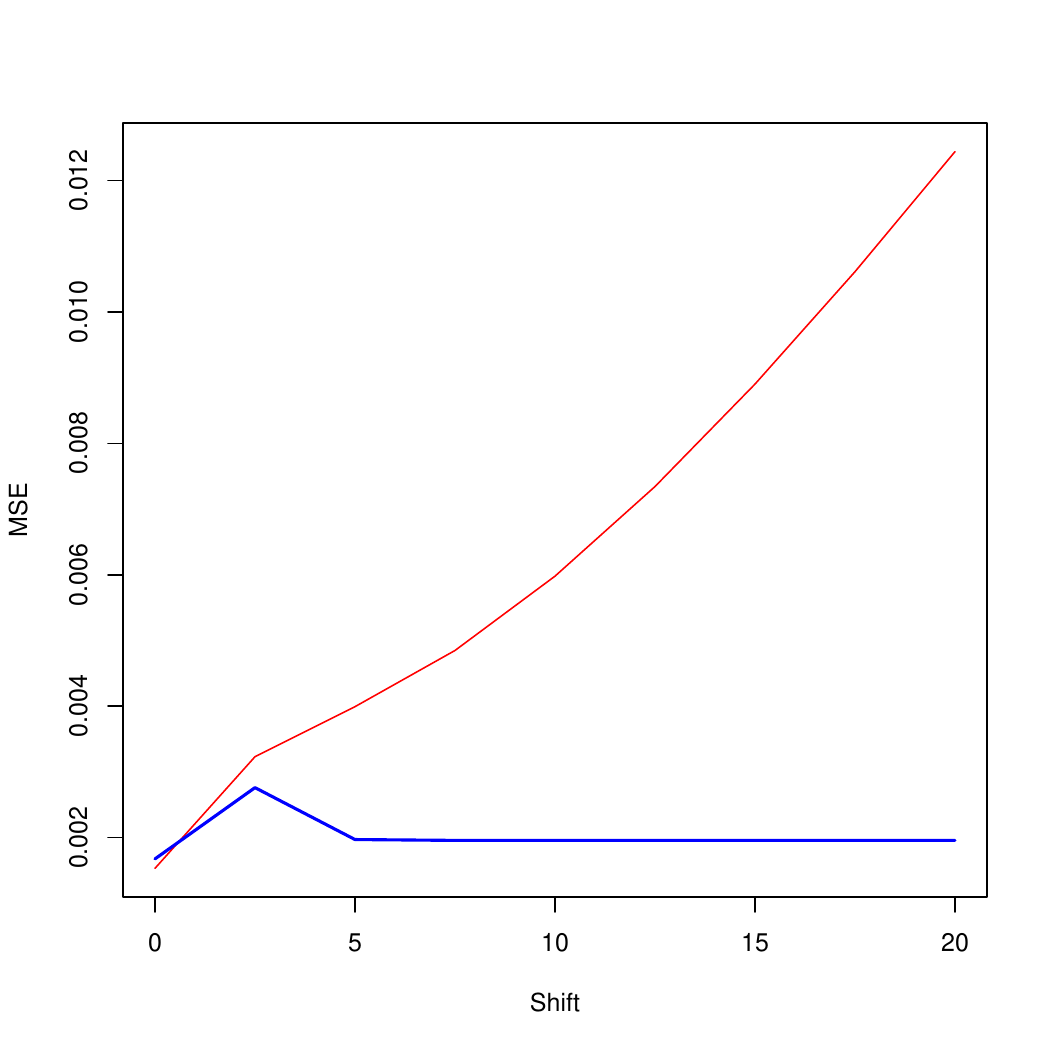}\\
\end{tabular}
\vskip-0.1in  \caption{\label{fig:MSE_CD} \footnotesize Sensitivity to the shift  size  $S$: $MSE$ under $C_{\delta}^D$ when $n_{D}=n_{H}=100$ and $n_{D}=n_{H}=200$ for $ \delta=0.05, 0.10 $. The red line corresponds to the classical procedure, while the blue one the robust one.}
\end{center} 
\end{figure}

To evaluate the effect of the considered contaminations, Tables \ref{tab:sensi_5_100_H} to \ref{tab:sensi_10_200_H},    report the summary measures under $C_{\delta}^H$ and $C_{\delta}^D$ for $\delta=0.05, 0.10$ and $n=100, 200$. It is worth noticing that $MSE$ and $KS$ take values between 0 and 1 and in this range, large deviations correspond to values close to $1$. The reported results show that the classical procedure to estimate the ROC curve is seriously affected by the introduced outliers. It should be taken into account that since the ROC curve varies between $0$ and $1$, the magnitude of the effect is not as evident as in other settings such as in linear regression models. However, when $n=100$, under $C_{0.05}^H$, the $MSE$ is $5.5$ larger when $S=20$ than for clean samples, while the robust procedure remains stable. This effect is more striking in Figures \ref{fig:MSE_CH} and  \ref{fig:MSE_CD} which  show the plot of the $MSE$ as a function of the level shift $S$ when $n=100$ and $200$ and for the two contamination percentages. The red and blue lines correspond to the classical  and robust proposed methods, respectively. Even though a slight influence is observed for the robust procedure under mild outliers ($S=2.5$), which are those more difficult to detect, the whole curve is stable when varying $S$, while the $MSE$ of the classical method quickly increases with the level shift.

\begin{table}[ht!]
\begin{center}
\small 
\begin{tabular}{|c|c|c|c|c|c|c|c|c|c|c|}
 \hline
 &&&\multicolumn{8}{c|}{$S$}\\
\hline
& Method  & $C_0$ &  2.5  &  5  &  7.5  &  10  &  12.5  &  15 & 17.5 & 20\\ 
\hline
& & & \multicolumn{8}{c|}{$C_{0.05}^H$}\\\hline
 $MSE$ & Robust & 0.0036 & 0.0040 & 0.0037 & 0.0037 & 0.0037 & 0.0037 & 0.0037 & 0.0037 & 0.0037 \\  
   & Classical   & 0.0032 & 0.0049 & 0.0099 & 0.0114 & 0.0122 & 0.0133 & 0.0145 & 0.0160 & 0.0176 \\ 
 \hline
 $KS$ & Robust & 0.1988 & 0.2156 & 0.2085 & 0.2054 & 0.2056 & 0.2024 & 0.2016 & 0.2016 & 0.2016 \\ 
 &  Classical & 0.1949 & 0.3567 & 0.7172 & 0.8189 & 0.8256 & 0.8256 & 0.8256 & 0.8257 & 0.8258 \\ 
\hline
& & & \multicolumn{8}{c|}{$C_{0.05}^D$}\\\hline
$MSE$ & Robust & 0.0036 & 0.0039 & 0.0038 & 0.0038 & 0.0039 & 0.0039 & 0.0039 & 0.0039 & 0.0039 \\  
    & Classical & 0.0032 & 0.0038 & 0.0045 & 0.0055 & 0.0067 & 0.0082 & 0.0098 & 0.0117 & 0.0137 \\ 
 \hline
 $KS$ & Robust & 0.1988 & 0.2041 & 0.2035 & 0.2034 & 0.2037 & 0.2040 & 0.2040 & 0.2040 & 0.2040 \\  
 &  Classical  & 0.1949 & 0.2007 & 0.2130 & 0.2279 & 0.2457 & 0.2640 & 0.2829 & 0.3015 & 0.3196 \\ 
 \hline
\end{tabular}
\end{center}
\caption{\small \label{tab:sensi_5_100_H} \footnotesize Sensitivity to the shift  size  $S$  for $C_{0.05}^H$ and $C_{0.05}^D$ when $n=100$.} 
\end{table}

\begin{table}[H]
\begin{center}
\small 
\begin{tabular}{|c|c|c|c|c|c|c|c|c|c|c|}
 \hline
 &&&\multicolumn{8}{c|}{$S$}\\
\hline
& Method  & $C_0$ & 2.5  &  5  &  7.5  &  10  &  12.5  &  15  &  17.5  &  20  \\ 
\hline
& & & \multicolumn{8}{c|}{$C_{0.10}^H$}\\\hline
 $MSE$ & Robust & 0.0036 & 0.0058 & 0.0041 & 0.0038 & 0.0038 & 0.0038 & 0.0038 & 0.0038 & 0.0038 \\ 
  & Classical & 0.0032 & 0.0086 & 0.0228 & 0.0277 & 0.0297 & 0.0317 & 0.0340 & 0.0365 & 0.0393 \\ 
 \hline
 $KS$ & Robust & 0.1988 & 0.2727 & 0.2411 & 0.2128 & 0.2128 & 0.2128 & 0.2128 & 0.2128 & 0.2128 \\ 
  &  Classical & 0.1949 & 0.4406 & 0.7832 & 0.8834 & 0.8946 & 0.8985 & 0.9007 & 0.9013 & 0.9015 \\ 
 \hline
 & & & \multicolumn{8}{c|}{$C_{0.10}^D$}\\\hline
  $MSE$ & Robust  & 0.0036 & 0.0048 & 0.0042 & 0.0041 & 0.0041 & 0.0041 & 0.0041 & 0.0041 & 0.0041 \\ 
 &  Classical & 0.0032 & 0.0052 & 0.0064 & 0.0080 & 0.0100 & 0.0123 & 0.0149 & 0.0176 & 0.0204 \\ 
\hline
 $KS$ & Robust & 0.1988 & 0.2135 & 0.2085 & 0.2083 & 0.2083 & 0.2083 & 0.2083 & 0.2083 & 0.2083 \\ 
  & Classical  & 0.1949 & 0.2125 & 0.2306 & 0.2522 & 0.2761 & 0.2998 & 0.3229 & 0.3450 & 0.3652 \\ 
 \hline
\end{tabular}
\end{center}
\caption{\small \label{tab:sensi_10_100_H} \footnotesize Sensitivity to the shift  size  $S$  for $C_{0.10}^H$ and $C_{0.10}^D$ when $n=100$.} 
\end{table}
\normalsize

\begin{table}[H]
\begin{center}
\small 
\begin{tabular}{|c|c|c|c|c|c|c|c|c|c|c|}
 \hline
 &&&\multicolumn{8}{c|}{$S$}\\
\hline
& Method  & $C_0$ &  2.5  &  5  &  7.5  &  10  &  12.5  &  15 & 17.5 & 20\\ 
\hline
& & & \multicolumn{8}{c|}{$C_{0.05}^H$}\\\hline
 $MSE$ & Robust & 0.0017 & 0.0021 & 0.0017 & 0.0017 & 0.0017 & 0.0017 & 0.0017 & 0.0017 & 0.0017 \\ 
 &  Classical  & 0.0015 & 0.0031 & 0.0083 & 0.0095 & 0.0099 & 0.0104 & 0.0110 & 0.0117 & 0.0125 \\ 
 \hline
  $KS$ & Robust &  0.1380 & 0.1654 & 0.1463 & 0.1419 & 0.1422 & 0.1413 & 0.1411 & 0.1411 & 0.1411 \\ 
 &  Classical & 0.1363 & 0.3248 & 0.7169 & 0.8207 & 0.8256 & 0.8256 & 0.8256 & 0.8256 & 0.8256 \\ 
 \hline
 & & & \multicolumn{8}{c|}{$C_{0.05}^D$}\\\hline
 $MSE$ & Robust  & 0.0017 & 0.0019 & 0.0018 & 0.0018 & 0.0018 & 0.0018 & 0.0018 & 0.0018 & 0.0018 \\ 
  & Classical & 0.0015 & 0.0020 & 0.0023 & 0.0028 & 0.0034 & 0.0042 & 0.0051 & 0.0061 & 0.0073 \\ 
\hline
 $KS$ & Robust & 0.1380 & 0.1428 & 0.1407 & 0.1406 & 0.1407 & 0.1408 & 0.1408 & 0.1408 & 0.1408 \\ 
  & Classical & 0.1363 & 0.1434 & 0.1514 & 0.1627 & 0.1759 & 0.1903 & 0.2049 & 0.2199 & 0.2349 \\ 
 \hline 
\end{tabular}
\end{center}
\caption{\small \label{tab:sensi_5_200_H} \footnotesize Sensitivity to the shift  size  $S$  for $C_{0.05}^H$  and $C_{0.05}^D$ when $n=200$.} 
\end{table}

\begin{table}[H]
\begin{center}
\small 
\begin{tabular}{|c|c|c|c|c|c|c|c|c|c|c|}
 \hline
 &&&\multicolumn{8}{c|}{$S$}\\
\hline
& Method  & $C_0$ &  2.5  &  5  &  7.5  &  10  &  12.5  &  15 & 17.5 & 20\\ 
\hline
& & & \multicolumn{8}{c|}{$C_{0.10}^H$}\\\hline
 $MSE$ & Robust  & 0.0017 & 0.0040 & 0.0020 & 0.0018 & 0.0018 & 0.0018 & 0.0018 & 0.0018 & 0.0018 \\ 
   & Classical  & 0.0015 & 0.0069 & 0.0211 & 0.0253 & 0.0265 & 0.0276 & 0.0288 & 0.0301 & 0.0316 \\ 
 \hline
 $KS$ & Robust & 0.1380 & 0.2470 & 0.1831 & 0.1465 & 0.1465 & 0.1465 & 0.1465 & 0.1465 & 0.1465 \\ 
 &  Classical  & 0.1363 & 0.4269 & 0.7829 & 0.8845 & 0.8937 & 0.8981 & 0.9008 & 0.9012 & 0.9013 \\ 
 \hline
 & & & \multicolumn{8}{c|}{$C_{0.10}^D$}\\\hline
  $MSE$ & Robust & 0.0017 & 0.0028 & 0.0020 & 0.0020 & 0.0020 & 0.0020 & 0.0020 & 0.0020 & 0.0020 \\ 
  & Classical  & 0.0015 & 0.0032 & 0.0040 & 0.0048 & 0.0060 & 0.0073 & 0.0089 & 0.0106 & 0.0124 \\ 
\hline
 $KS$ & Robust & 0.1380 & 0.1585 & 0.1461 & 0.1458 & 0.1458 & 0.1458 & 0.1458 & 0.1458 & 0.1458 \\ 
 &  Classical  & 0.1363 & 0.1620 & 0.1766 & 0.1930 & 0.2109 & 0.2296 & 0.2482 & 0.2662 & 0.2842 \\ 
 \hline
\end{tabular}
\end{center}
\caption{\small \label{tab:sensi_10_200_H} \footnotesize Sensitivity to the shift  size  $S$  for $C_{0.10}^H$  and $C_{0.10}^D$ when $n=200$.} 
\end{table}
\normalsize

\begin{figure}[ht!]
\begin{center}
\footnotesize
\renewcommand{\arraystretch}{0.4}
\newcolumntype{M}{>{\centering\arraybackslash}m{\dimexpr.1\linewidth-1\tabcolsep}}
  \newcolumntype{G}{>{\centering\arraybackslash}m{\dimexpr.5\linewidth-1\tabcolsep}}
\begin{tabular}{G G}
$n=100$ & $n=200$\\
\includegraphics[scale=0.45]{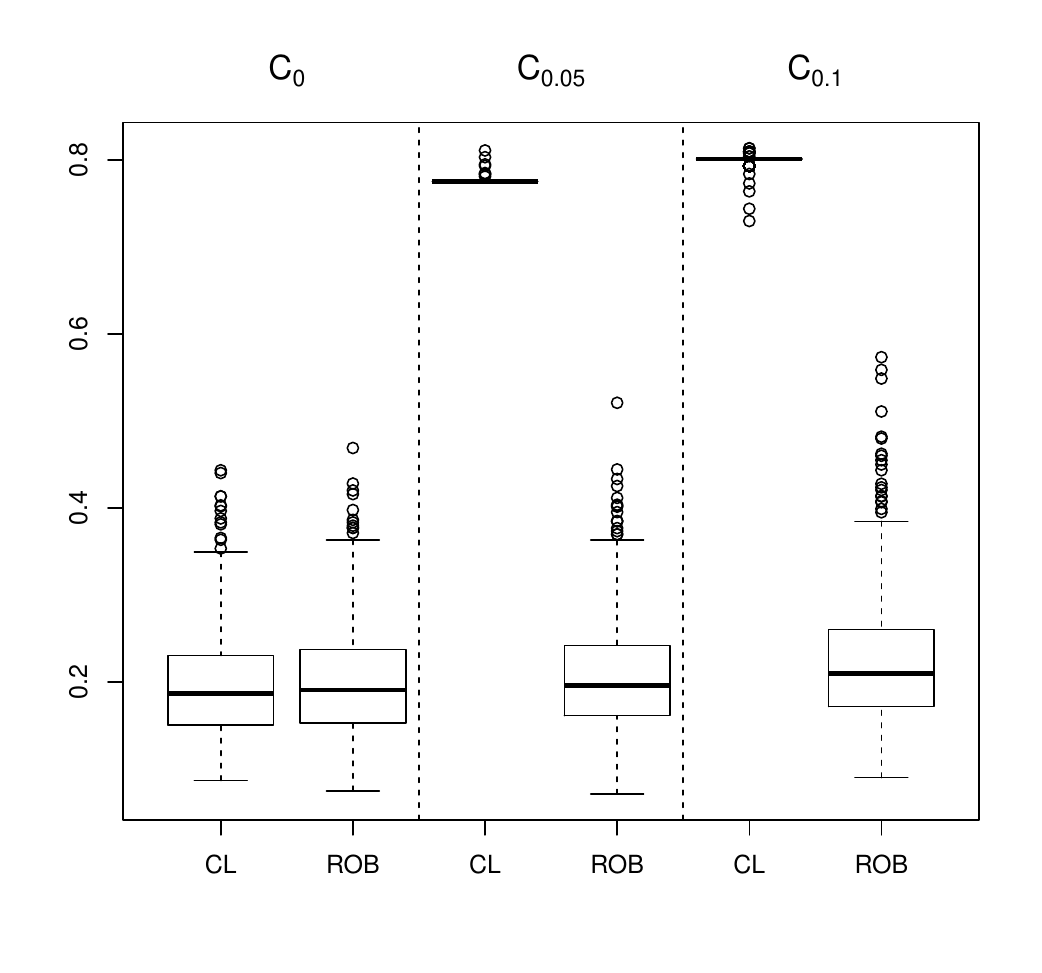} & 
\includegraphics[scale=0.45]{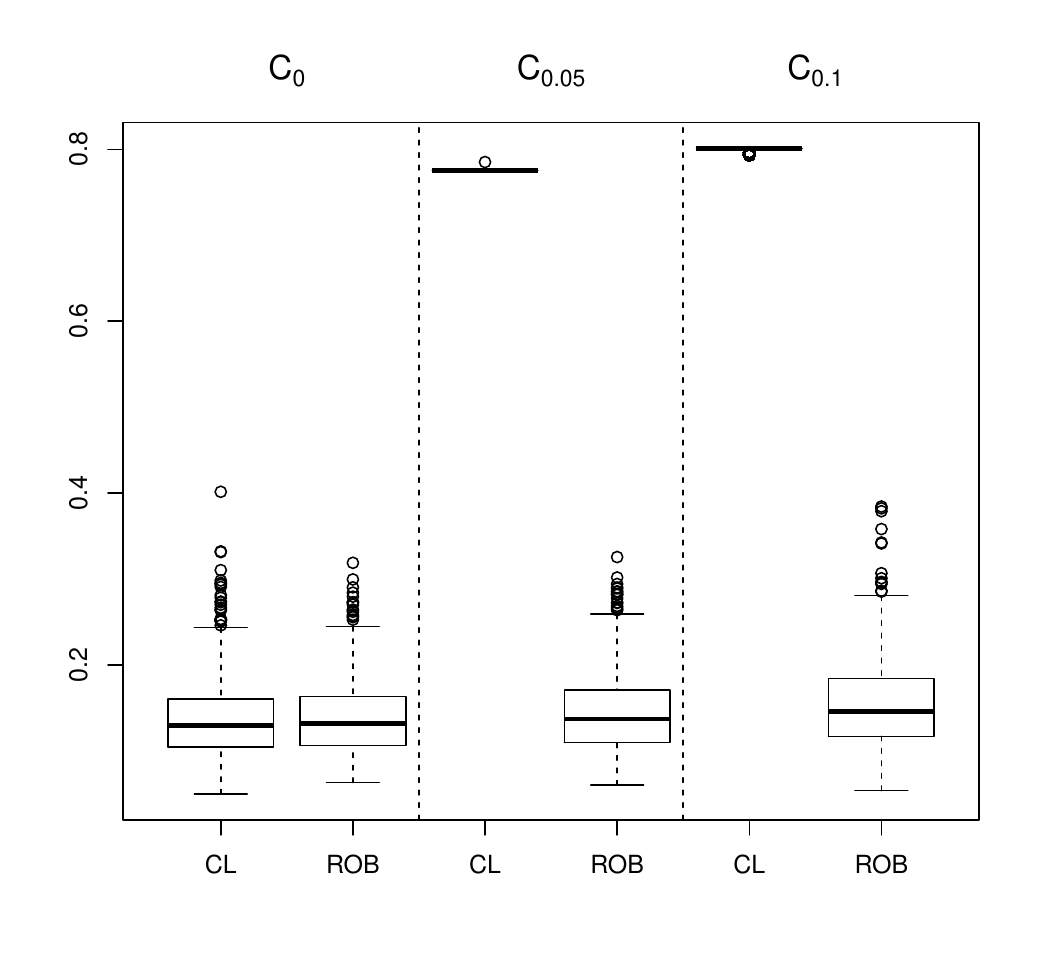}
\end{tabular}
\caption{\label{fig:KS_bxp} Boxplots of the measure  $KS$ obtained from 1000 replications using the classical and robust estimators  under the linear model \eqref{trued} and \eqref{trueh}.}
\end{center}
\end{figure}

Tables \ref{tab:mse}   summarizes the  results obtained when both samples are contaminated. As above, the reported results correspond to   the mean  of $MSE$ and $KS$ over 1000 replications.  As when contaminating only one population, the mean over replications of  $MSE$ for  the classical procedure is clearly enlarged under $C_\delta$, while those corresponding  to the robust procedure are stable for shifted outliers. 
It should be noticed that, when considering the discrepancy measure $KS$ of the classical procedure, the median over replications under $C_{0.05}$ equals $0.7756$ when the sample size is $n=100$ and the Absolute Median Deviation (\textsc{mad}) is $0$, meaning that for more than half of the samples the obtained global measure equals   $0.7756$, which  is close to the maximal possible value. This behaviour is also reflected in    Figure \ref{fig:KS_bxp} that shows the boxplots of   $KS$ for $n=100$ and $n=200$. The boxplot of the classical estimator is completely shifted away under contamination attaining values close to $0.8$.

\normalsize

\normalsize

\begin{table}[ht!]
\begin{center}
\begin{tabular}{|c|  c|c|c|c|c|c|c|}
\hline 
$n$ & &\multicolumn{2}{c|}{$C_0$}& \multicolumn{2}{c|}{$C_{0.05}$}& \multicolumn{2}{c|}{$C_{0.10}$}\\ 
\hline
 &   & Classical & Robust& Classical & Robust& Classical & Robust\\ 
\hline
100  
& $MSE$ & 0.0032 & 0.0036 & 0.0205 & 0.0040 & 0.0349 & 0.0043 \\ 
 & $KS$ & 0.1949 & 0.1988 & 0.7757 & 0.2060 & 0.7993 & 0.2215 \\ 
\hline 
200   & $MSE$ & 0.0015 & 0.0017 & 0.0134 & 0.0018 & 0.0269 & 0.0021 \\ 
 & $KS$  & 0.1363 & 0.1380 & 0.7756 & 0.1436 & 0.7997 & 0.1538 \\ 
\hline
\end{tabular} 
\end{center}
\caption{\small \label{tab:mse} Mean    of $MSE$ and $KS$  over replications  under the linear model \eqref{trued} and \eqref{trueh}, for clean and samples contaminated as in $C_{\delta}$.}
\end{table}
 
\begin{figure}[H]
 \begin{center}
 \footnotesize
 \renewcommand{\arraystretch}{0.4}
 \newcolumntype{M}{>{\centering\arraybackslash}m{\dimexpr.1\linewidth-1\tabcolsep}}
   \newcolumntype{G}{>{\centering\arraybackslash}m{\dimexpr.4\linewidth-1\tabcolsep}}
\begin{tabular}{M GG}
 & Classical & Robust \\
 $C_0$ & 
 \includegraphics[scale=0.35]{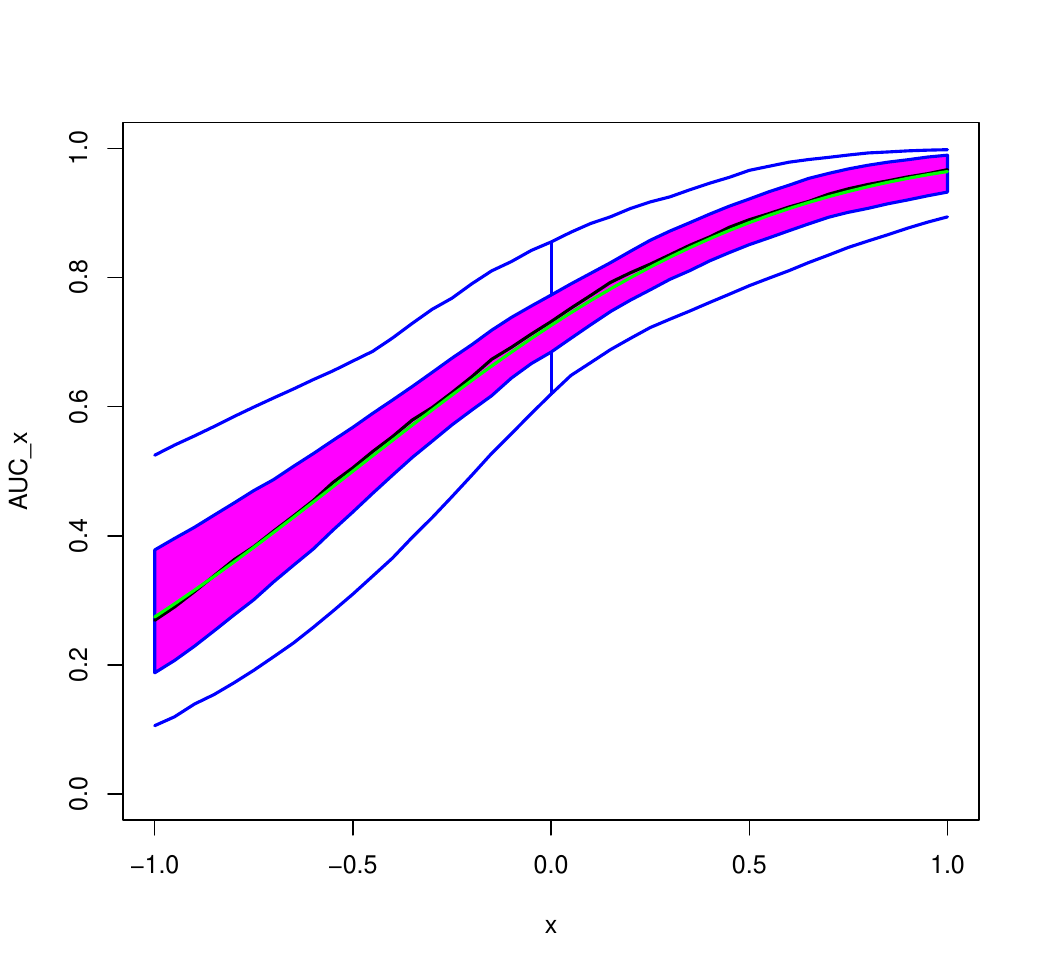} & 
\includegraphics[scale=0.35]{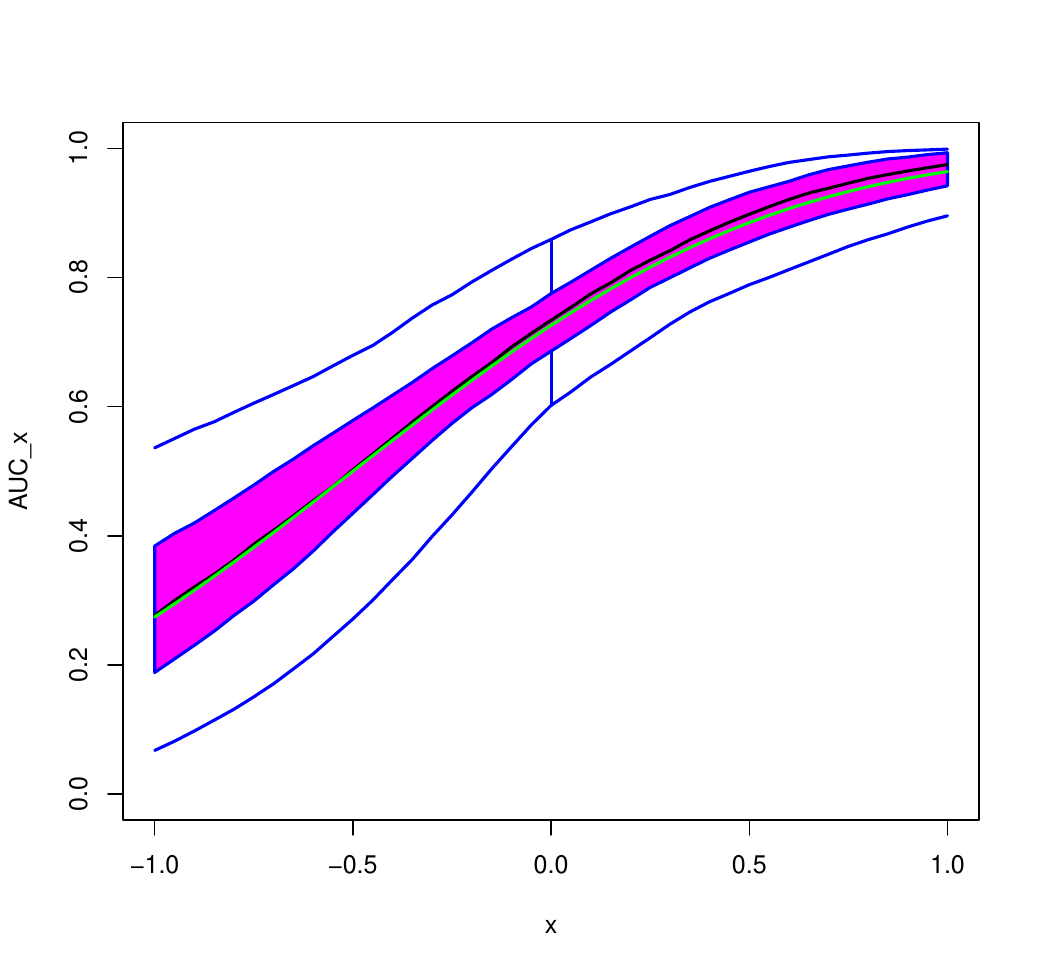} \\
$C_{0.05}^H$ &  
\includegraphics[scale=0.35]{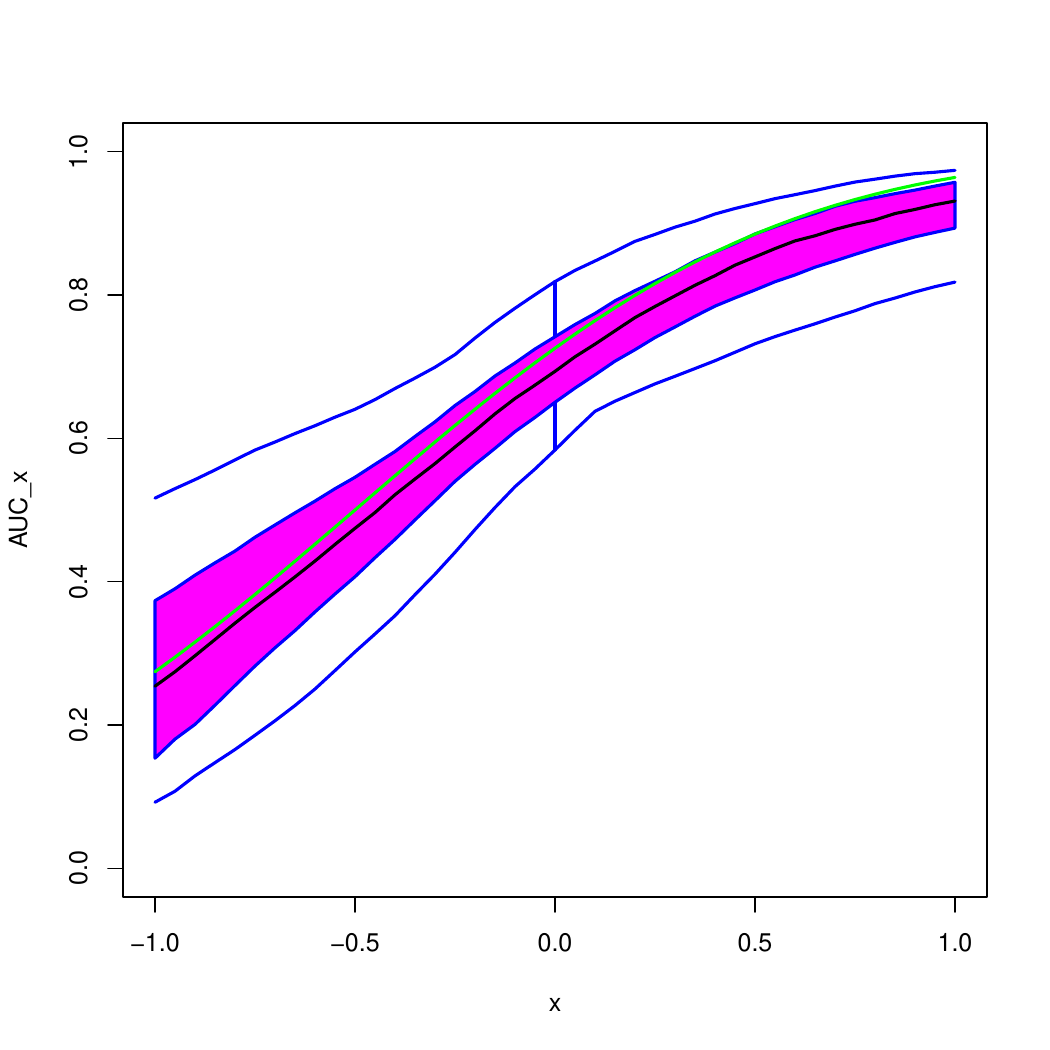} & 
\includegraphics[scale=0.35]{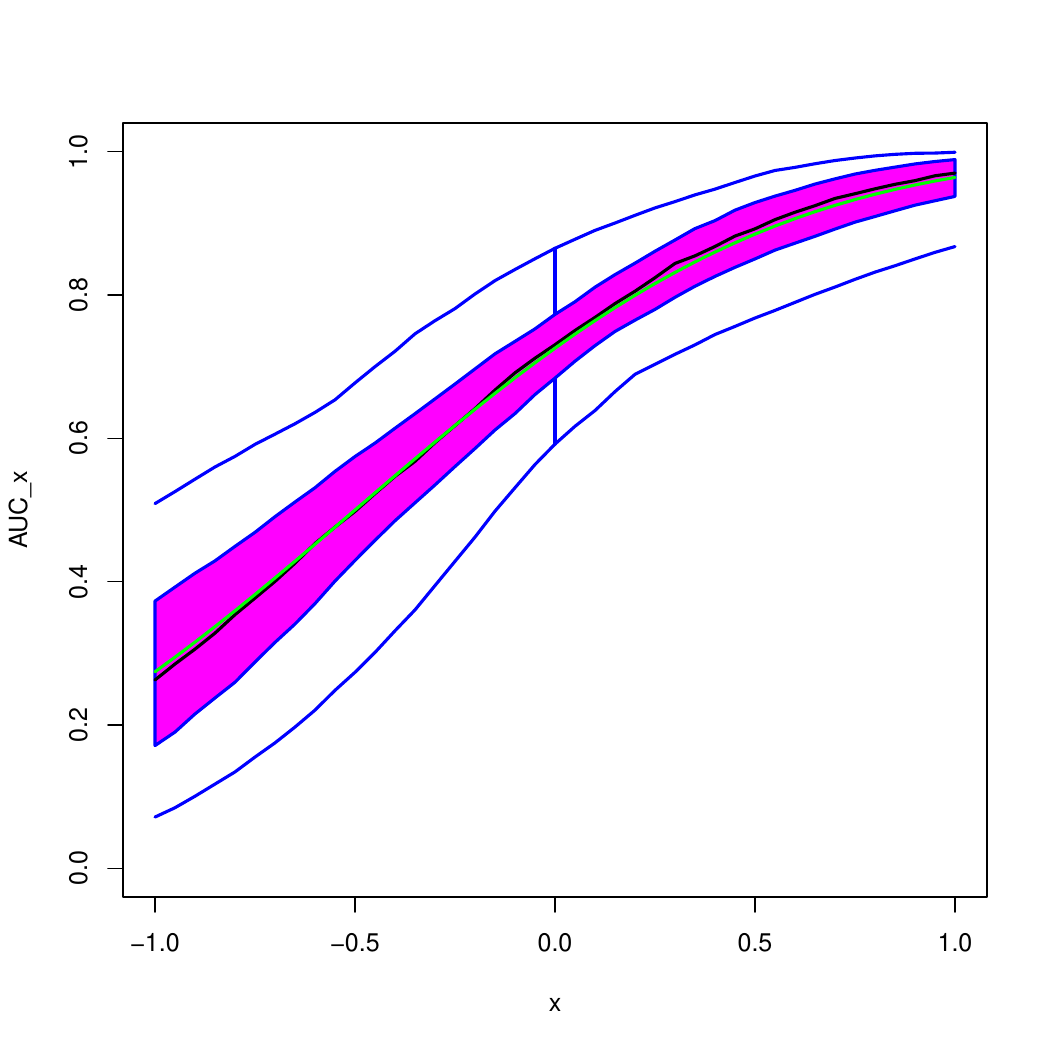}\\
$C_{0.10}^H$ & 
\includegraphics[scale=0.35]{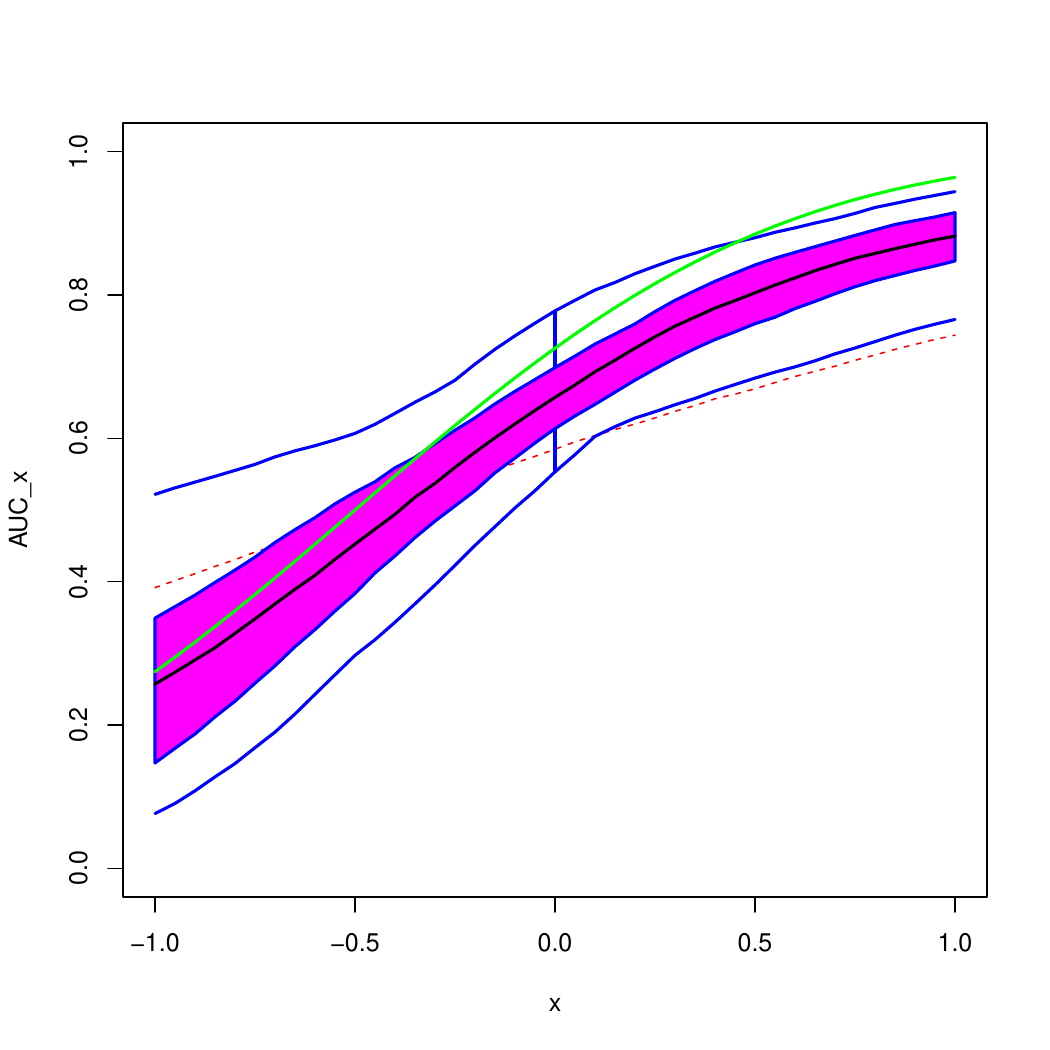} & 
\includegraphics[scale=0.35]{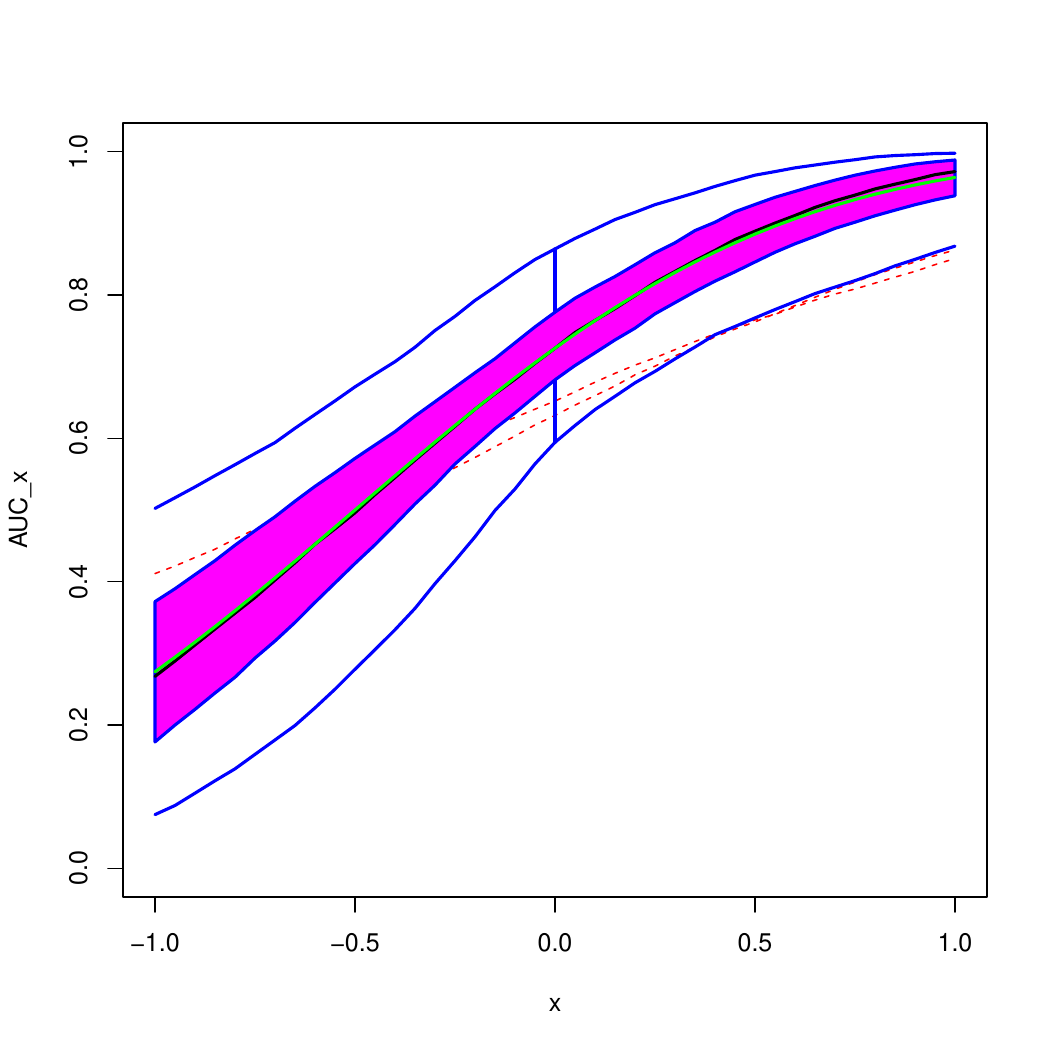}\\
\end{tabular}
\vskip-0.1in  \caption{\footnotesize \label{fig:fbx_sensi_100_S5_C3}   Functional boxplots of  $\widehat{\AUC}_x$   for $n=100$ under the linear model \eqref{trued} and \eqref{trueh} for clean samples and when the samples are contaminated according to  $C_{\delta}^H$ for  $S=5$ and $ n_H=n_D=100$. The green line corresponds to the true $\AUC_x$ and the dotted red lines to the outlying curves detected by the functional boxplot.}
\end{center} 
\end{figure}
\begin{figure}[H]
 \begin{center}
 \footnotesize
 \renewcommand{\arraystretch}{0.4}
 \newcolumntype{M}{>{\centering\arraybackslash}m{\dimexpr.1\linewidth-1\tabcolsep}}
   \newcolumntype{G}{>{\centering\arraybackslash}m{\dimexpr.4\linewidth-1\tabcolsep}}
\begin{tabular}{M GG}
 & Classical & Robust \\
$C_{0.05}^H$ &  
\includegraphics[scale=0.3]{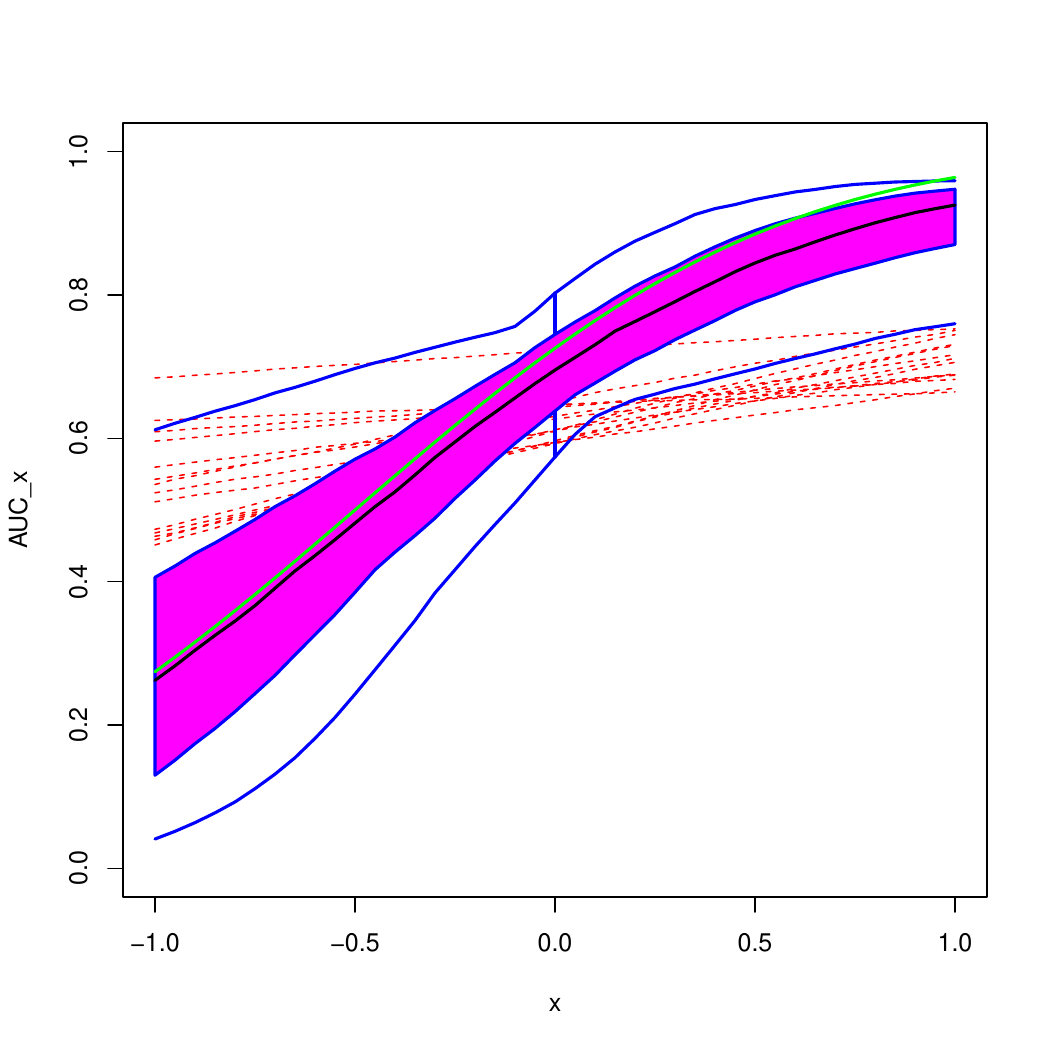} & 
\includegraphics[scale=0.3]{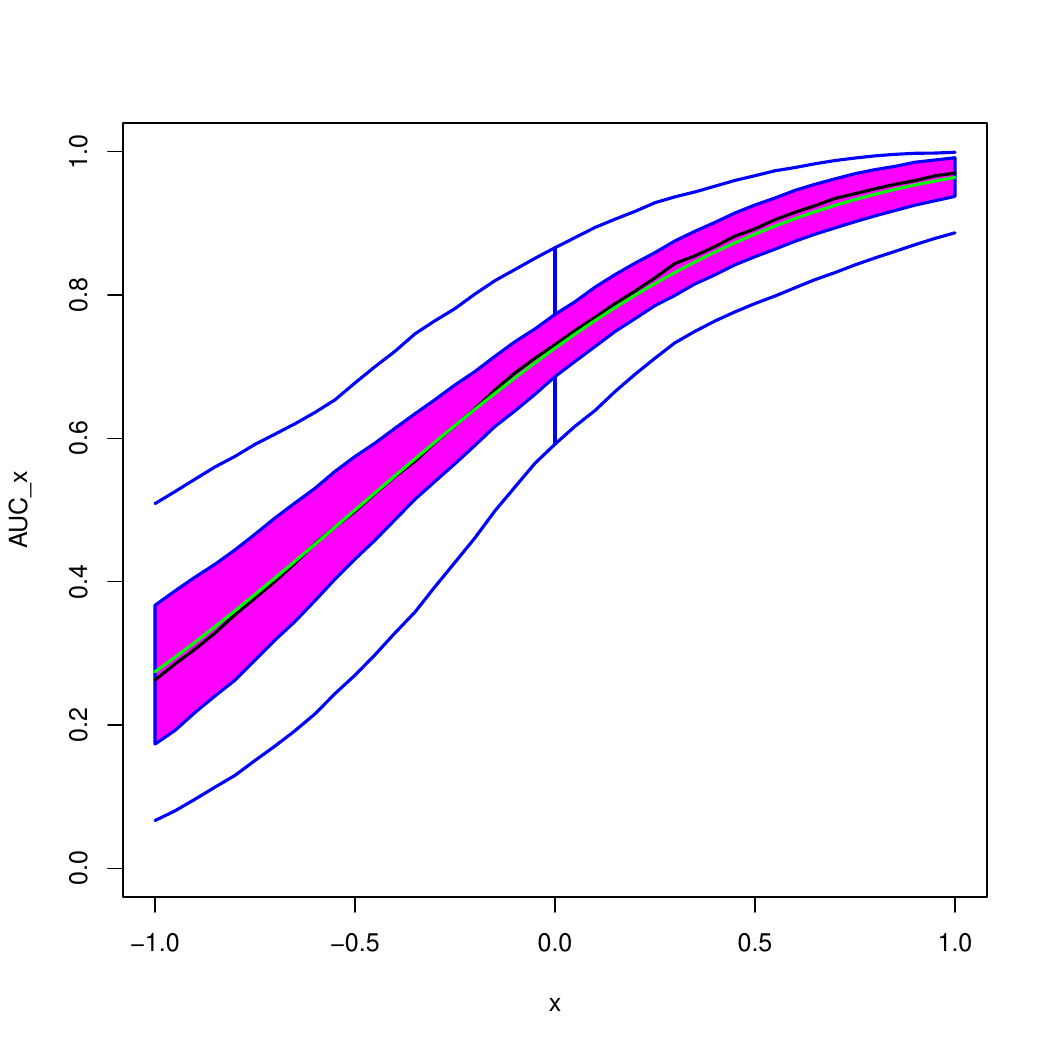}\\
$C_{0.10}^H$ &  
\includegraphics[scale=0.3]{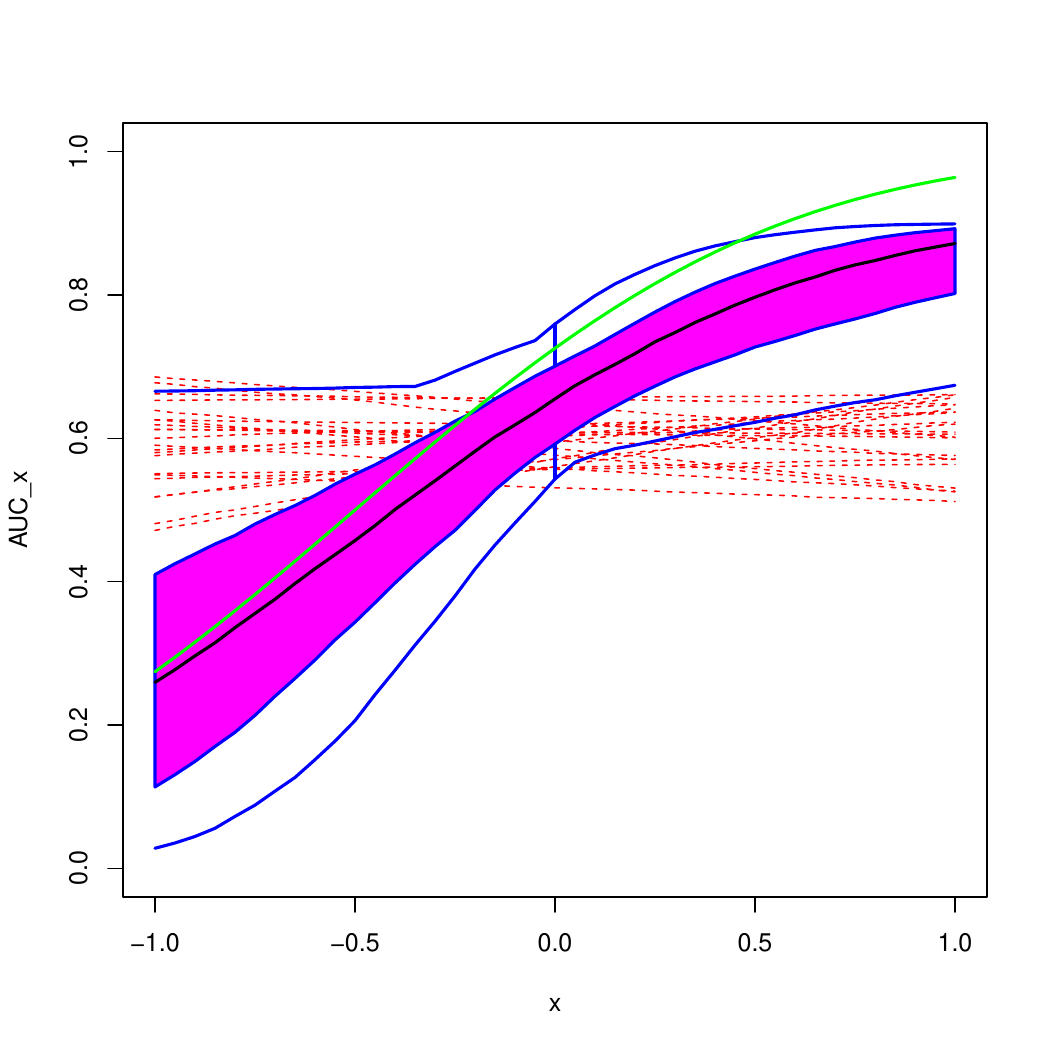} & 
\includegraphics[scale=0.3]{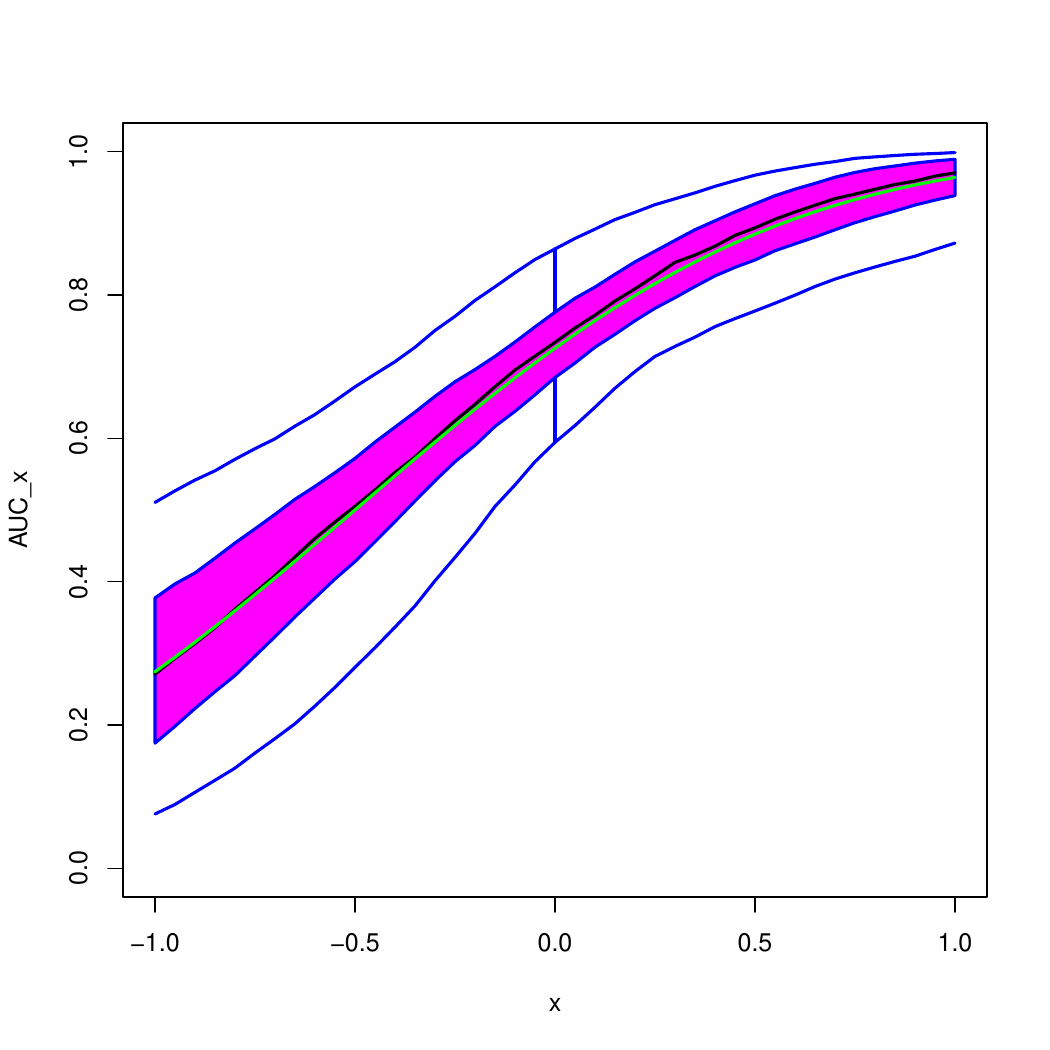}\\
\end{tabular}
\vskip-0.1in  \caption{\footnotesize \label{fig:fbx_sensi_100_S15_C3} Functional boxplots of  $\widehat{\AUC}_x$   for $n=100$  under the linear model \eqref{trued} and \eqref{trueh} when the samples are contaminated according to $C_{\delta}^H$ for  $S=15$ and $ n_H=n_D=100$ . The green line corresponds to the true $\AUC_x$ and the dotted red lines to the outlying curves detected by the functional boxplot.}
\end{center} 
\end{figure}
\begin{figure}[H]
 \begin{center}
 \footnotesize
 \renewcommand{\arraystretch}{0.4}
 \newcolumntype{M}{>{\centering\arraybackslash}m{\dimexpr.1\linewidth-1\tabcolsep}}
   \newcolumntype{G}{>{\centering\arraybackslash}m{\dimexpr.4\linewidth-1\tabcolsep}}
\begin{tabular}{M GG}
 & Classical & Robust \\
$C_{0.05}^D$ &  
\includegraphics[scale=0.3]{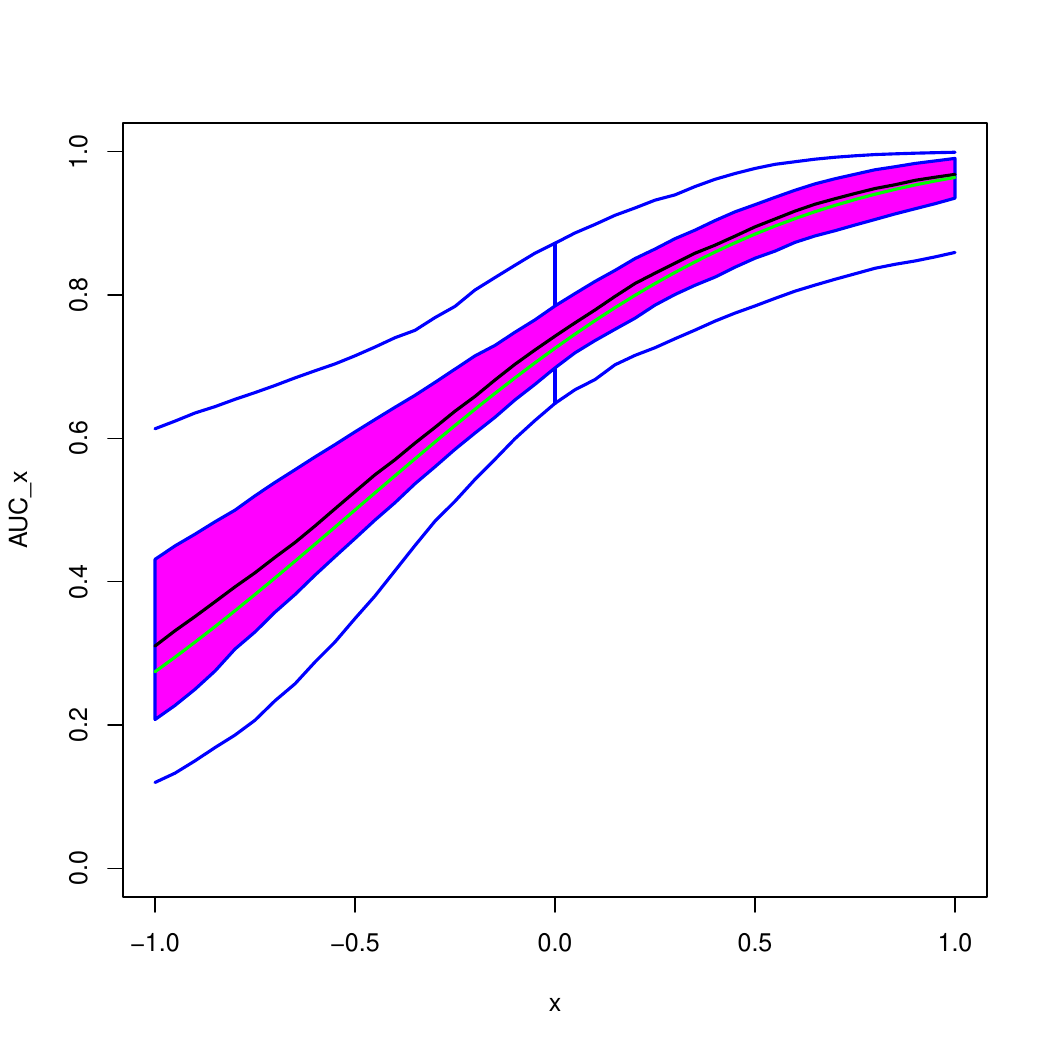} & 
\includegraphics[scale=0.3]{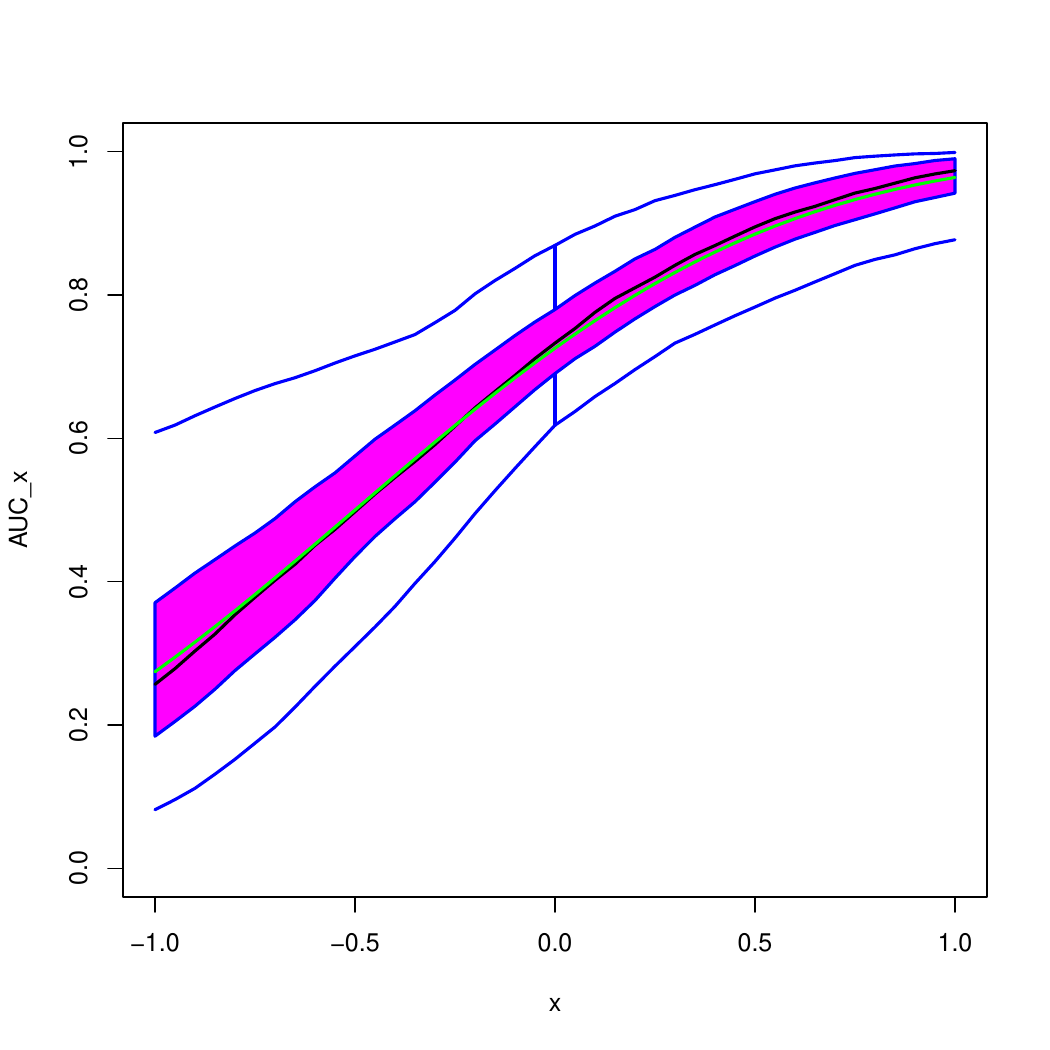}\\
$C_{0.10}^D$ &  
\includegraphics[scale=0.3]{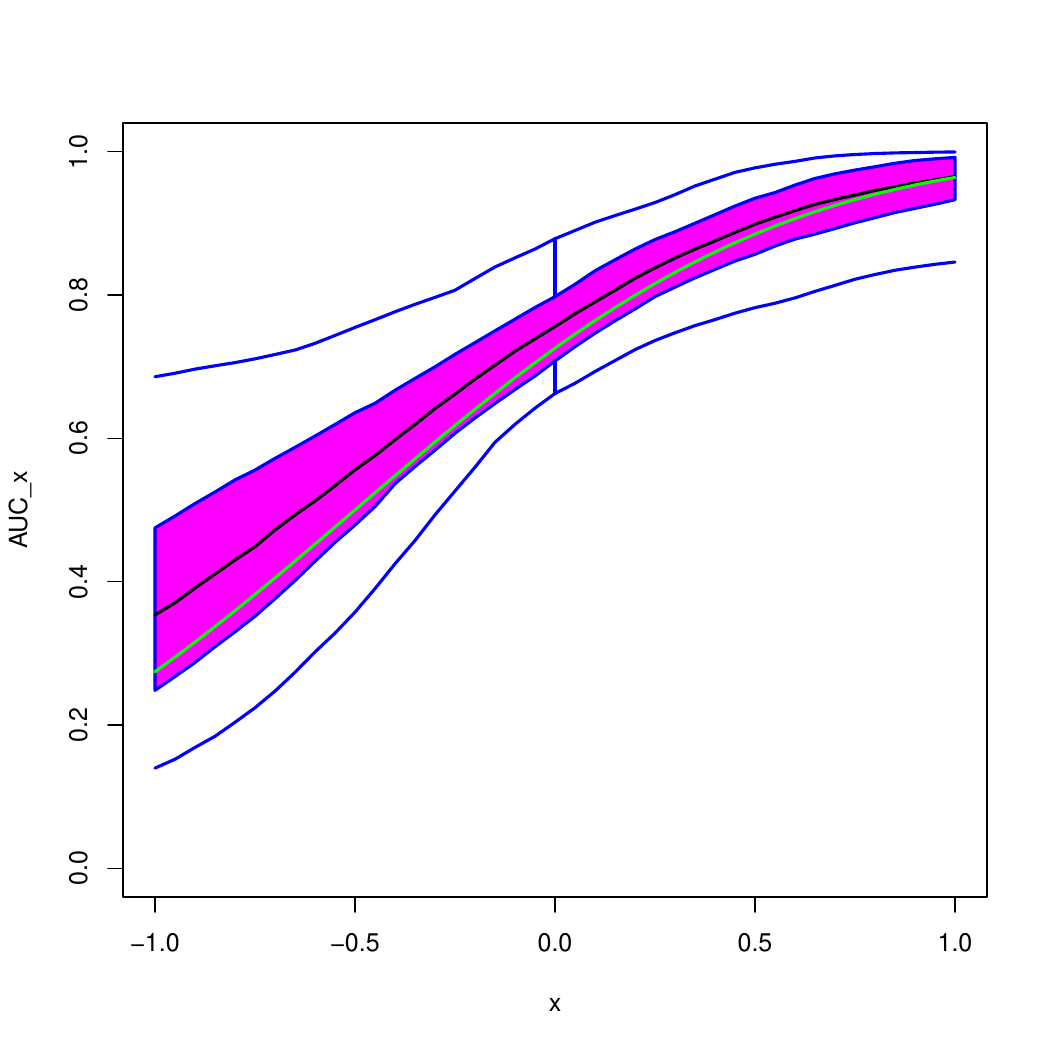} & 
\includegraphics[scale=0.3]{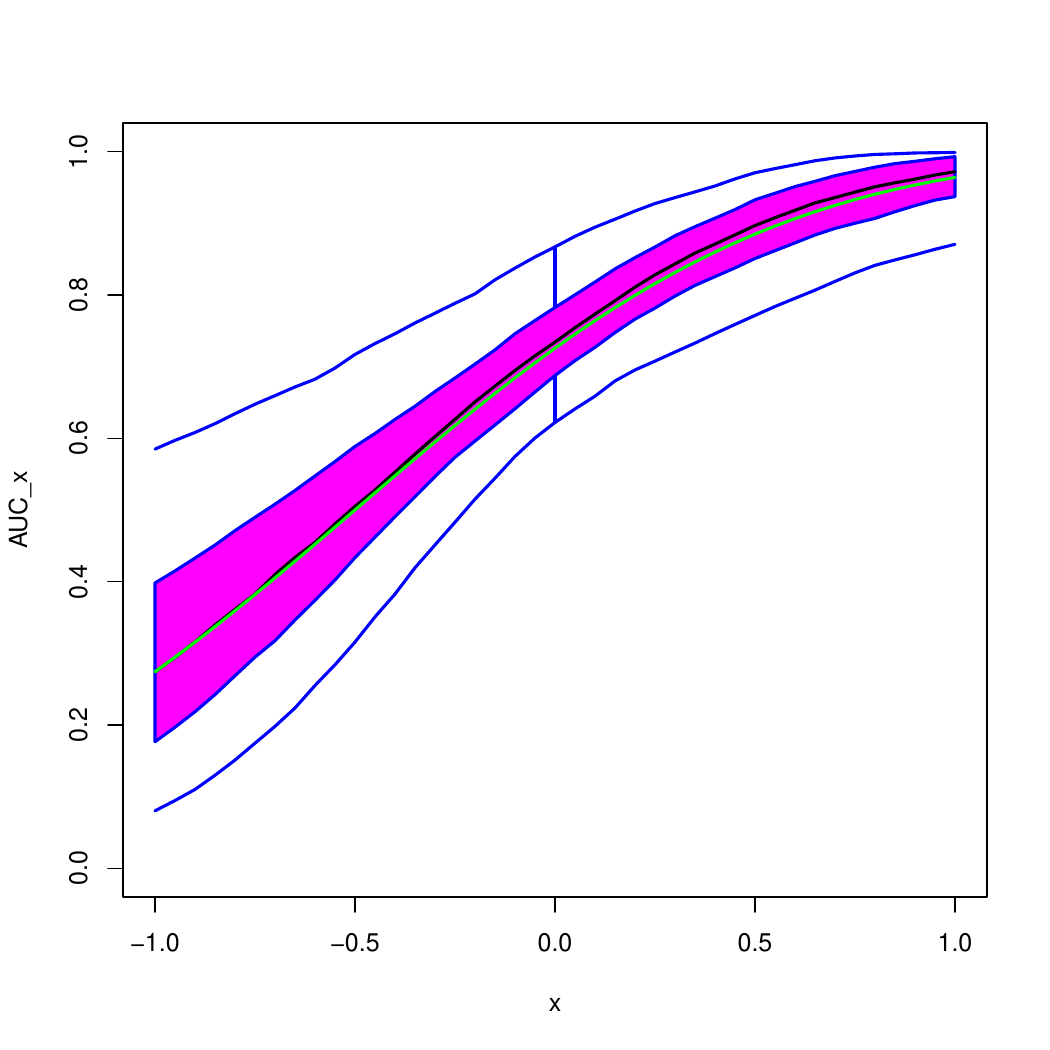}\\
\end{tabular}
\vskip-0.1in  \caption{\footnotesize \label{fig:fbx_sensi_100_S5_C4}  Functional boxplots of  $\widehat{\AUC}_x$   for $n=100$  under the linear model \eqref{trued} and \eqref{trueh} when the samples are contaminated according to $C_{\delta}^D$ for  $S=5$ and $ n_H=n_D=100$. The green line corresponds to the true $\AUC_x$ and the dotted red lines to the outlying curves detected by the functional boxplot.}
\end{center} 
\end{figure}
\begin{figure}[H]
 \begin{center}
 \footnotesize
 \renewcommand{\arraystretch}{0.4}
 \newcolumntype{M}{>{\centering\arraybackslash}m{\dimexpr.1\linewidth-1\tabcolsep}}
   \newcolumntype{G}{>{\centering\arraybackslash}m{\dimexpr.4\linewidth-1\tabcolsep}}
\begin{tabular}{M GG}
 & Classical & Robust \\
$C_{0.05}^D$ &  
\includegraphics[scale=0.3]{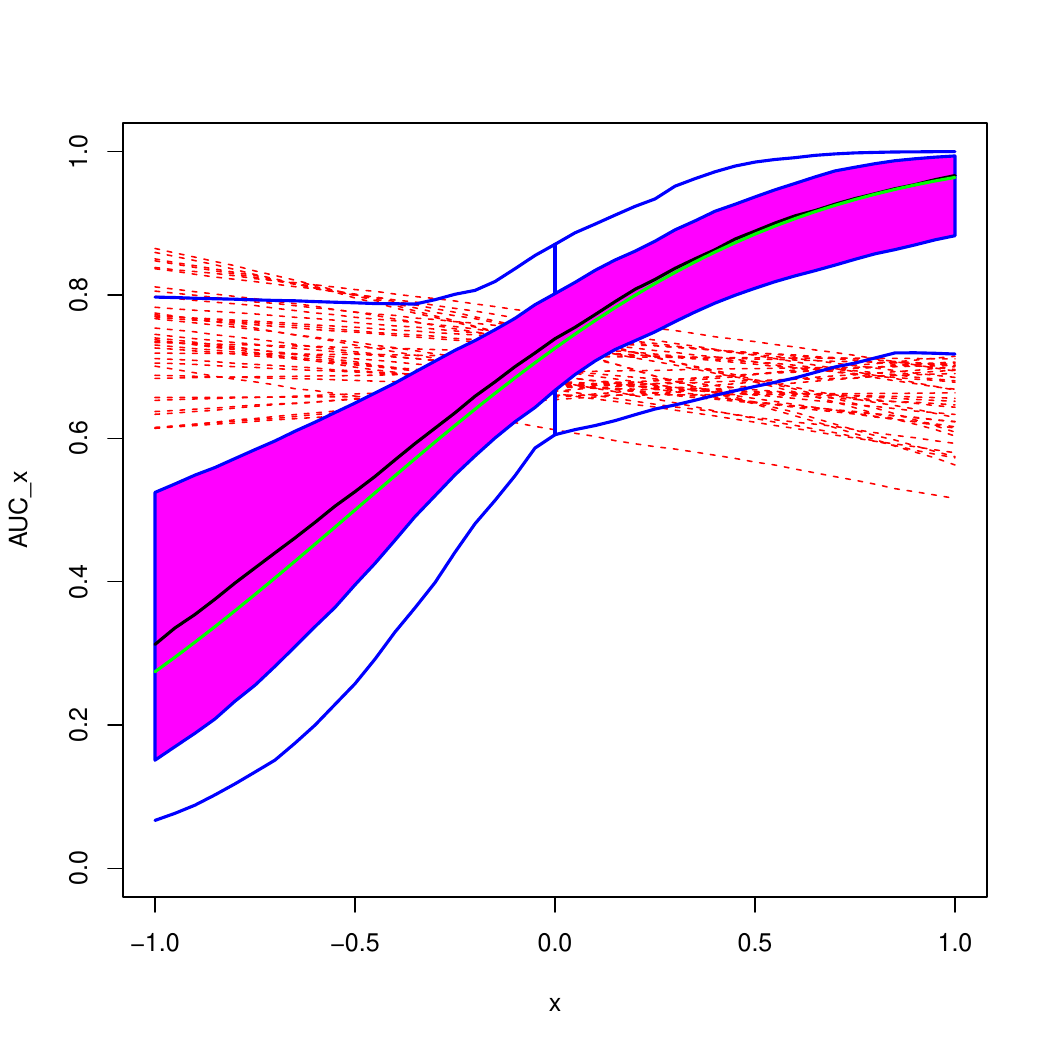} & 
\includegraphics[scale=0.3]{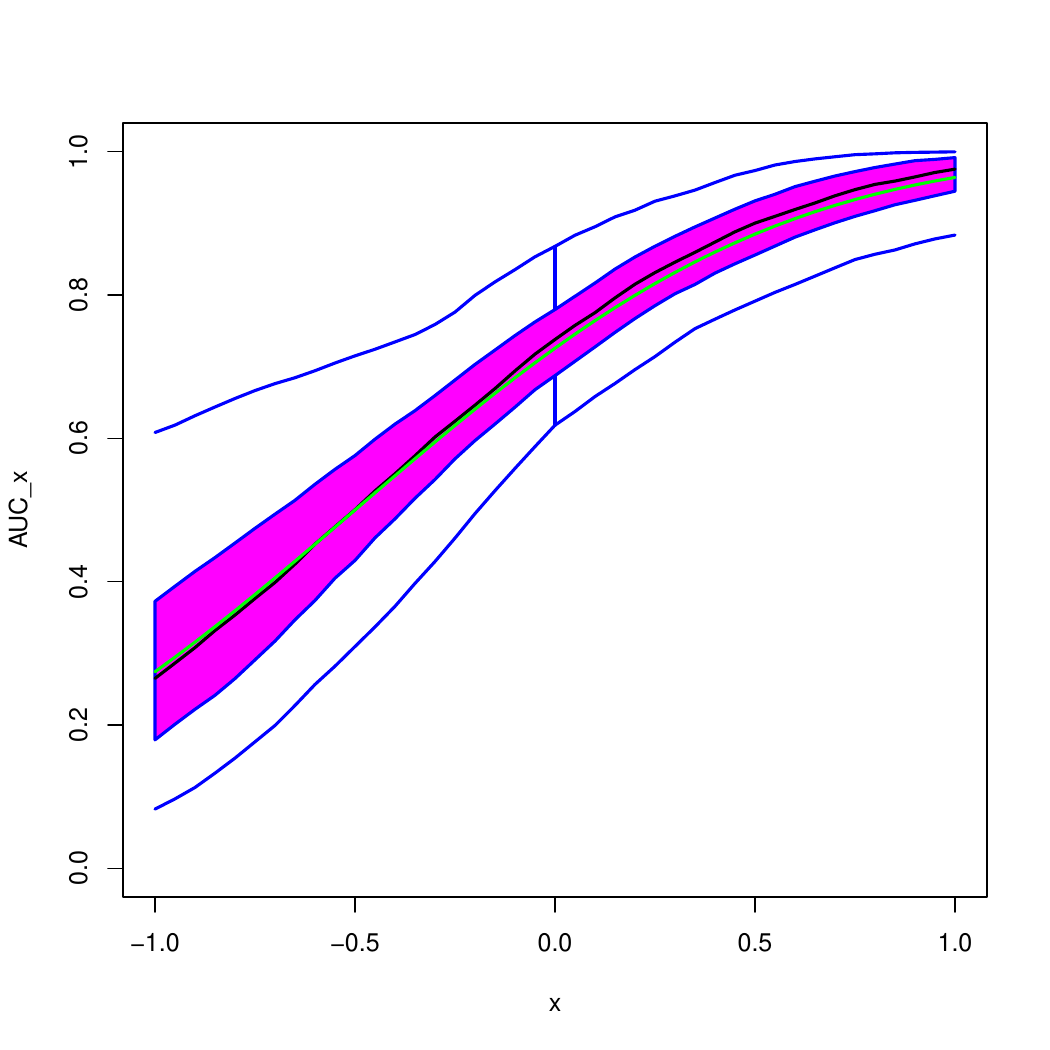}\\ 
$C_{0.10}^D$ &  
\includegraphics[scale=0.3]{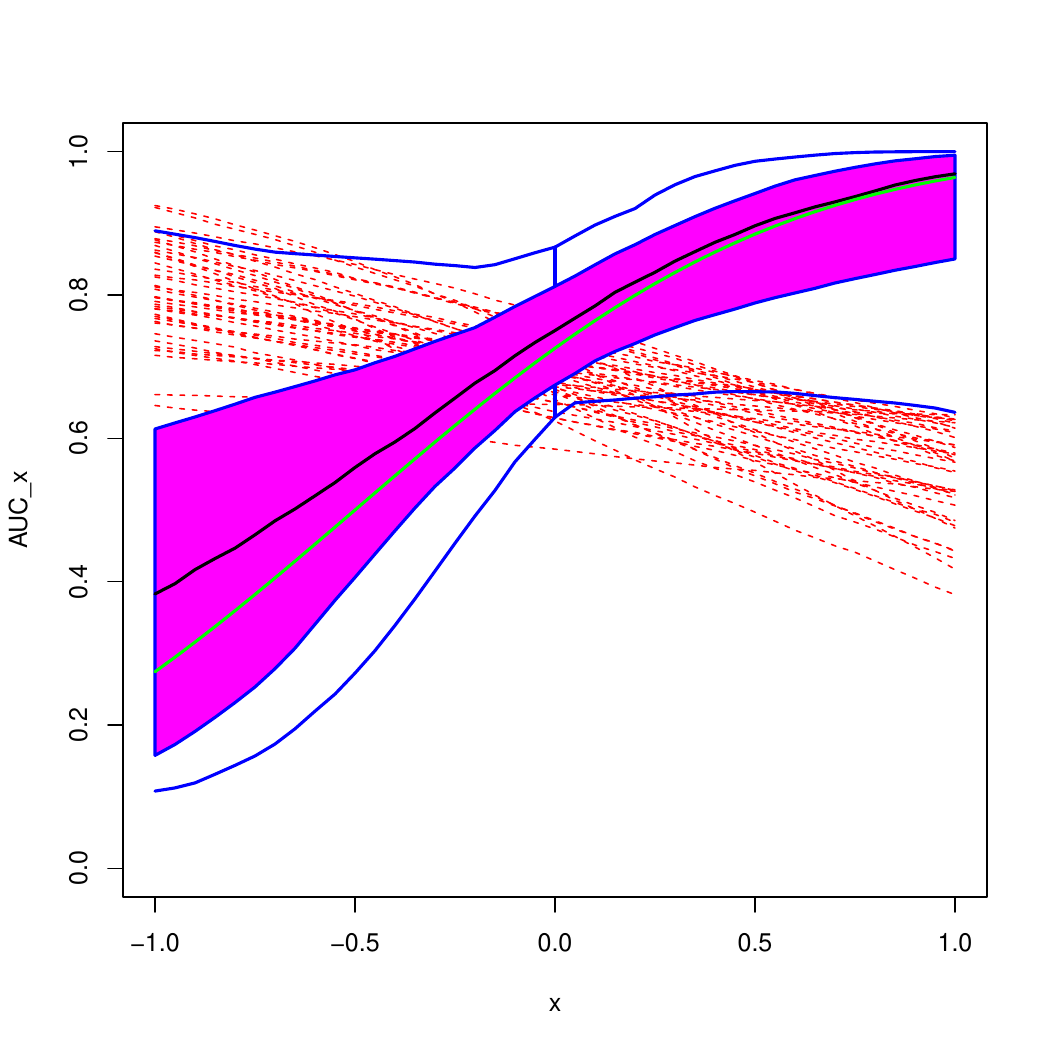} & 
\includegraphics[scale=0.3]{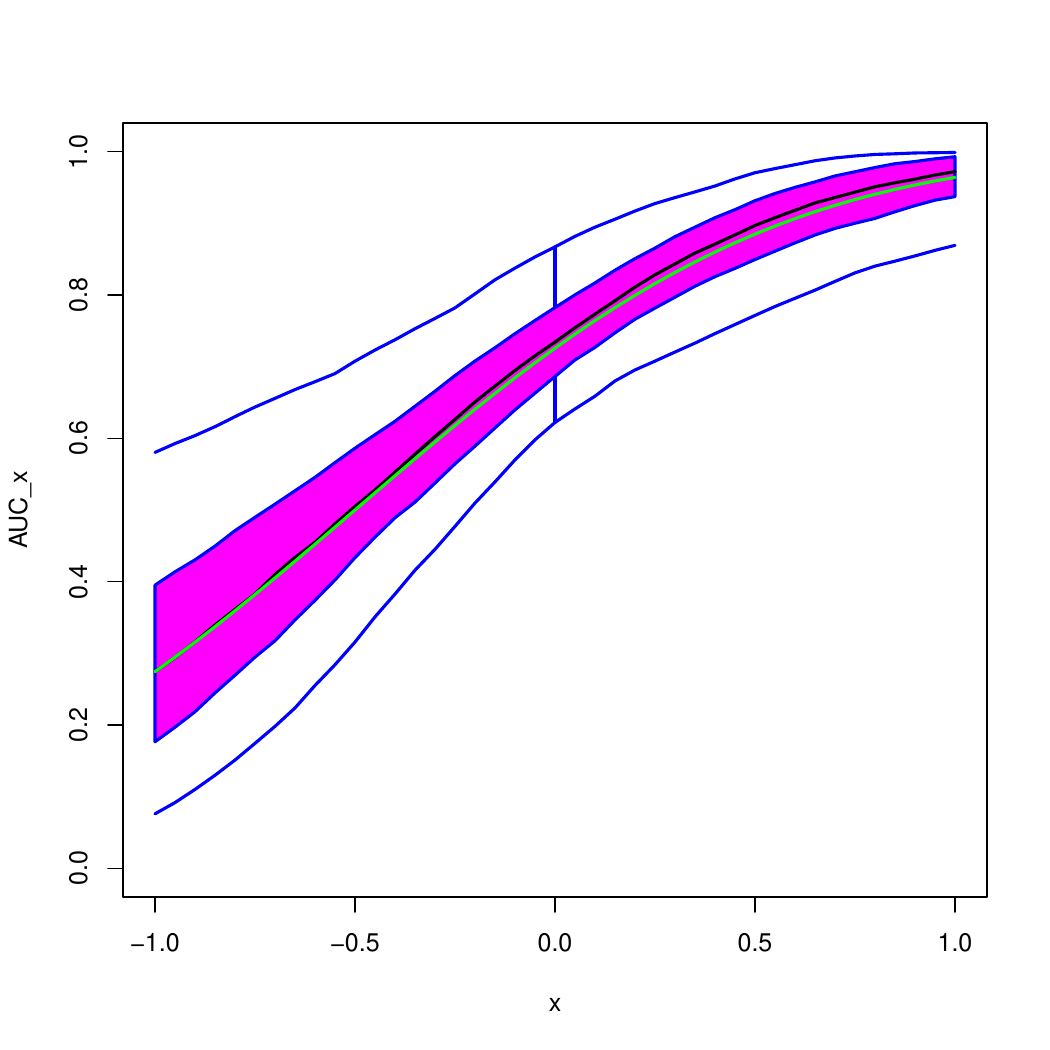}\\
\end{tabular}
\vskip-0.1in  \caption{\footnotesize \label{fig:fbx_sensi_100_S20_C4}  Functional boxplots of  $\widehat{\AUC}_x$   for $n=100$  under the linear model \eqref{trued} and \eqref{trueh} when the samples are contaminated according to $C_{\delta}^D$ for  $S=20$ and $ n_H=n_D=100$ . The green line corresponds to the true $\AUC_x$ and the dotted red lines to the outlying curves detected by the functional boxplot.}
\end{center} 
\end{figure}
\begin{figure}[H]
\begin{center}
\footnotesize
\renewcommand{\arraystretch}{0.4}
\newcolumntype{M}{>{\centering\arraybackslash}m{\dimexpr.1\linewidth-1\tabcolsep}}
 \newcolumntype{G}{>{\centering\arraybackslash}m{\dimexpr.4\linewidth-1\tabcolsep}}
\begin{tabular}{M GG}
 & Classical Estimators & Robust Estimators \\
$C_0$ & 
\includegraphics[scale=0.35]{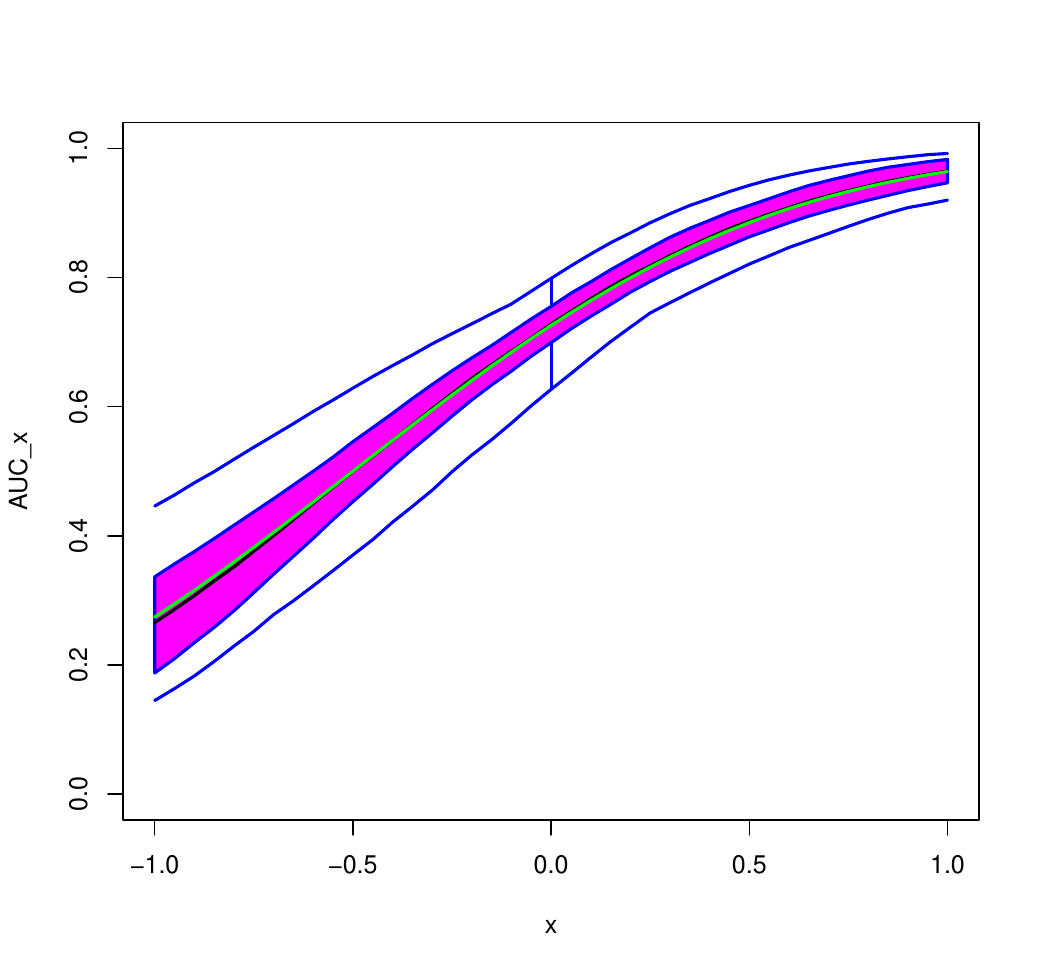} & 
\includegraphics[scale=0.35]{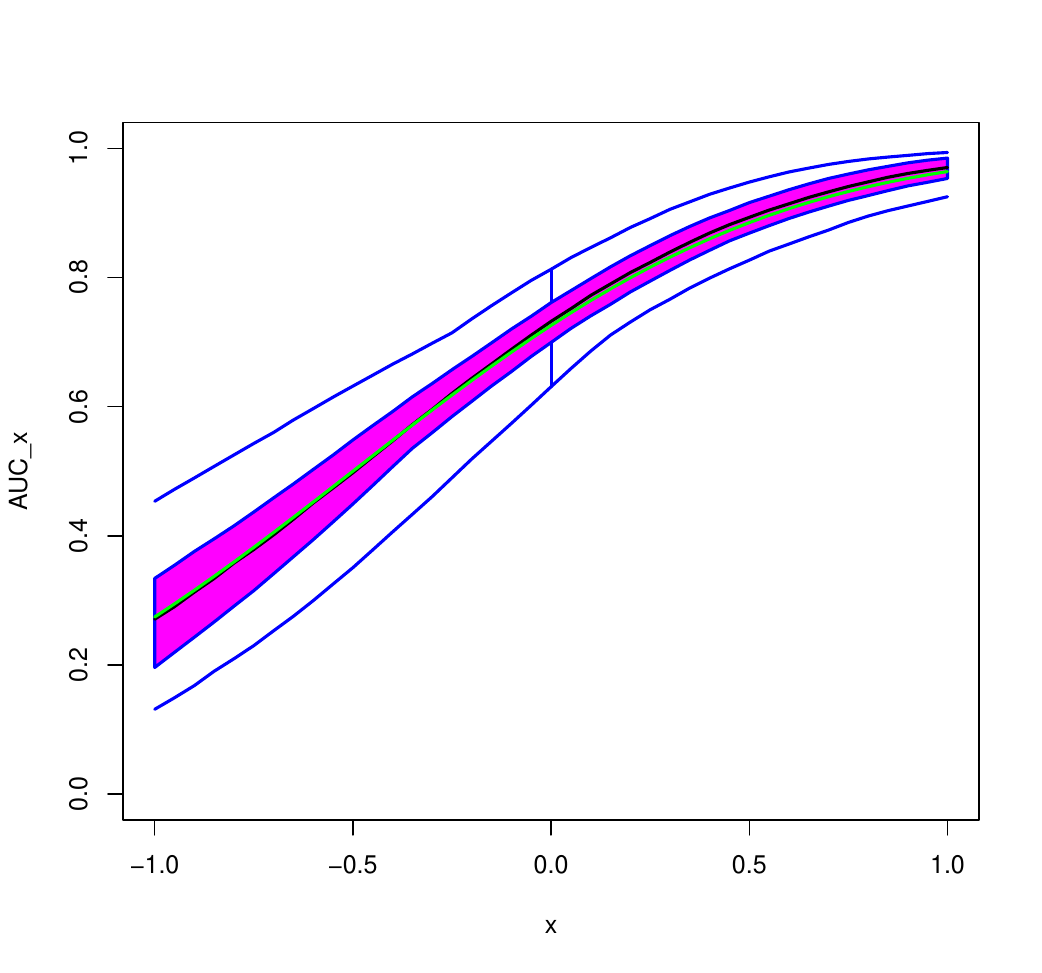} \\
$C_{0.05}$ &  
\includegraphics[scale=0.35]{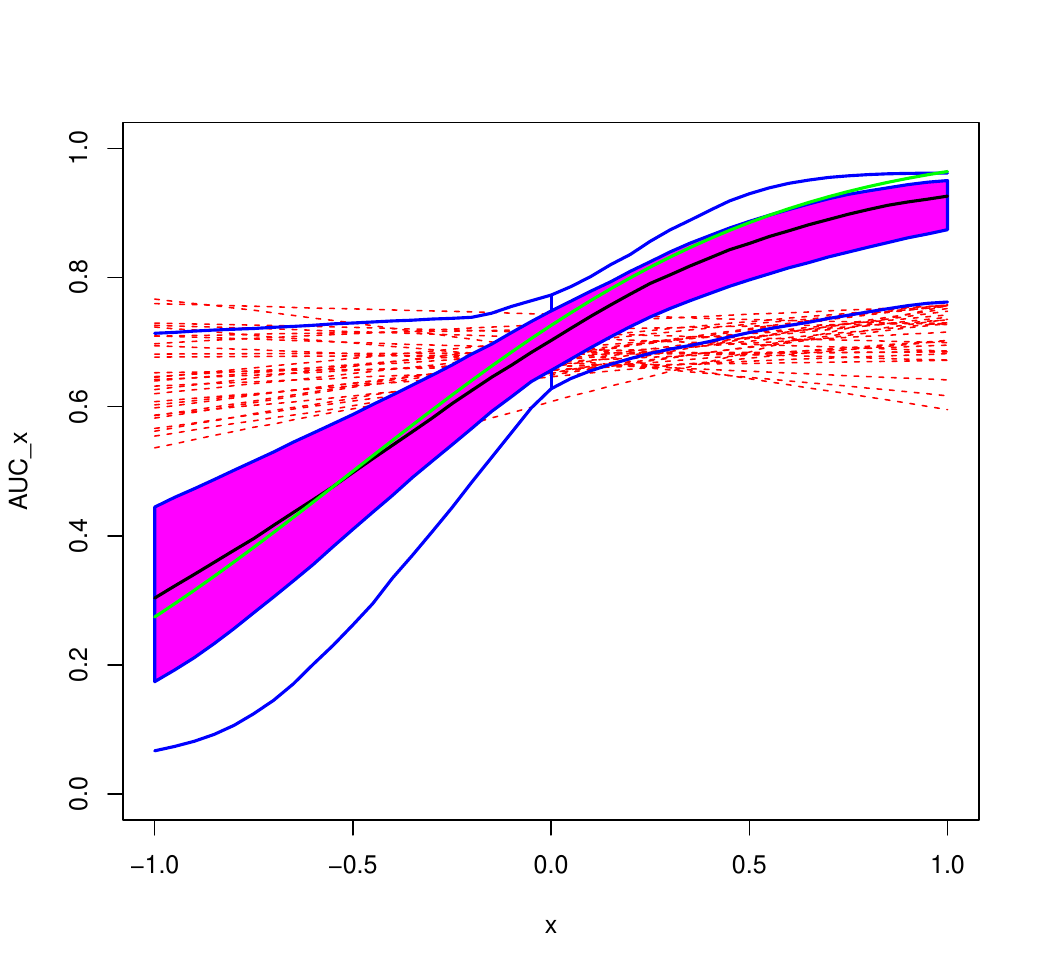} & 
\includegraphics[scale=0.35]{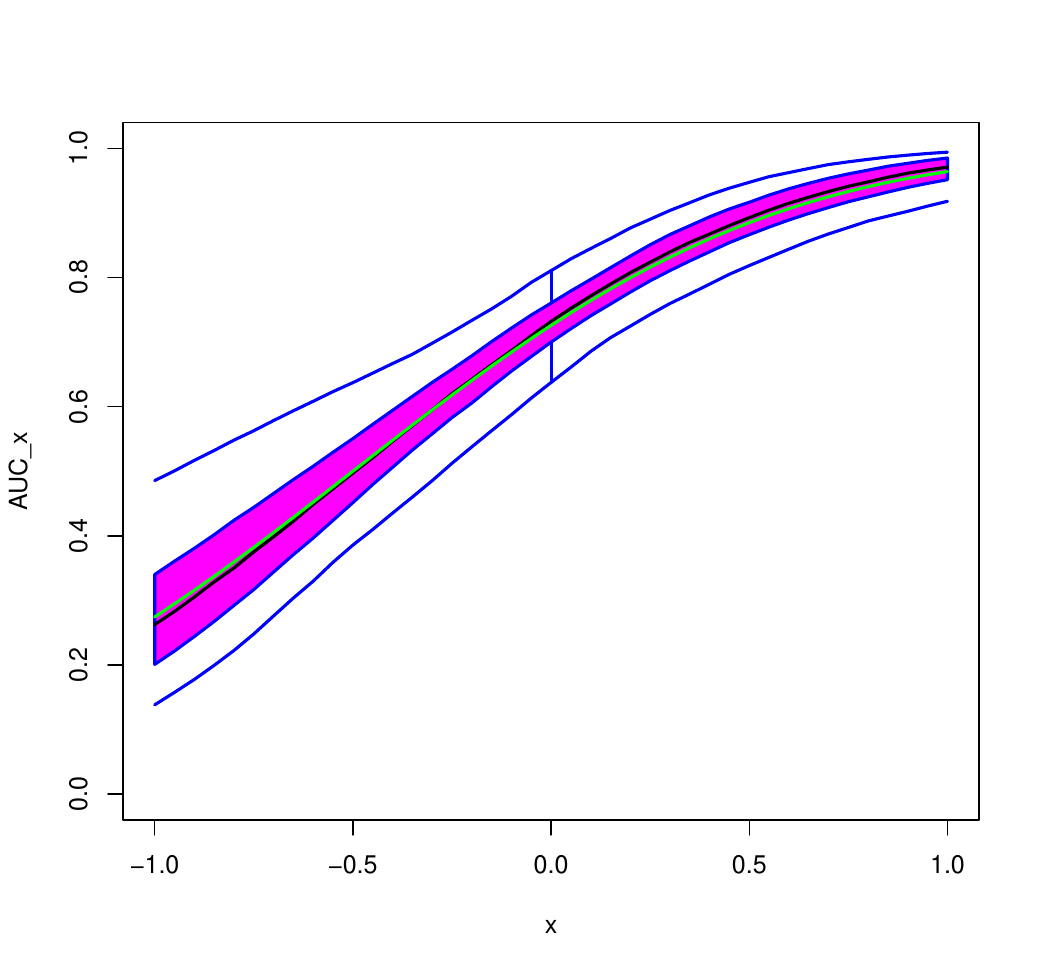}\\
$C_{0.10}$ &  
\includegraphics[scale=0.35]{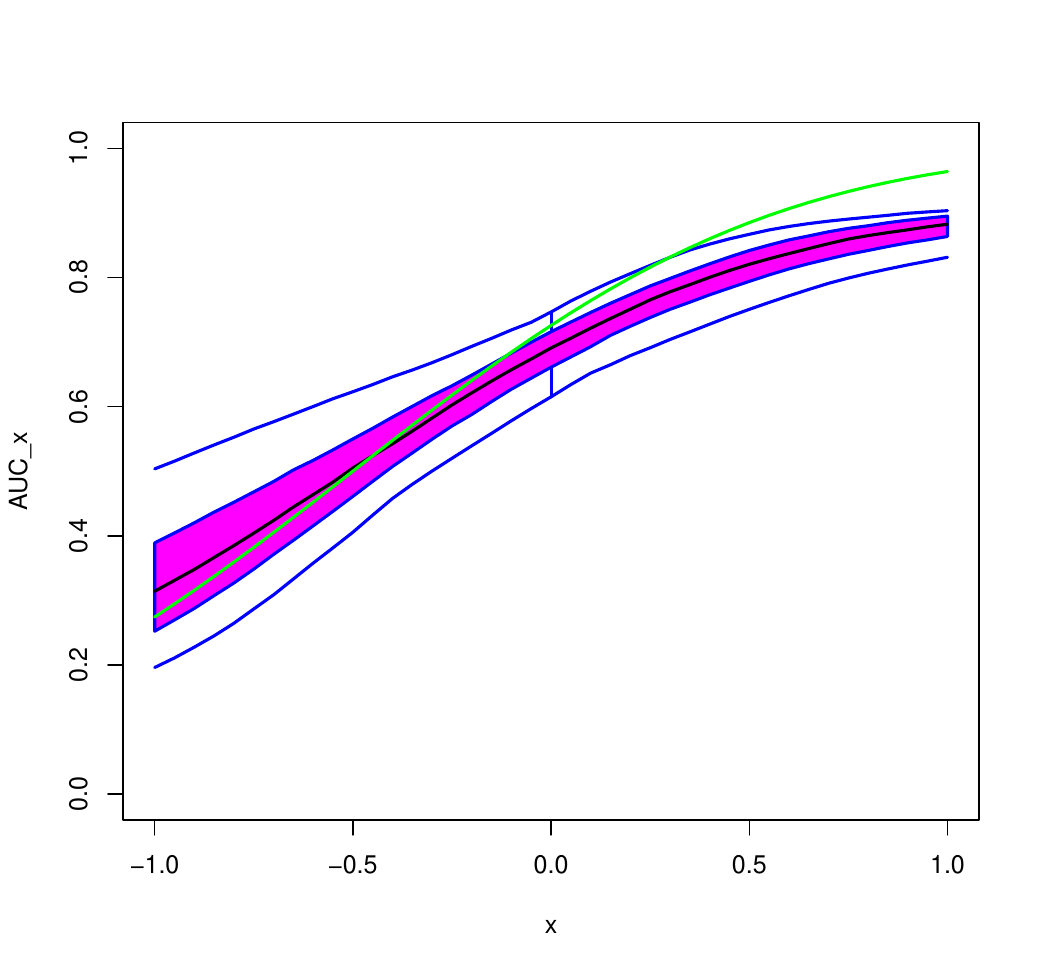} &  
\includegraphics[scale=0.35]{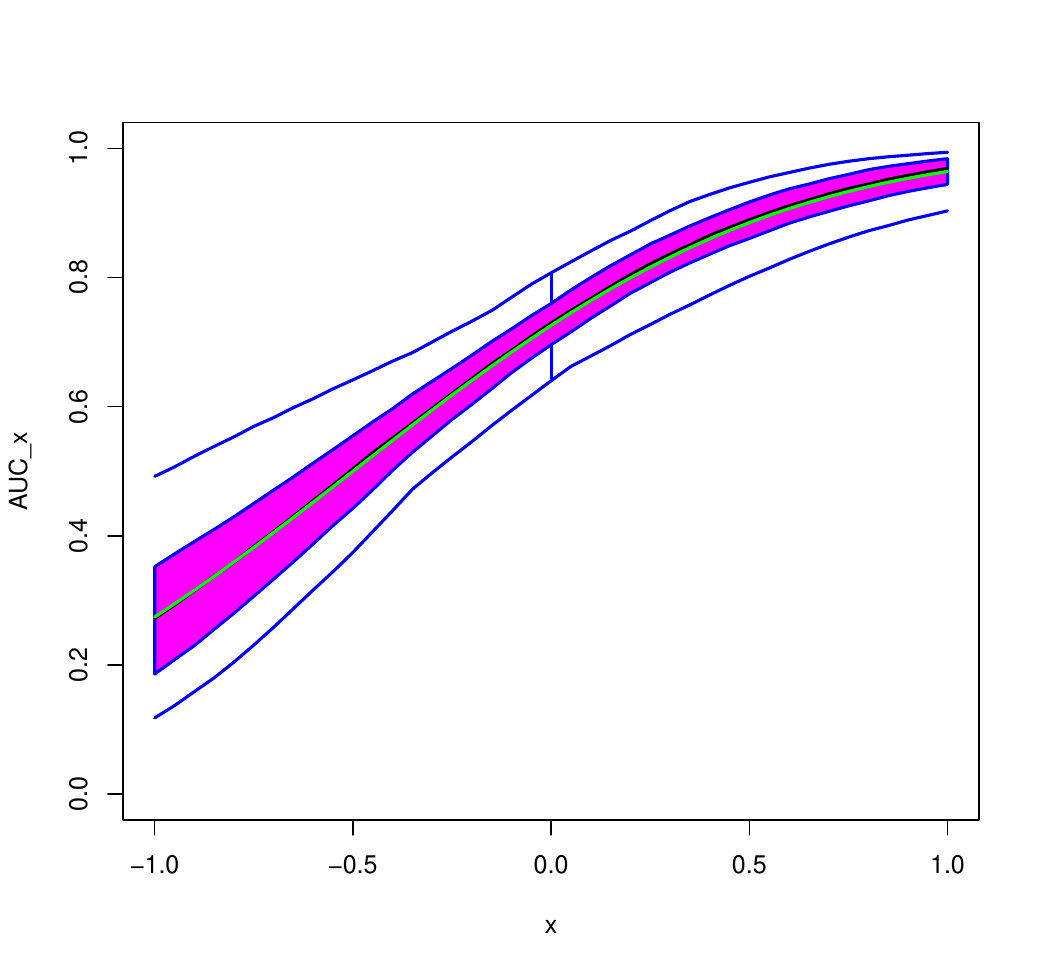}\\
\end{tabular}
\vskip-0.1in  
\caption{\footnotesize \label{fig:fbx_200}Functional boxplots of  $\widehat{\AUC}_x$   for $n=200$, under the linear model \eqref{trued} and \eqref{trueh}  when the samples are contaminated according to  $C_{\delta}$ . The green line corresponds to the true $\AUC_x$ and the dotted red lines to the outlying curves detected by the functional boxplot.}
\end{center} 
\end{figure}

As mentioned in the Introduction,  one of the most popular indices is the \textsl{area under the curve}, AUC, which is a summary measure usually considered to evaluate the discriminating effect of the biomarker. When covariates are present, the conditional area under the curve is also used as index of the marker accuracy. It is defined as $\AUC_x= \int_0^1 \ROC_x(p) dp $. Note that in this case, we obtain a single value for each $x$, hence, the function $x\to \widehat{\AUC}_x$ can be plotted for each sample. Taking into account the observed sensitivity of the classical estimators to outliers, it seems natural that this effect will be inherited by the estimators of the conditional area under the curve, ${\AUC}_x$. 
To evaluate this effect, Figures \ref{fig:fbx_sensi_100_S5_C3}  to \ref{fig:fbx_sensi_100_S20_C4}   show the functional boxplots of the estimators  $\widehat{\AUC}_x$ obtained with the classical and robust procedures,  when the sample sizes are $ n_H=n_D=100$, under contaminations $C_{\delta}^H$ and $C_{\delta}^D$ with  $\delta=0.05$ and $ 0.10$ and   different values of $S$. 
To facilitate comparisons, in Figure \ref{fig:fbx_sensi_100_S5_C3} we also give the plots corresponding to clean samples. Functional boxplots were introduced by  Sun and Genton (2011) and are a  useful visualization tool to give a whole picture of the behaviour of a collection of curves.  
The area in purple represents the 50\% inner band of curves, the dotted red lines correspond to outlying curves, the black line indicates the central (deepest) function, while the green line in the plot corresponds to the  true  $\AUC_x$ curve. 
As shown in Figure \ref{fig:fbx_sensi_100_S5_C3},  when $\delta=0.05$ and the healthy population is contaminated, the shift  causes a  bias in the classical estimator of $\AUC_x$, so that the central region of the functional boxplot fails to contain the true function   for much of its domain. This effect is more striking when $S=15$, where also some outlying curves completely distorted appear (see Figure \ref{fig:fbx_sensi_100_S15_C3}). On the other hand, the effect when contaminating the diseased population is not so devastating for the classical procedure. As shown in Figure \ref{fig:fbx_sensi_100_S5_C4}, even though the true curve is not close to the deepest curve it is still within the central region. However, when $S=20$, the  50\% inner band  is completely enlarged (see Figure \ref{fig:fbx_sensi_100_S20_C4}). As expected, the robust proposal is stable across the considered contaminations. Moreover, by comparing  the upper  panel  in Figure  \ref{fig:fbx_sensi_100_S5_C3}, we observe   that the classical and robust estimators of $\AUC_x$ are quite similar for clean samples when $n=100$ and a similar conclusion holds for $n=200$ (see Figure \ref{fig:fbx_200}).

\begin{figure}[ht!]
 \begin{center}
 \footnotesize
 \renewcommand{\arraystretch}{0.35}
 \newcolumntype{M}{>{\centering\arraybackslash}m{\dimexpr.1\linewidth-1\tabcolsep}}
   \newcolumntype{G}{>{\centering\arraybackslash}m{\dimexpr.4\linewidth-1\tabcolsep}}
\begin{tabular}{M GG}
 & Classical Estimators & Robust Estimators \\
$C_{0.05}$ &  
\includegraphics[scale=0.35]{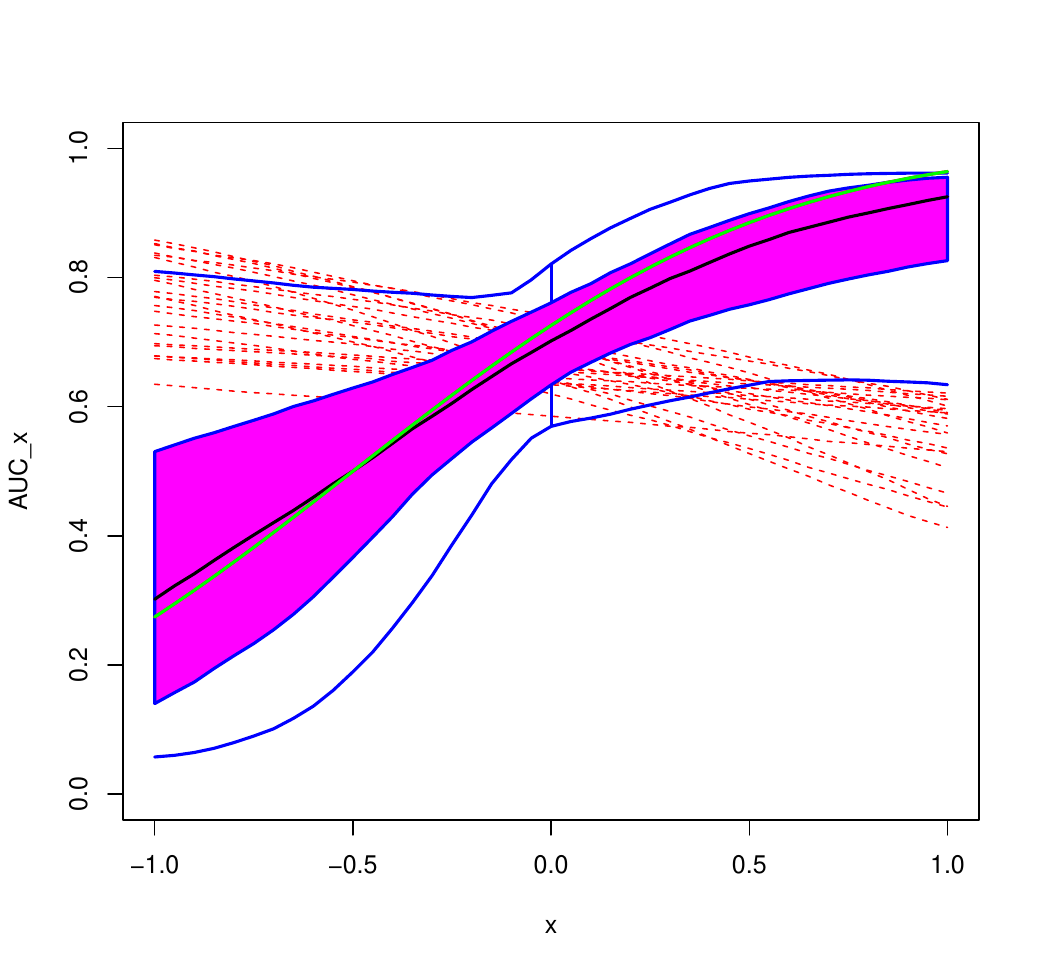} & 
\includegraphics[scale=0.35]{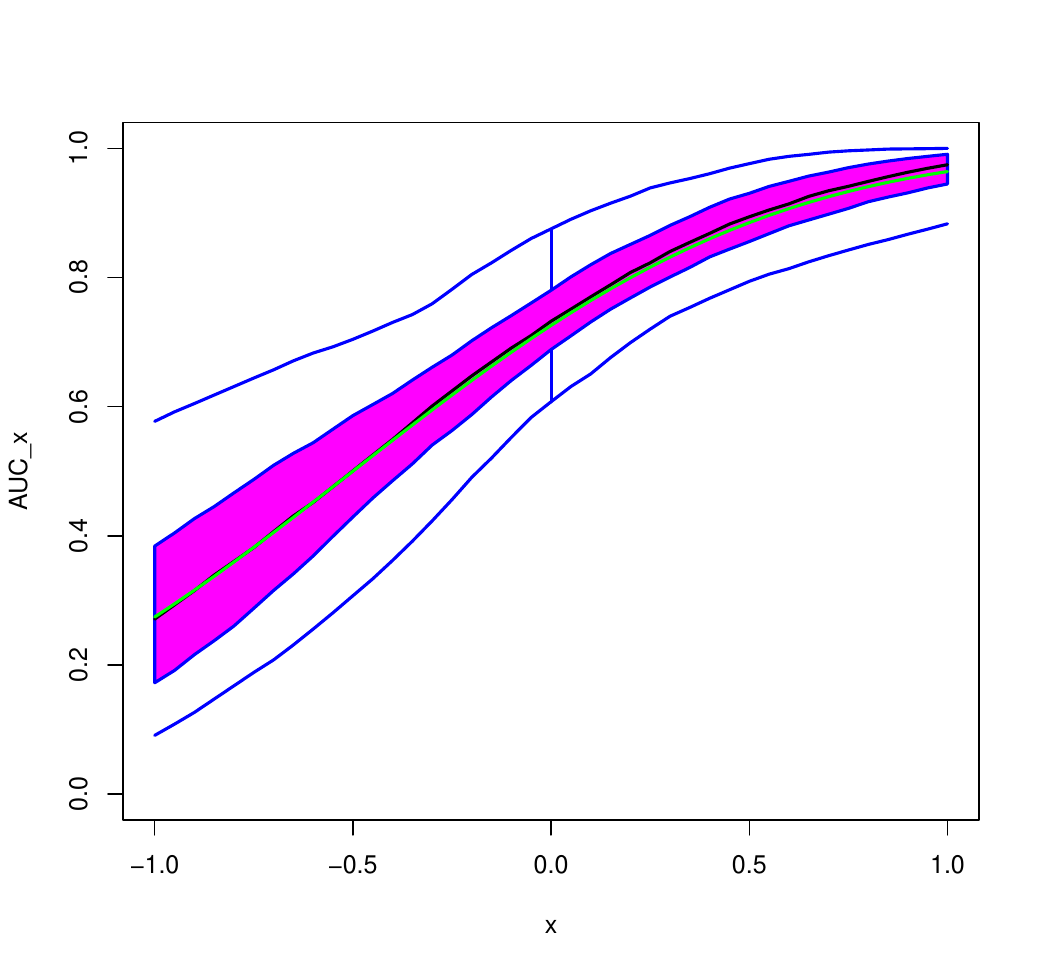}\\
$C_{0.10}$ &  
\includegraphics[scale=0.35]{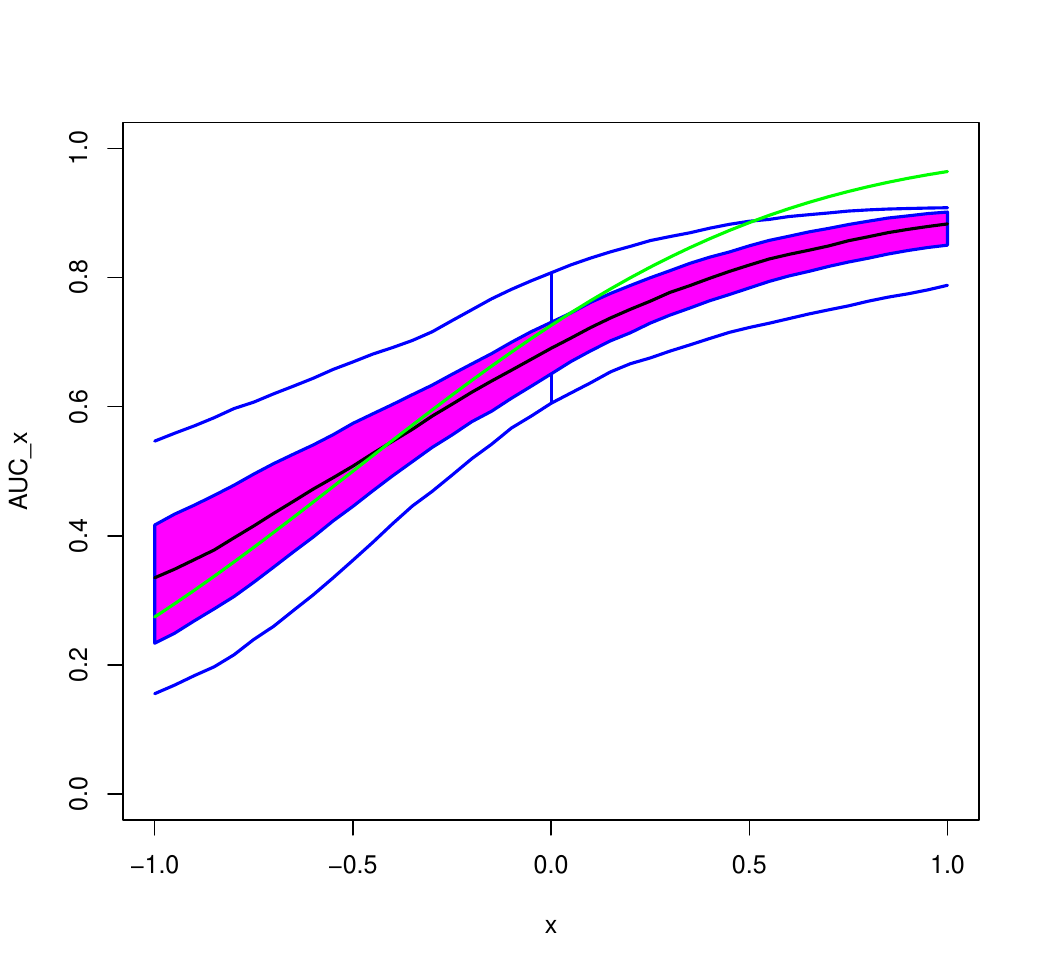} &  
\includegraphics[scale=0.35]{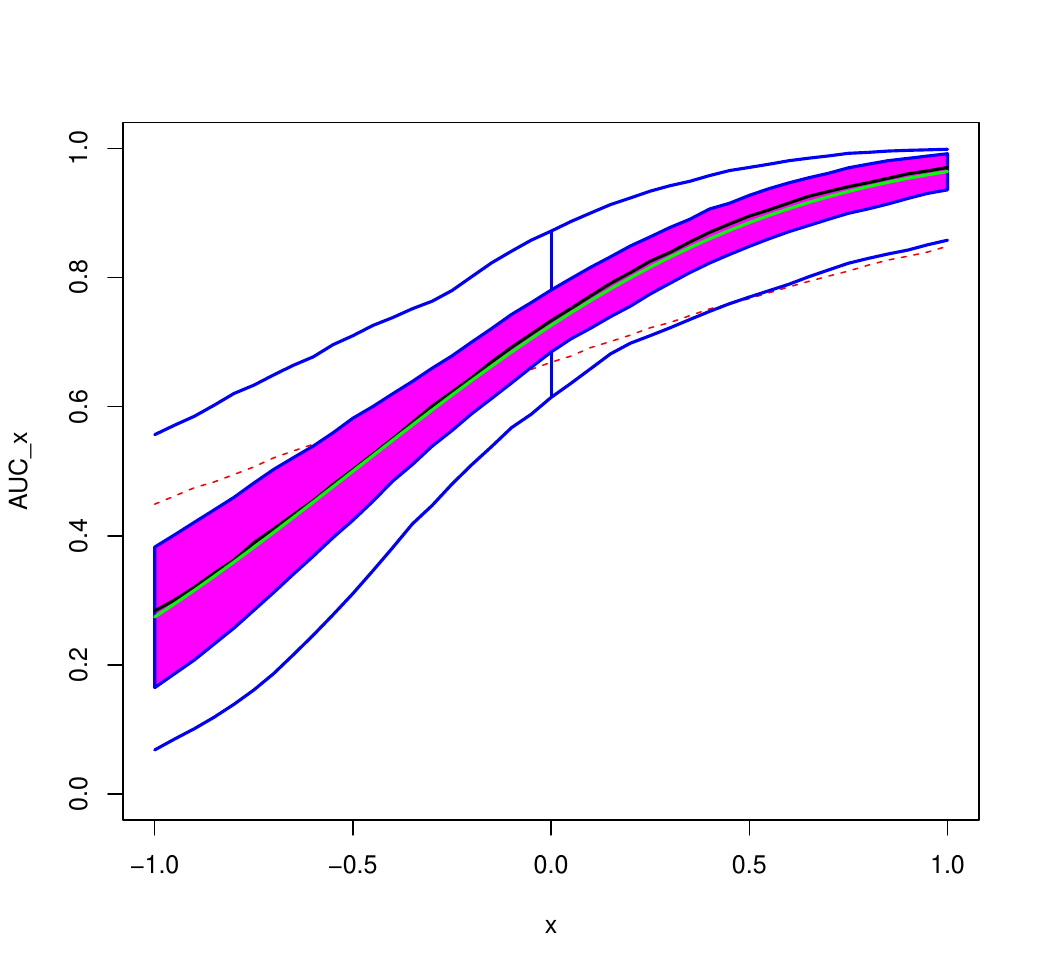}\\
\end{tabular}
\vskip-0.1in  
\caption{\footnotesize \label{fig:fbx_100}Functional boxplots of  $\widehat{\AUC}_x$   for $n=100$, under the linear model \eqref{trued} and \eqref{trueh}  when the samples are contaminated according to  $C_{\delta}$. The green line corresponds to the true $\AUC_x$ and the dotted red lines to the outlying curves detected by the functional boxplot.}
\end{center} 
\end{figure}

Figures \ref{fig:fbx_100} and  \ref{fig:fbx_200}  show  the functional boxplots  of $\widehat{\AUC}_x$ for both the classical and robust estimators when the samples are contaminated according to $C_\delta$, when $n=100$ and $n=200$, respectively.   These figures  reveal that the effect of outliers on the classical estimator of the ROC curve is  inherited by the estimated area under the curve,   which is reflected not only by the presence of a great number of outlying curves, but also by the enlargement of the width of the bars of the functional boxplots, as when contaminating only the diseased population. It should be noted that for $n=200$ and for values of $x$ in the range $[0.5,1]$, the true curve is on the limit of the central region.   As mentioned above, the robust procedure is stable for the considered contamination.  To conclude, these figures make evident the dramatic effect of  the introduced outliers  on the classical estimates of the area under the ROC curve, while at the same time the robust estimators look very stable.

\begin{figure}[H]
 \begin{center}
 \footnotesize
 \renewcommand{\arraystretch}{0.4}
 \newcolumntype{M}{>{\centering\arraybackslash}m{\dimexpr.1\linewidth-1\tabcolsep}}
   \newcolumntype{G}{>{\centering\arraybackslash}m{\dimexpr.4\linewidth-1\tabcolsep}}
\begin{tabular}{M GG}
& $n=100$ &  $n=200$\\
$C_{0}$ & 
\includegraphics[scale=0.35]{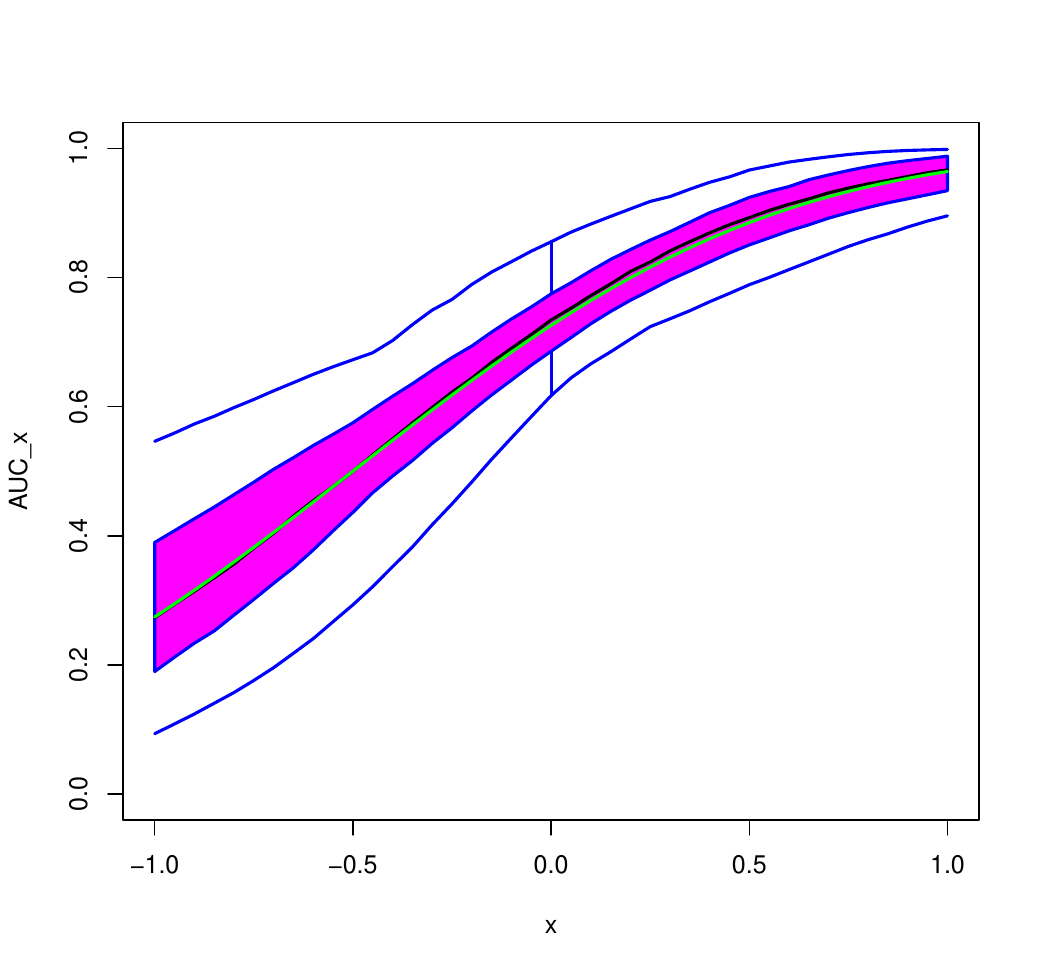} & 
\includegraphics[scale=0.35]{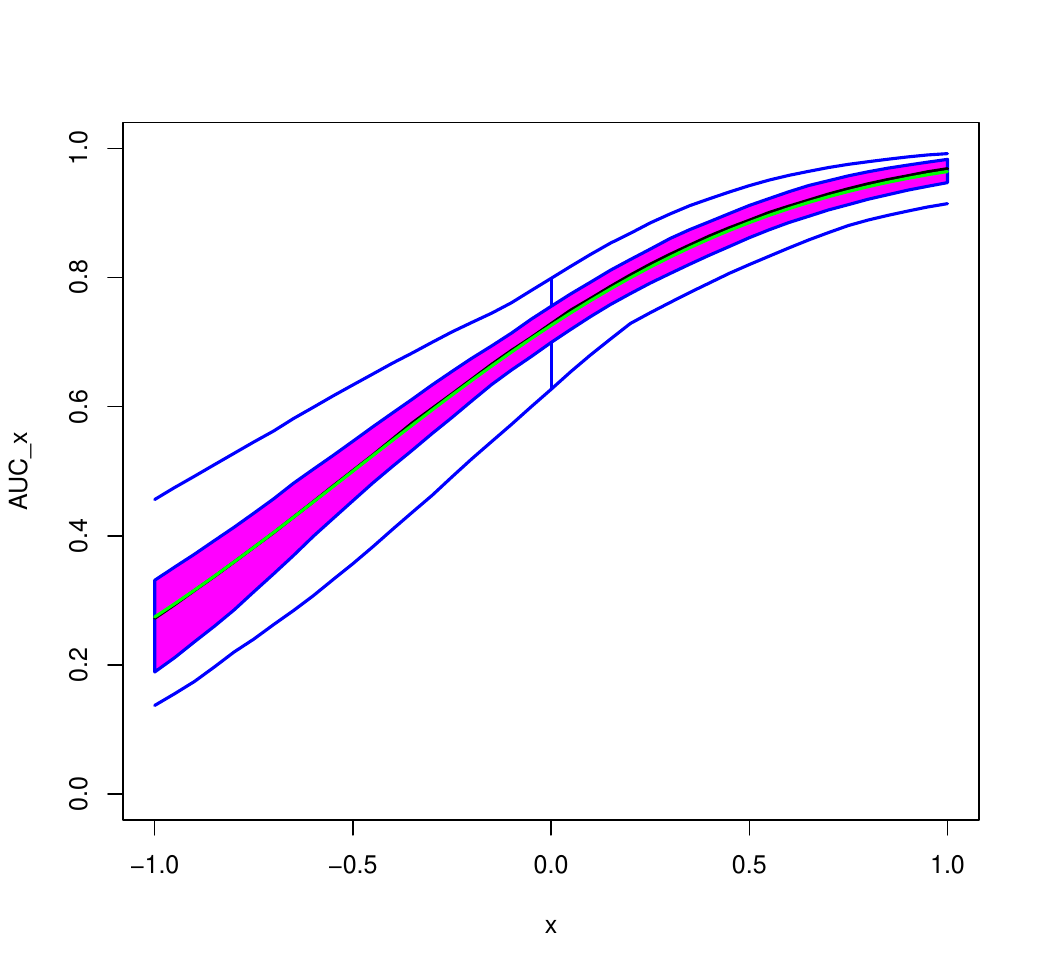}\\
$C_{0.05}$ & 
\includegraphics[scale=0.35]{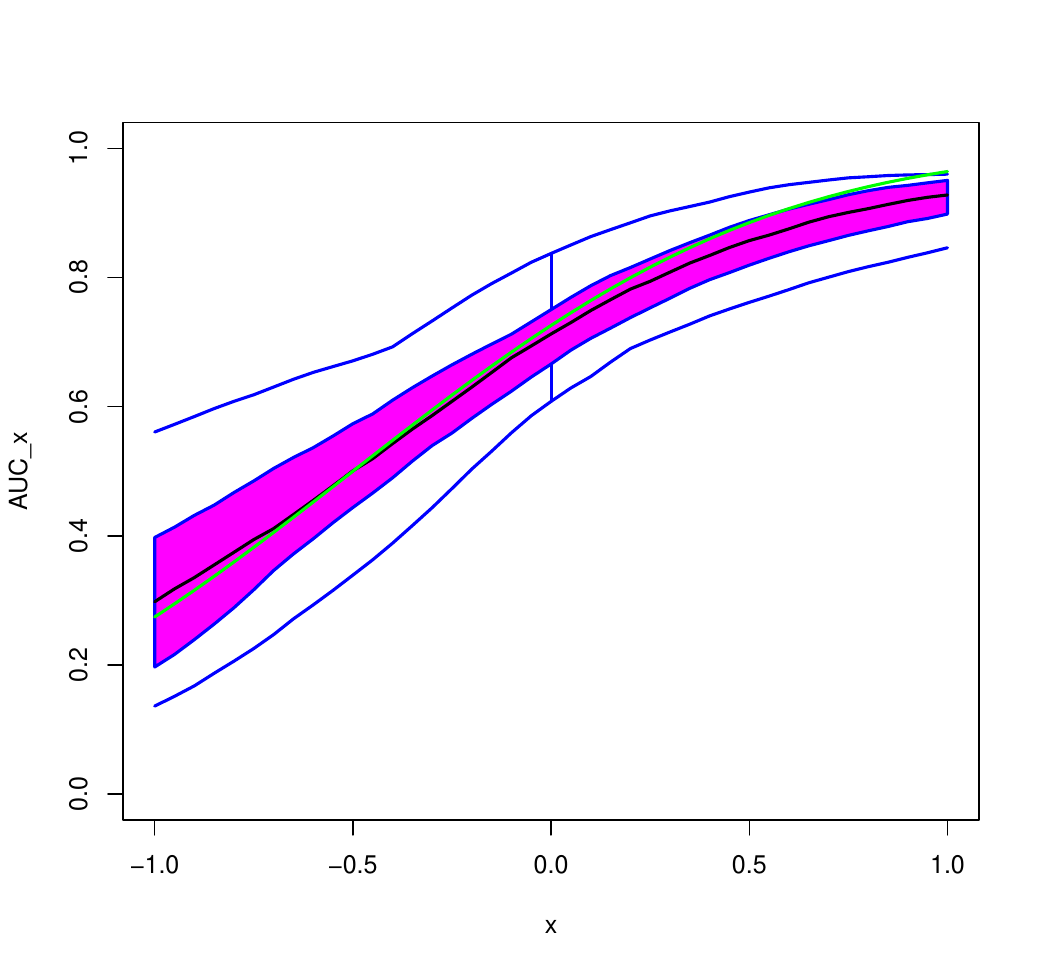} & 
\includegraphics[scale=0.35]{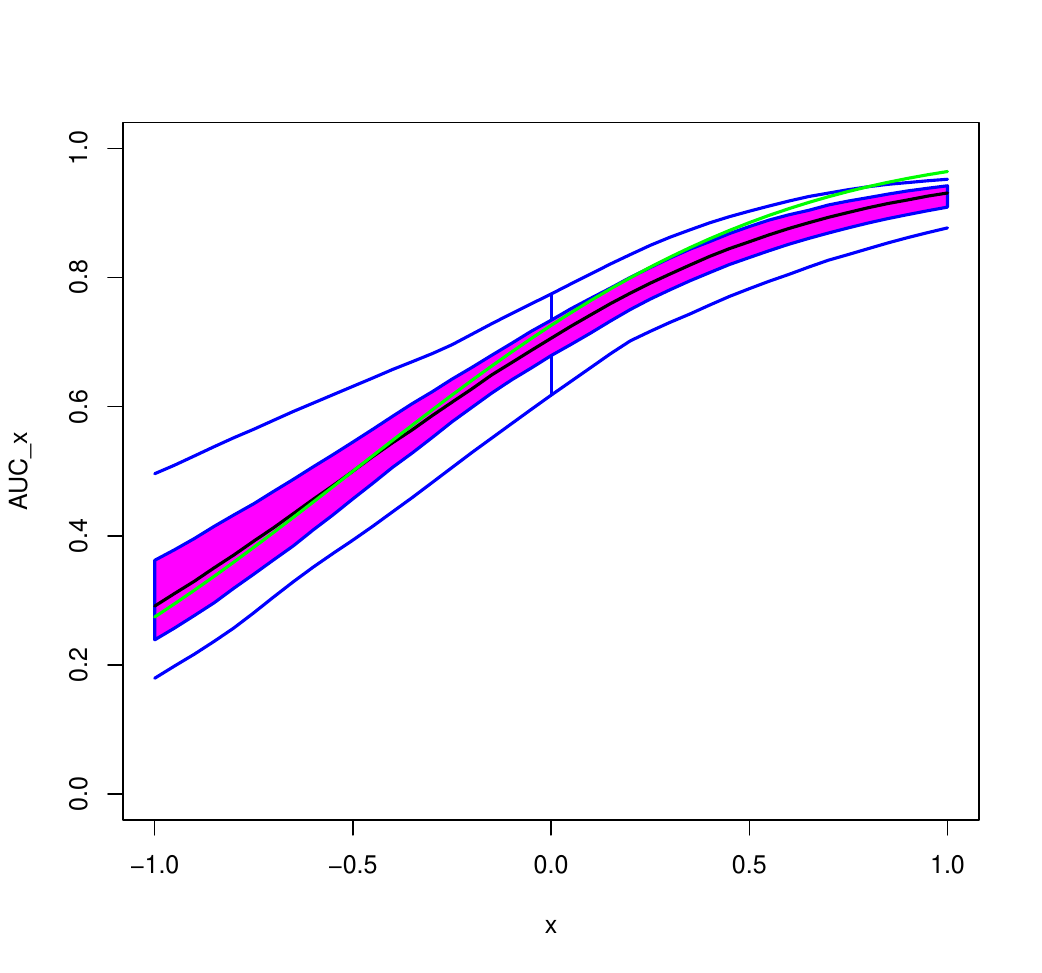}\\
$C_{0.10}$ & 
\includegraphics[scale=0.35]{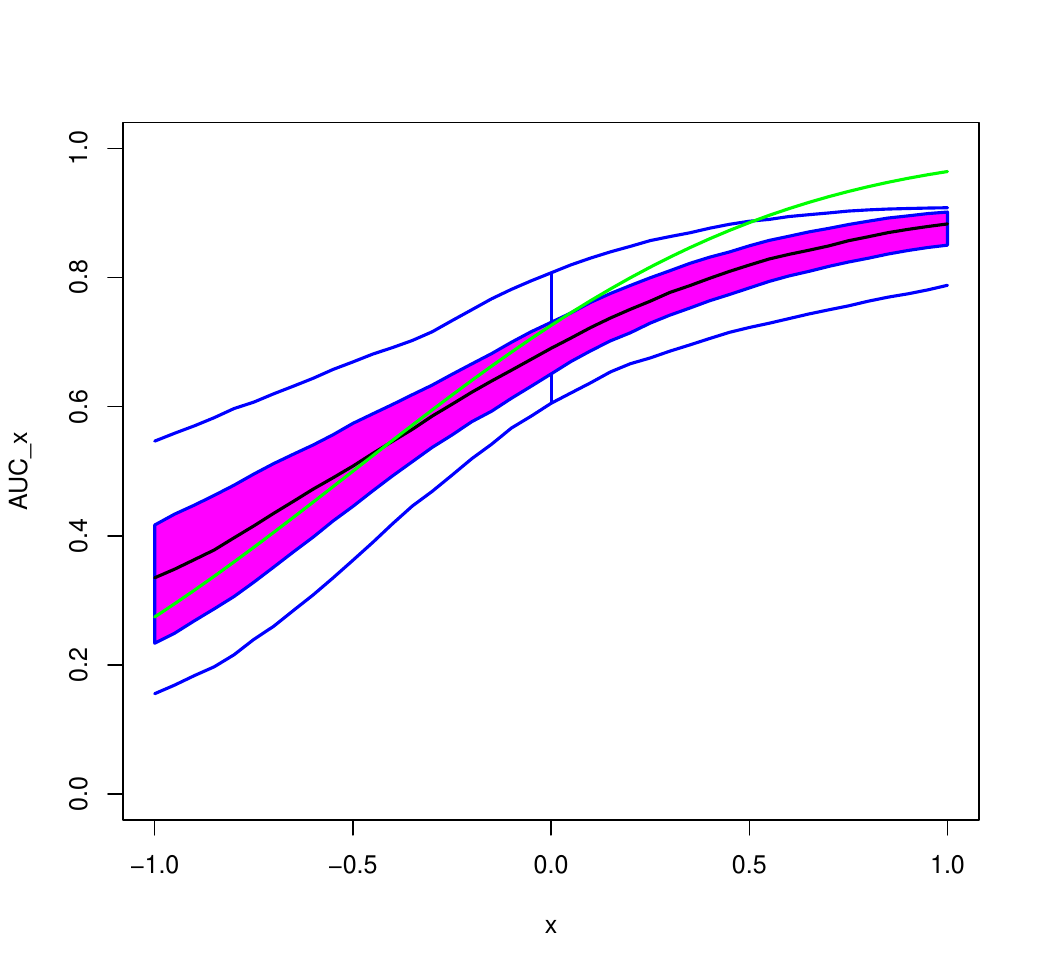} &  
\includegraphics[scale=0.35]{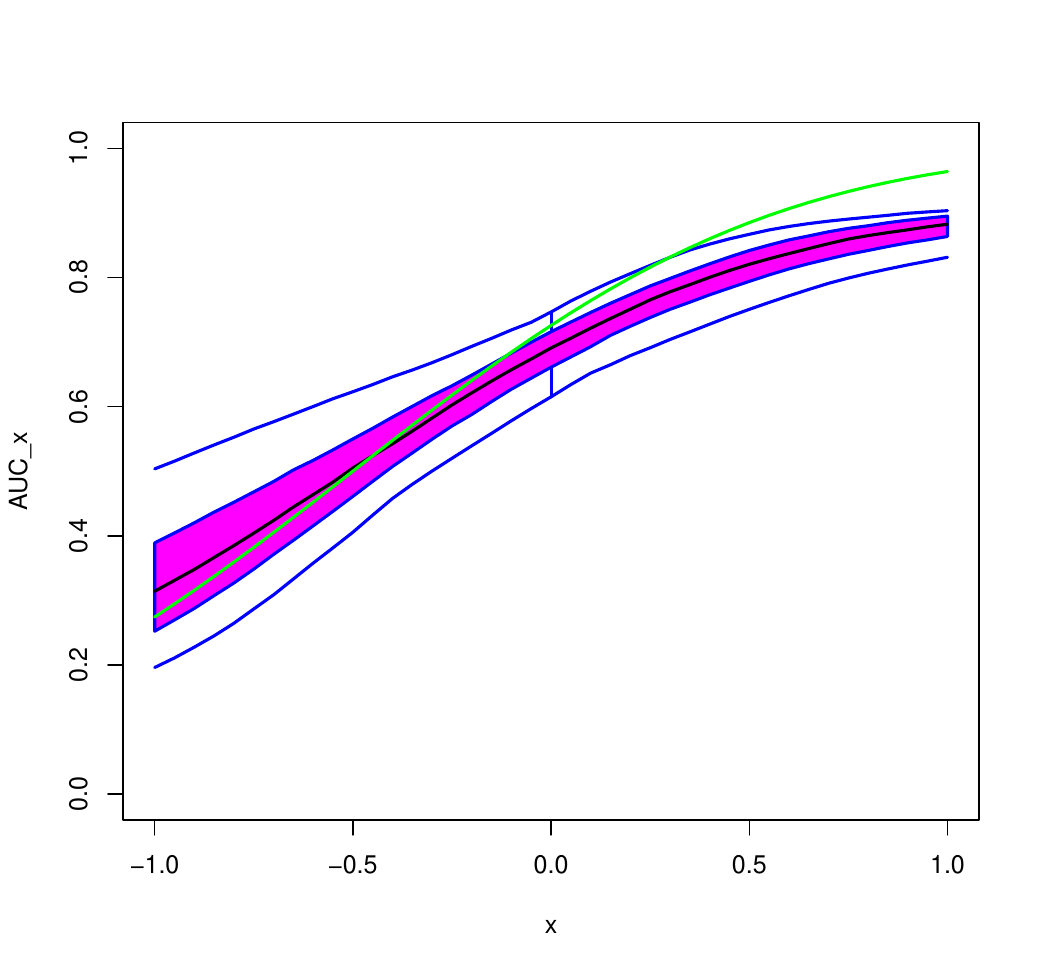}
\end{tabular}
\vskip-0.1in \caption{\footnotesize \label{fig:mix}Functional boxplots of  $\widehat{\AUC}_x$  obtained with the hybrid estimator for $n=100$ and  $n=200$ with clean samples and under 5\%  and 10\% of contamination under the first linear model. The green line corresponds to the true $\AUC_x$ and the dotted red lines to the outlying curves detected by the functional boxplot.}
\end{center} 
\end{figure}

To have a deeper comprehension of the proposal, it is also of interest to see what would happen if in the stepwise procedure described in Section \ref{sec:general}, robust estimators were considered only in the first step, i.e., only when computing the regression parameters, while    the usual  empirical distribution and quantile  function estimators are used in Step 2. The resulting hybrid procedure is illustrated through the functional boxplots of  $\widehat{\AUC}_x$  obtained  for $n=100$ and  $n=200$ in  Figure \ref{fig:mix}. These boxplots show that, even when the contamination is less harmful for these estimators than for the classical ones,  the true curve lies   beyond the functional boxplot 50\% inner band of curves when $x\in [0.5,1]$ and $\delta=0.05$ and beyond the limits of the functional boxplot when $\delta=0.10$.

\normalsize

\normalsize

\normalsize

\newpage

\clearpage
\normalsize

\normalsize

\subsection{Numerical study under a non--linear regression model}{\label{sec:nonlinearmodel}}

In this second scenario, we consider an exponential  model as in Bianco and Spano (2019), that is,  we assume that the observations follow  the  non--linear regression models 
\begin{eqnarray}
y_{D,i} &=&  \beta_{D,1}\; \exp(\beta_{D,2}\, x_{D,i} ) + \epsilon_{D,i} \,,\label{truednl}\\
y_{H,i} &=&   \beta_{H,1}\; \exp(\beta_{H,2} \, x_{H,i} ) +  \epsilon_{H,i}\; , \label{truehnl}
\end{eqnarray}
with $(\beta_{D,1},\beta_{D,2})\trasp=(5,2)$, $(\beta_{H,1},\beta_{H,2})\trasp=(3,1)$ for all $i=1,\dots,n$ $\epsilon_{j,i} \sim N(0,1)$ are independent and independent from   $x_{j,i} \sim U(-0.5,0.5)$, for $j=D,H$. Besides, the sample from one population was generated independently from that of the other one.  

In this case, in \textbf{Step 1},  the robust regression estimators correspond to  the weighted $MM-$estimators  defined in Bianco and Spano (2019), while   the classical ones to the usual least squares estimators for   nonlinear regression models. 

 To assess the impact of anomalous data on the estimation of the conditional ROC curve, we introduce   \textsl{shift outliers} in both populations. To explore the sensitivity of the studied methods to the size of the shift, we  vary its magnitude. To this end,  the first $m$ observations of each sample were replaced by observations following the models 
\begin{eqnarray*}
y_{D,i} &=&  \beta_{D,1}\; \exp(\beta_{D,2}\, x_{D,i} ) + z_{D,i} + 0.01 \epsilon_{D,i}\,, \label{contaminadnl}\\
y_{H,i} &=&  \beta_{H,1}\; \exp(\beta_{H,2} \, x_{H,i} )+ z_{H,i} + 0.01 \epsilon_{H,i}\; ,  \label{contaminahnl} 
\end{eqnarray*}
where $x_{j,i}\sim U(0.49,0.5)$  and $\epsilon_{j,i}$  are as above,  for $j=D,H$. The shift variables are taken as $ z_{j,i}=S +u_{j,i}$, with $S= 2.5, 5, 7.5, 10, 12.5, 15$, $u_{j,i} \sim N(0,0.01^2)$ for $j=D,H$, $i=1,\dots,m$.

We consider similar proportions of anomalous points as in Section \ref{sec:linearmodel}, that is, we replace  $m=n \delta$ points, $\delta=0.05$ and $0.10$, which correspond to a 5\% or a 10\% of replaced observations.  As above, we denote this contamination $C_{\delta}$, while $C_0$ stands for clean samples.
Table  \ref{tab:sens_5_100_NL}   summarizes the discrepancy between the true and estimated ROC curves in terms of  the mean over replications of the $MSE$. The damage of shift outliers on the conditional ROC curve is striking, since  the $MSE$ increases more than 10 times when $n=100$ and more than $20$ times when $n=200$ when $S$ takes the largest values. 

\begin{table}[ht!]
\begin{center}
\begin{tabular}{|c|c|c|c|c|c|c|c|c|c|}
  \hline
  & & & & \multicolumn{6}{c|}{$S$}\\
  \hline
$ \delta$ & $n$ & & $C_0$ &  2.5  &  5  &  7.5  &  10  &  12.5  &  15  \\ 
\hline
$0.05$ & $100$  
&  Robust & 0.0023 & 0.0028 & 0.0023 & 0.0023 & 0.0024 & 0.0024 & 0.0024 \\ 
& & Classical & 0.0019 & 0.0023 & 0.0066 & 0.0127 & 0.0183 & 0.0231 & 0.0269 \\ 
\hline
$0.10$ & $100$ 
&  Robust &  0.0023 & 0.0032 & 0.0031 & 0.0024 & 0.0024 & 0.0024 & 0.0024 \\ 
&  &  Classical  & 0.0019 & 0.0029 & 0.0092 & 0.0180 & 0.0240 & 0.0248 & 0.0268 \\ 
 \hline
 $0.05$ & $200$  
 &  Robust &  0.0011 & 0.0018 & 0.0011 & 0.0011 & 0.0011 & 0.0011 & 0.0011 \\ 
&  &  Classical   &  0.0010 & 0.0015 & 0.0062 & 0.0124 & 0.0181 & 0.0229 & 0.0267 \\ 
 \hline
 $0.10$ & $200$  
 &  Robust &  0.0011 & 0.0022 & 0.0014 & 0.0011 & 0.0011 & 0.0011 & 0.0011 \\ 
&  &  Classical  & 0.0010 & 0.0021 & 0.0088 & 0.0178 & 0.0239 & 0.0242 & 0.0249 \\ 
 \hline
\end{tabular}
\end{center}
\caption{\small \label{tab:sens_5_100_NL} Sensitivity of   $MSE$ to the shift  size  $S$ for $C_{\delta}$, when $n=100$ and $200$, under the nonlinear model \eqref{truednl} and \eqref{truehnl}.} 
\end{table}

Henceforth, we focus on the particular case of outliers with shift value $S=10$, a mild value among those considered,  so as to have a deeper comprehension of the effect of the introduced anomalous points.
 Table  \ref{tab:mse_NL}   summarizes the results through the mean  of the measures $MSE$  and $KS$. Note that   the mean of the summary measures are distorted for the classical procedure.  In particular, when considering the measure $KS$ based on the Kolmogorov distance,   the mean is enlarged almost 7 times, under $C_{0.05}$ when $n=200$. 
 
\begin{figure}[H]
 \begin{center}
 \footnotesize
 \renewcommand{\arraystretch}{0.4}
 \newcolumntype{M}{>{\centering\arraybackslash}m{\dimexpr.1\linewidth-1\tabcolsep}}
   \newcolumntype{G}{>{\centering\arraybackslash}m{\dimexpr.4\linewidth-1\tabcolsep}}
\begin{tabular}{M GG}
& Classical &  Robust\\
$C_{0}$ & 
\includegraphics[scale=0.35]{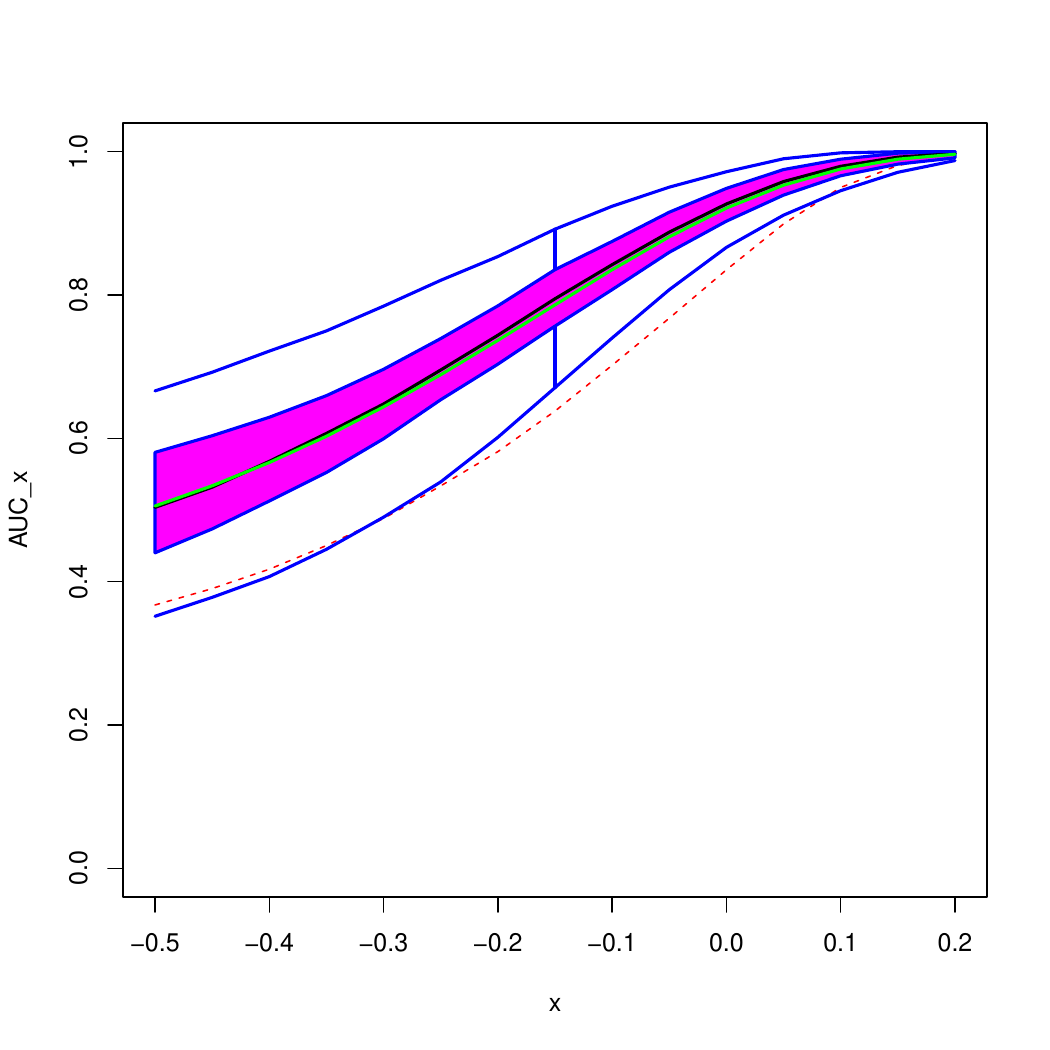} & 
\includegraphics[scale=0.35]{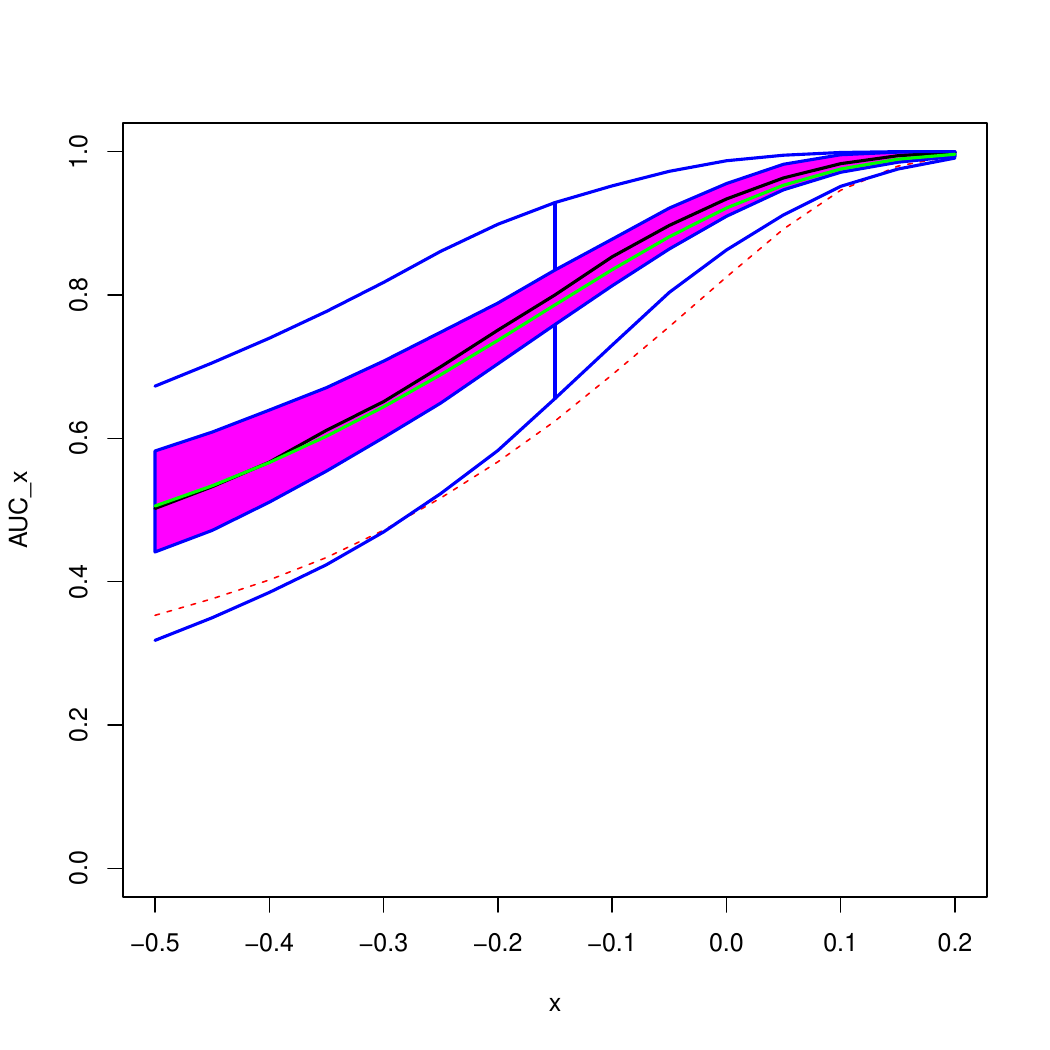}\\
$C_{0.05}$ & 
\includegraphics[scale=0.35]{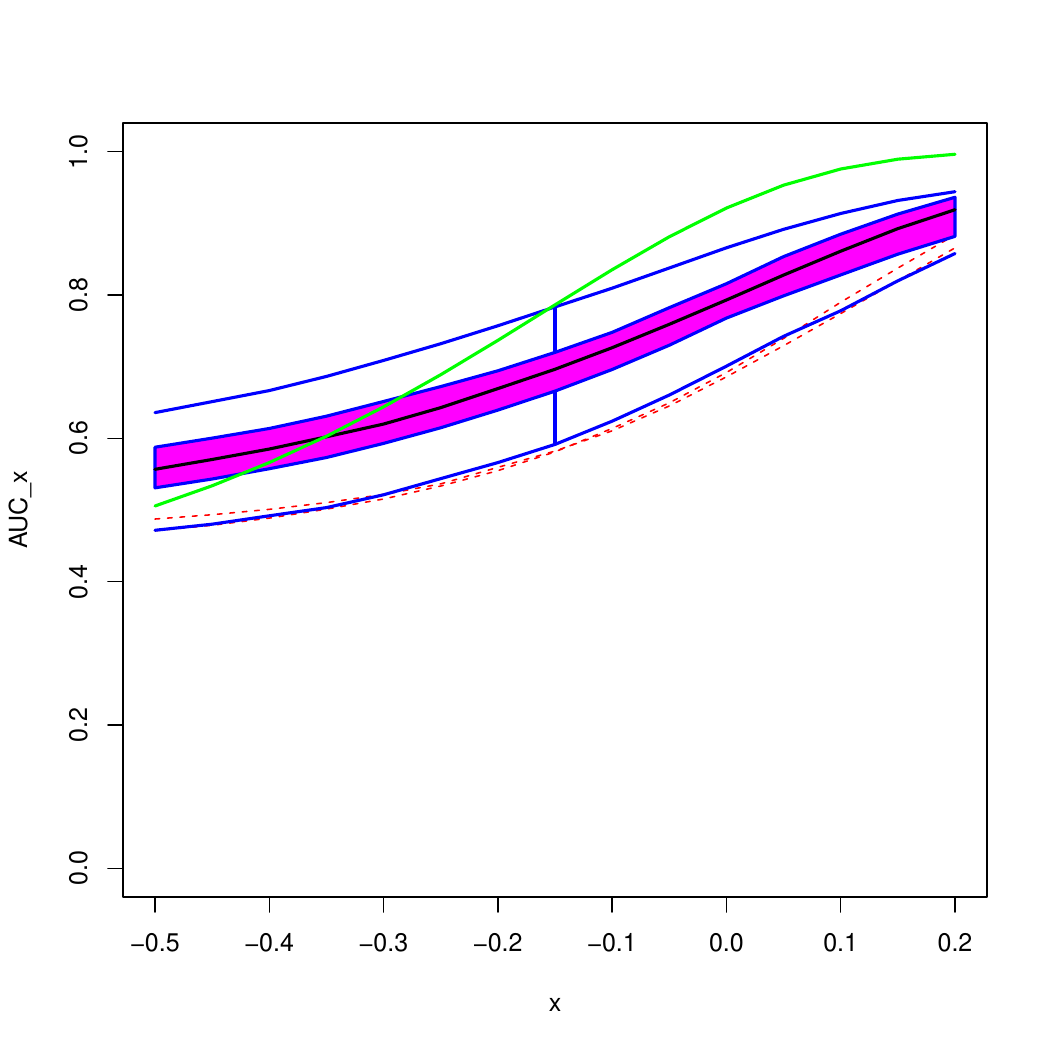} & 
\includegraphics[scale=0.35]{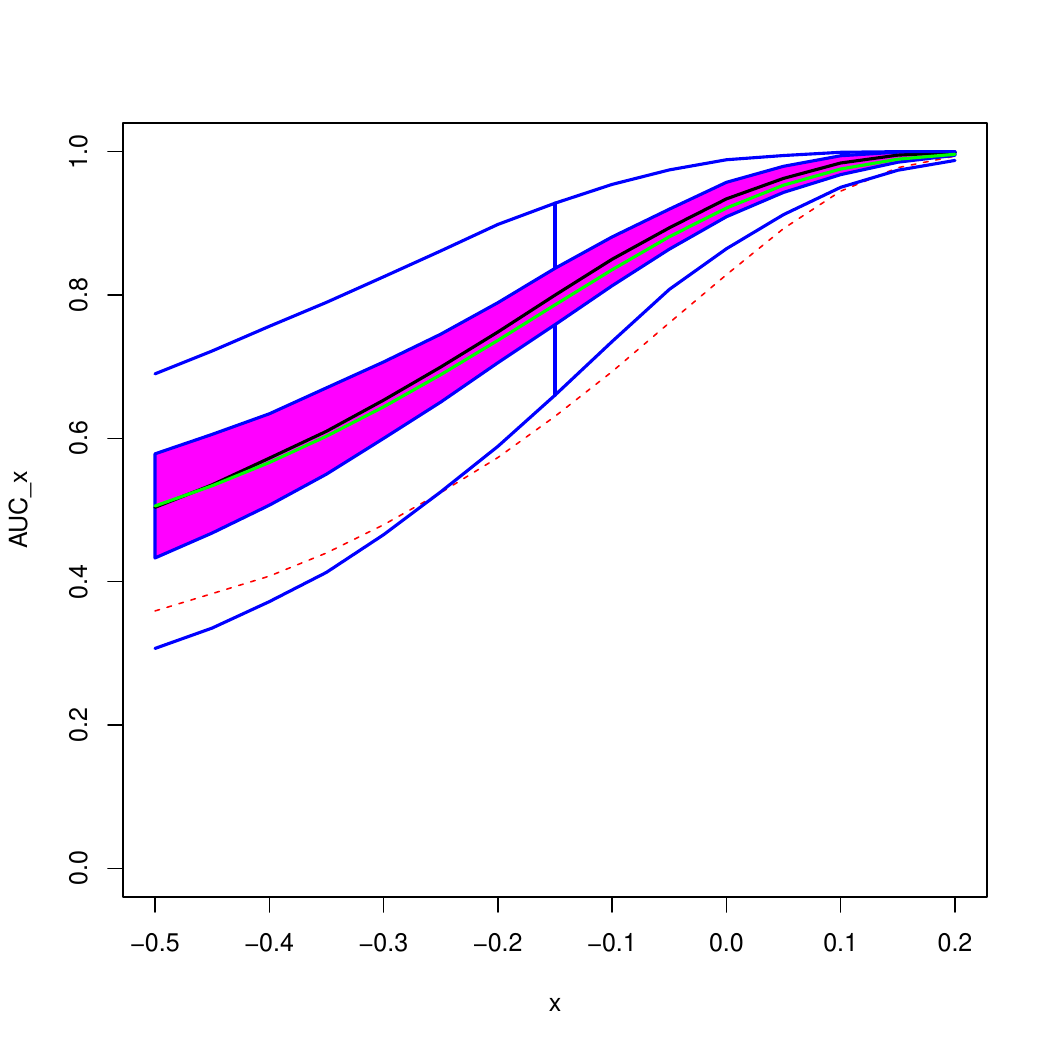}\\
$C_{0.10}$ &
\includegraphics[scale=0.35]{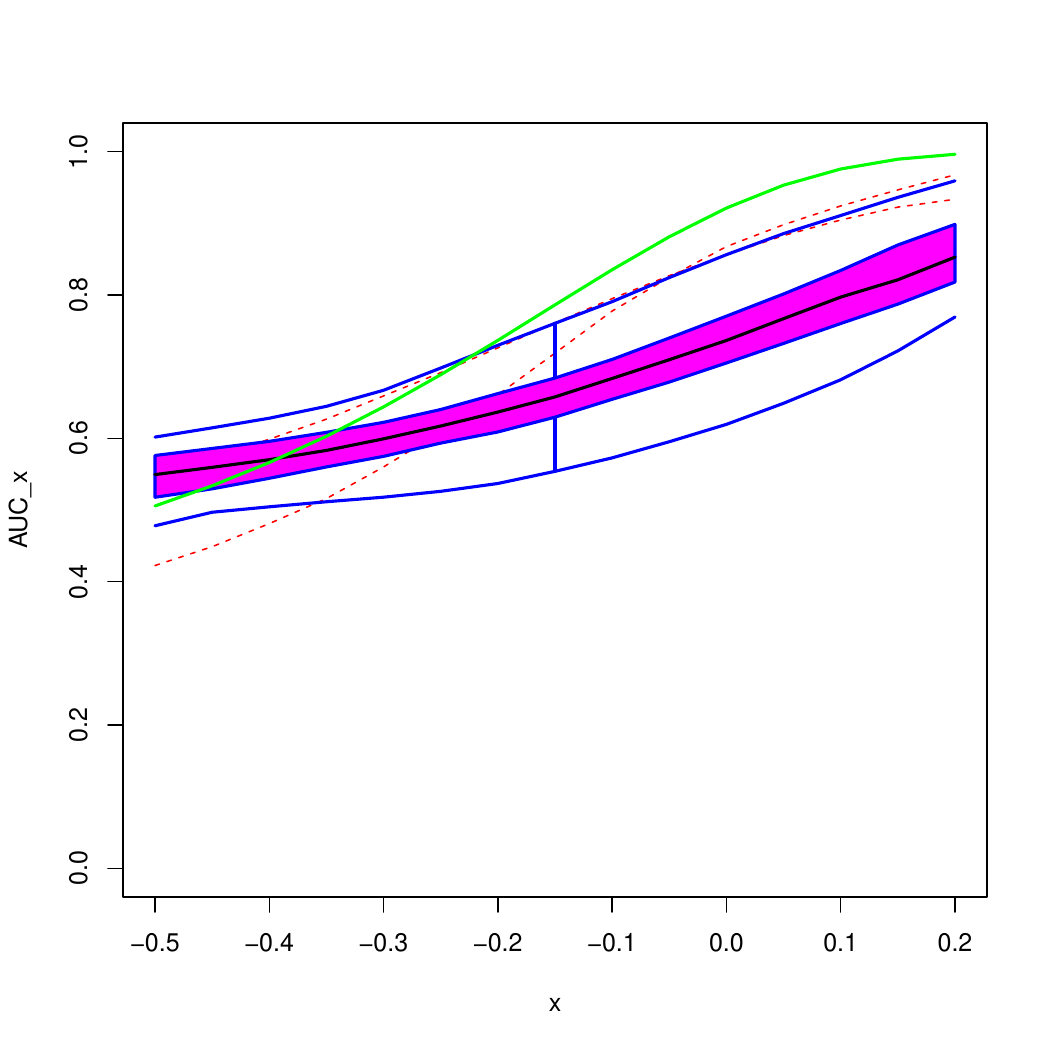} & 
\includegraphics[scale=0.35]{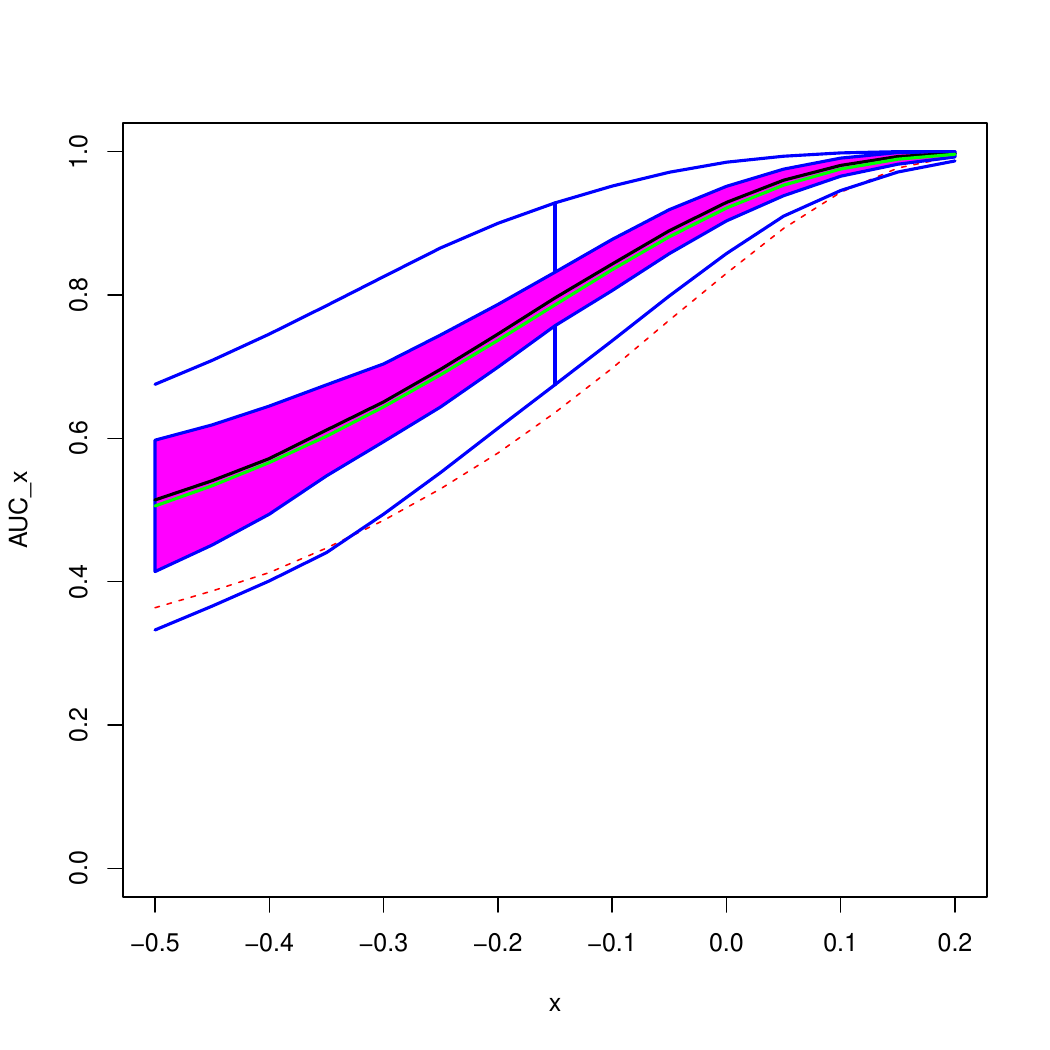}
\end{tabular}
\vskip-0.1in \caption{\footnotesize \label{fig:n100-NL} \footnotesize Functional boxplots of  $\widehat{\AUC}_x$  obtained with the classical and robust estimators for $n=100$ with clean samples and under 5\%  and 10\% of contamination  with level shift $S=10$, under the nonlinear model \eqref{truednl} and \eqref{truehnl}. The green line corresponds to the true $\AUC_x$ and the dotted red lines to the outlying curves detected by the functional boxplot.}
\end{center} 
\end{figure}

\begin{figure}[H]
 \begin{center}
 \footnotesize
 \renewcommand{\arraystretch}{0.4}
 \newcolumntype{M}{>{\centering\arraybackslash}m{\dimexpr.1\linewidth-1\tabcolsep}}
   \newcolumntype{G}{>{\centering\arraybackslash}m{\dimexpr.4\linewidth-1\tabcolsep}}
\begin{tabular}{M GG}
& Classical &  Robust\\
$C_{0}$ & 
\includegraphics[scale=0.35]{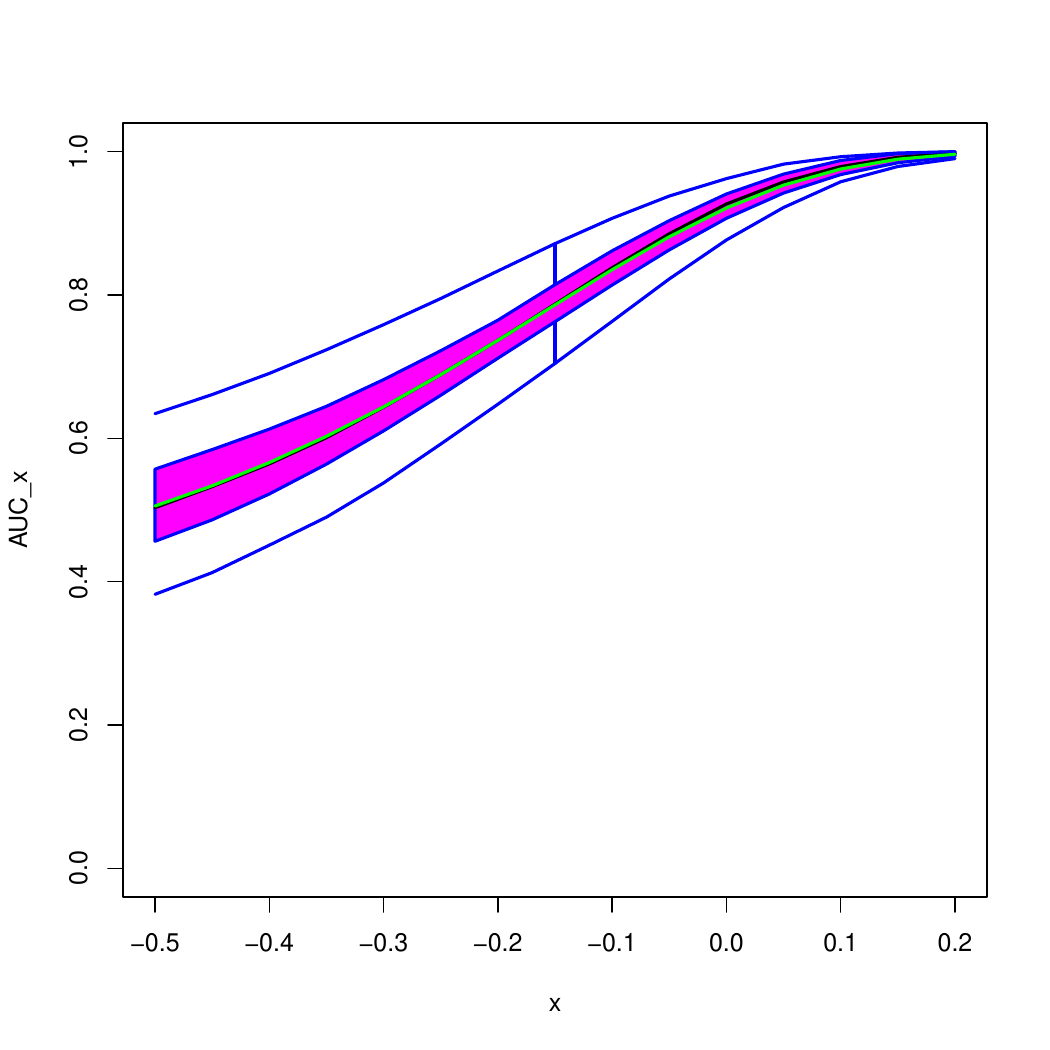} & 
\includegraphics[scale=0.35]{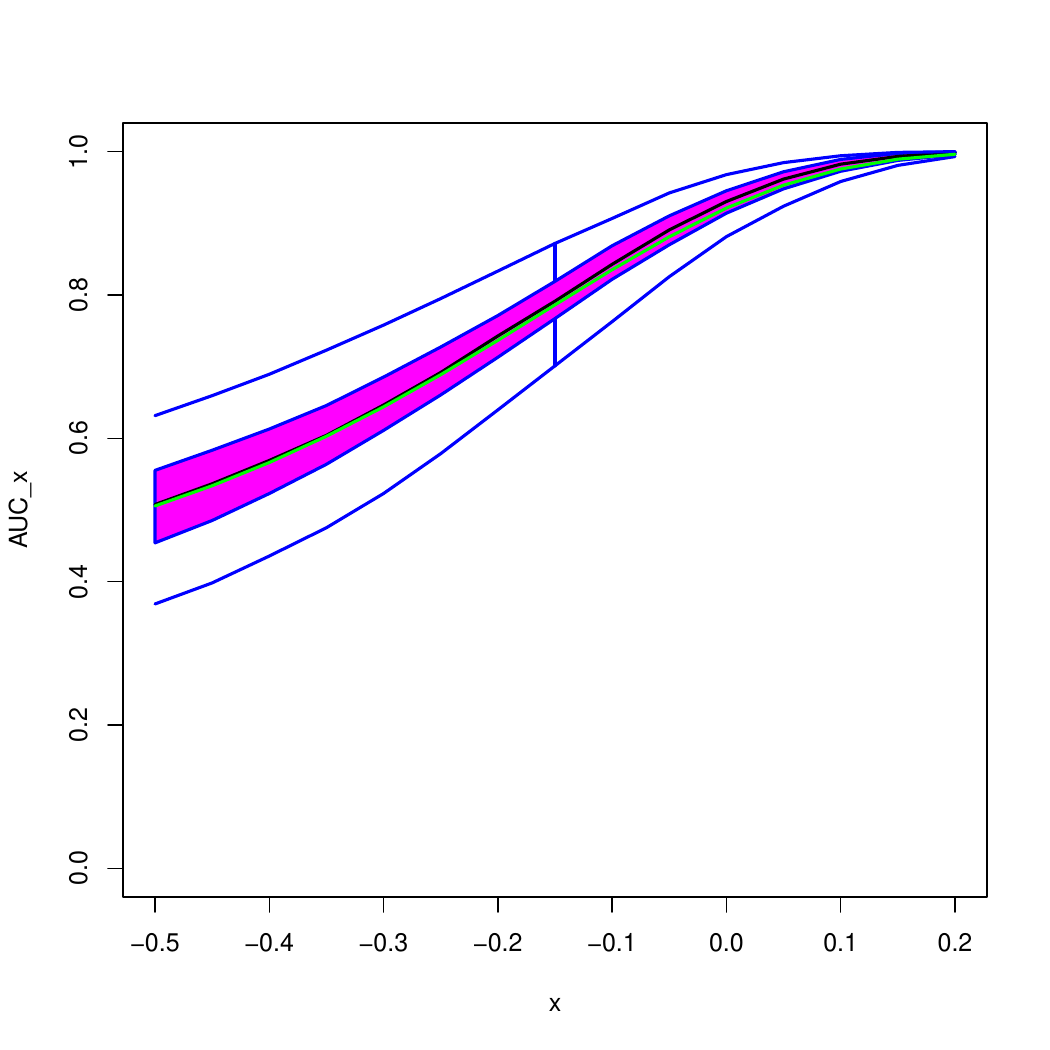}\\
$C_{0.05}$ & 
\includegraphics[scale=0.35]{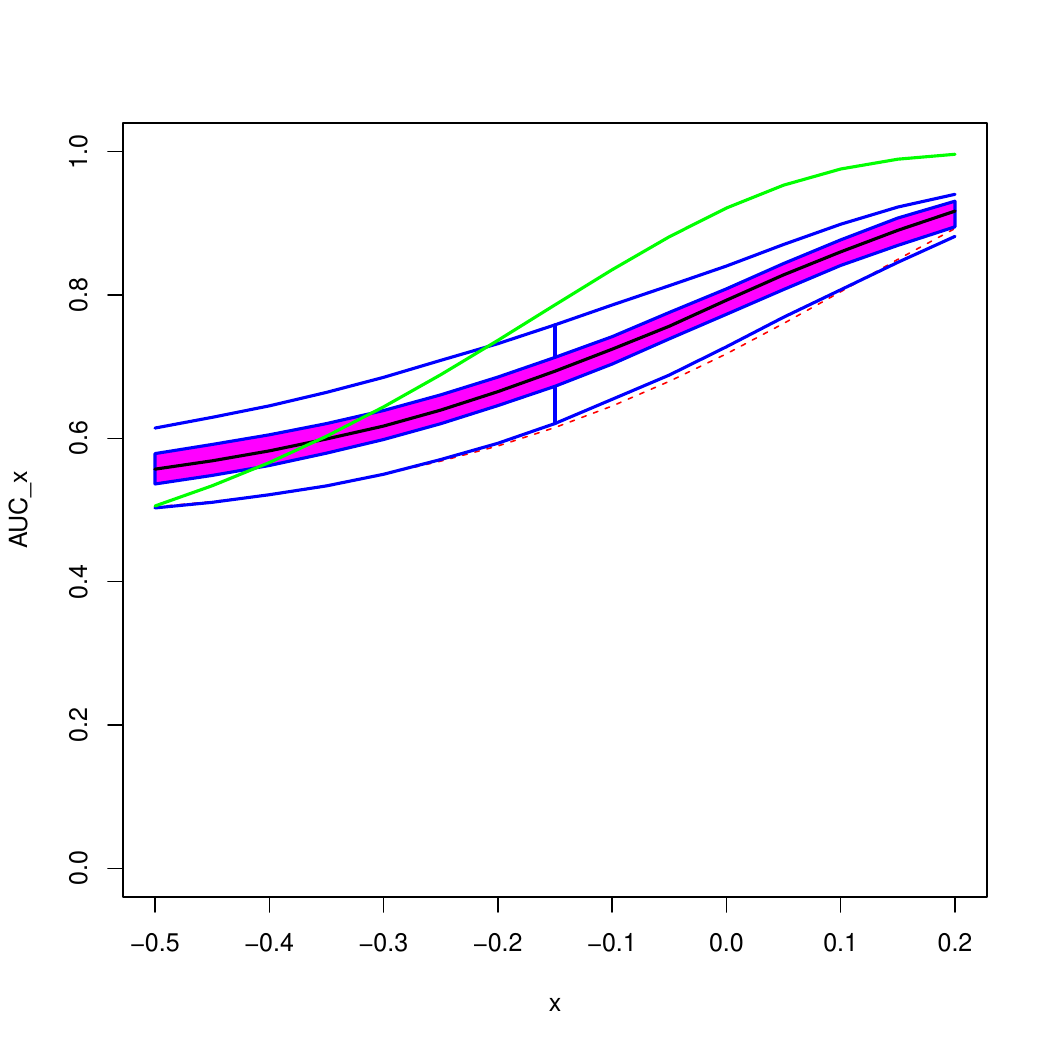} & 
\includegraphics[scale=0.35]{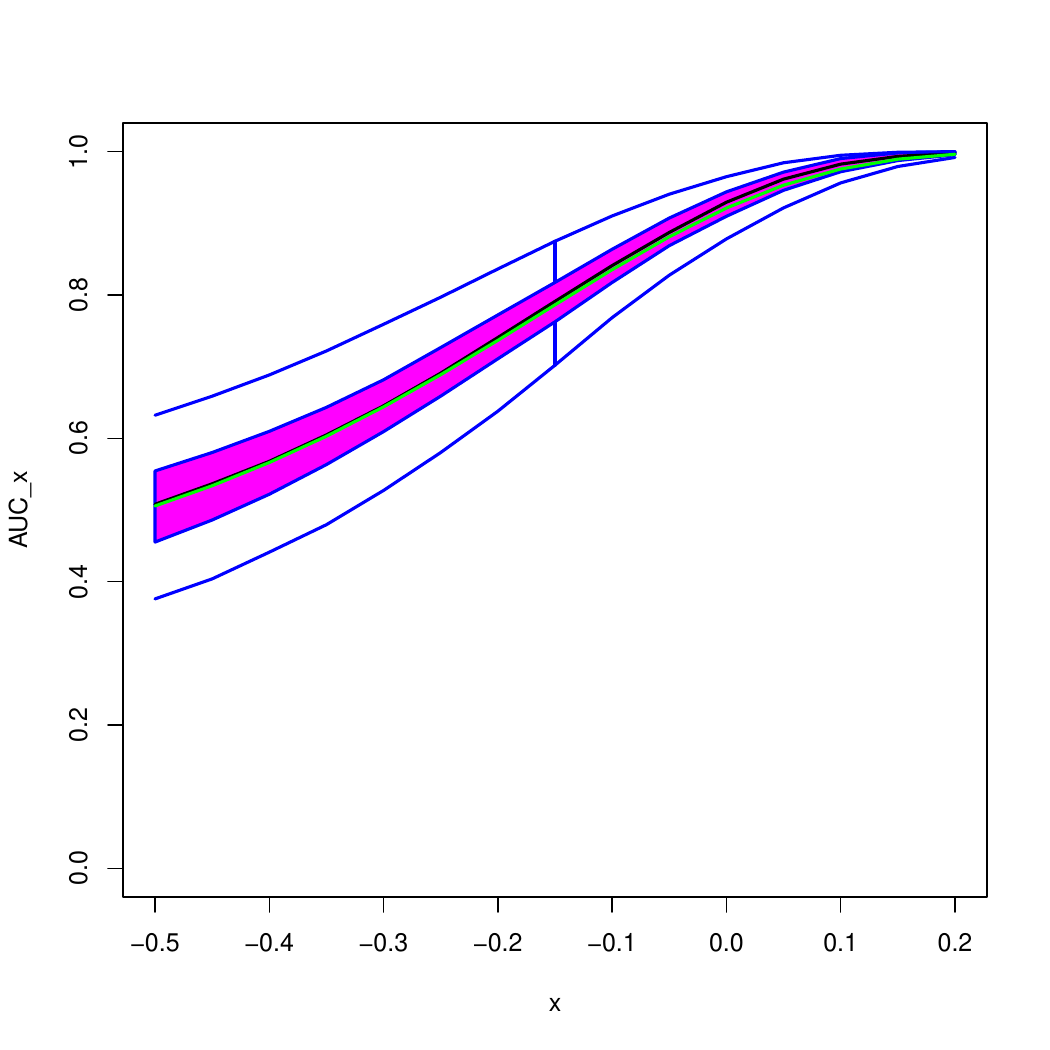}\\
$C_{0.10}$ &
\includegraphics[scale=0.35]{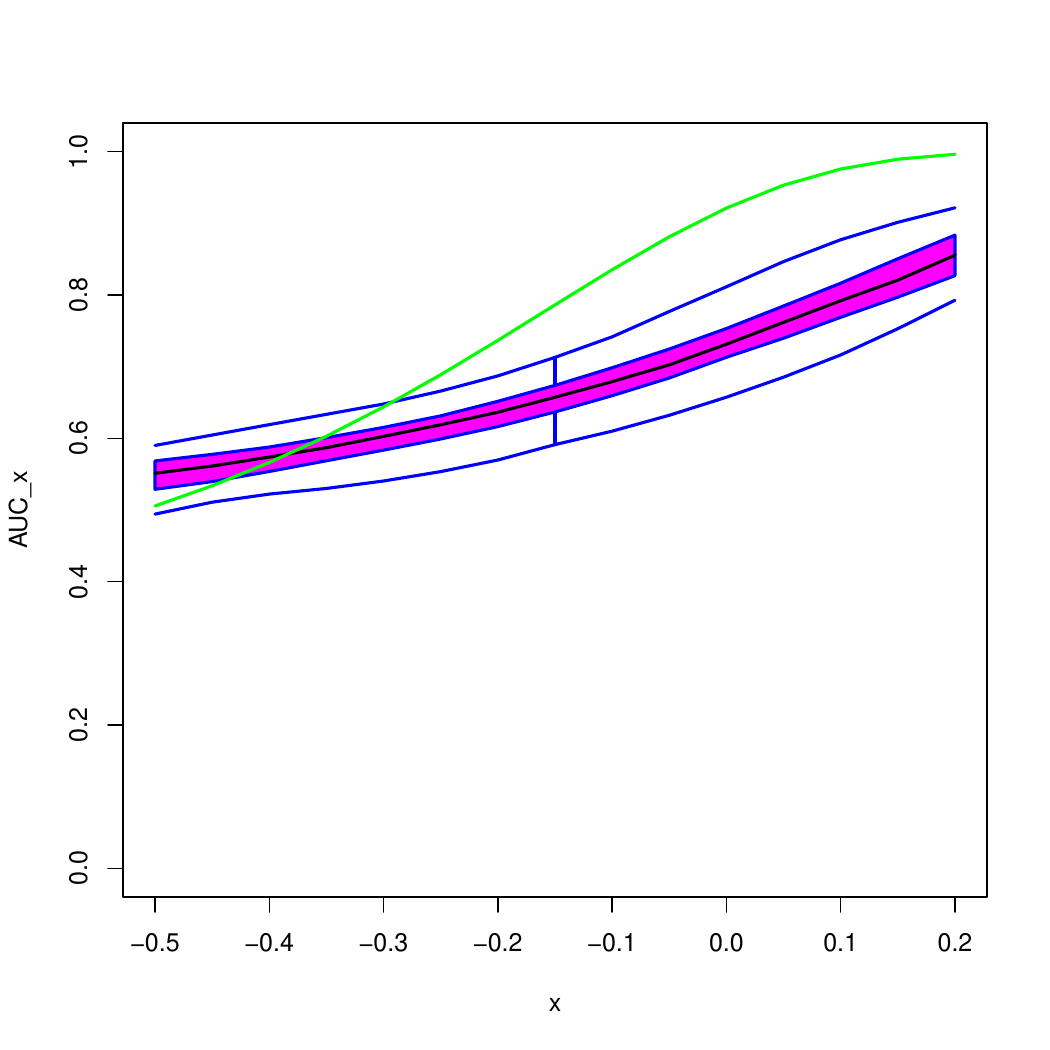} & 
\includegraphics[scale=0.35]{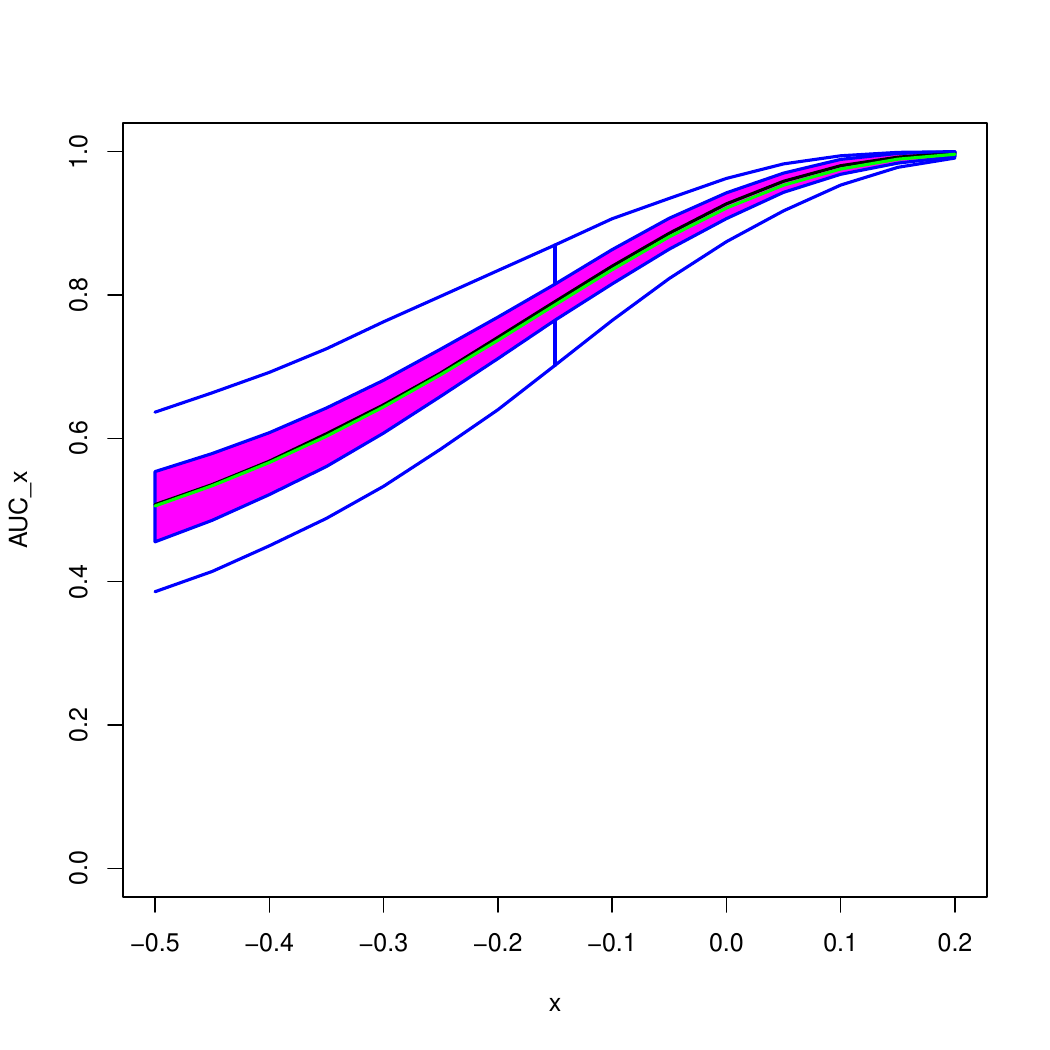}
\end{tabular}
\vskip-0.1in \caption{\footnotesize \label{fig:n200-NL} \footnotesize Functional boxplots of  $\widehat{\AUC}_x$  obtained with the classical and robust estimators for $n=200$ with clean samples and under 5\%  and 10\% of contamination with level shift $S=10$, under the nonlinear model \eqref{truednl} and \eqref{truehnl}. The green line corresponds to the true $\AUC_x$ and the dotted red lines to the outlying curves detected by the functional boxplot.}
\end{center} 
\end{figure}

Figures \ref{fig:n100-NL} and \ref{fig:n200-NL} show the functionals boxplots of the classical and robust ${\AUC}_x$  obtained  for $n=100$ and  $n=200$, respectively. Notice that in these boxplots, the estimators of conditional area under the curve were plotted  in the range $(-0.5,0.2)$, since for this simulation scheme the ${\AUC}_x$ is almost 1 when the covariate takes values from 0.2 to 0.5. Once again, it becomes evident that the classical estimator suffers from the introduced contamination and that the classical estimator of ${\AUC}_x$ is completely deviated from the true conditional area under the curve, which is plotted in green, while the robust $ {\AUC}_x$ estimator remains very stable.

\begin{table}[ht!]
\begin{center}
\begin{tabular}{|c|  c|c|c|c|c|c|c|}
\hline 
$n$ & &\multicolumn{2}{c|}{$C_0$}& \multicolumn{2}{c|}{$C_{0.05}$}& \multicolumn{2}{c|}{$C_{0.10}$}\\ 
\hline
 &   & Classical & Robust& Classical & Robust& Classical & Robust\\ 
\hline
100   & $MSE$ & 0.0019 & 0.0023 & 0.0183 & 0.0024 & 0.0240 & 0.0024 \\ 
  &  $KS$ & 0.1881 & 0.1944 & 0.9334 & 0.2001 & 0.6893 & 0.2048 \\ 
\hline 
200  & $MSE$ & 0.0010 & 0.0011 & 0.0181 & 0.0011 & 0.0239 & 0.0011 \\  
& $KS$ & 0.1352 & 0.1367 & 0.9364 & 0.1395 & 0.7102 & 0.1445 \\ 
\hline
\end{tabular} 
\end{center}
\caption{\small \label{tab:mse_NL} Mean    of $MSE$ and $KS$ over replications for clean and contaminated samples, under the nonlinear model \eqref{truednl} and \eqref{truehnl}, for the level shift $S=10$.}
\end{table}

\normalsize

\normalsize

\section{Analysis of real data set}{\label{sec:realdata}}

In this section, we  illustrate  the benefits of  the robust proposed methodology by means of the  diabetes real dataset described in the Introduction.

\begin{figure}[ht!]
	\begin{center}
		\footnotesize
		\begin{tabular}{ccc}
			\includegraphics[scale=0.3]{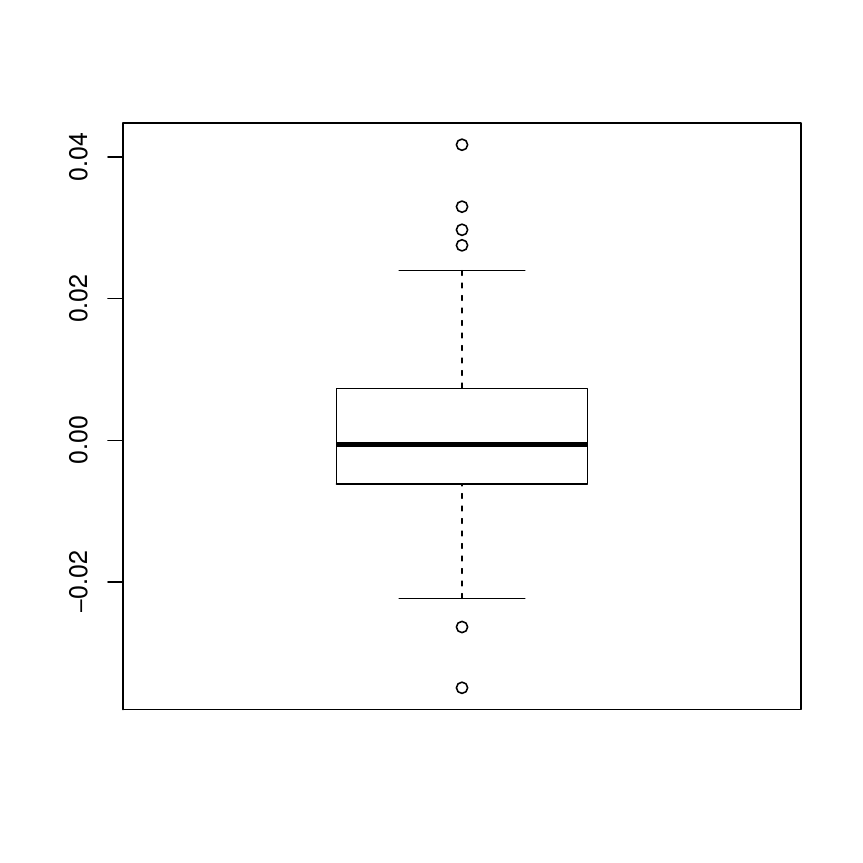}&
			\includegraphics[scale=0.3]{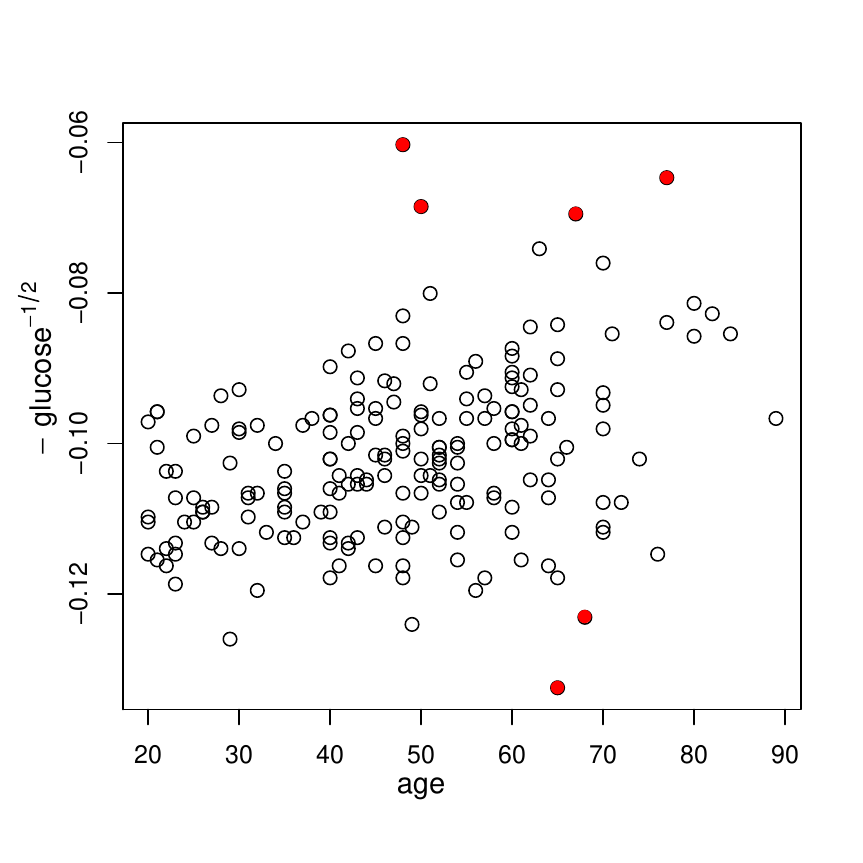}&
			\includegraphics[scale=0.3]{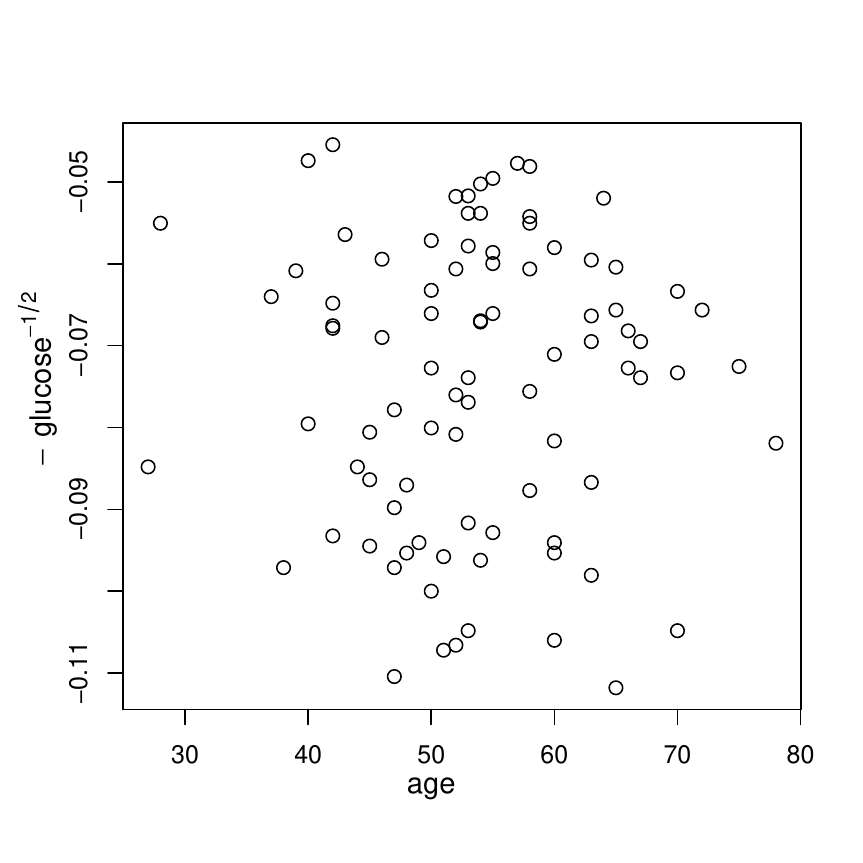} 
		\end{tabular}
		\vskip-0.1in \caption{\label{fig:boxplotejemplo}  The left panel corresponds to the boxplots of the residuals obtained after a robust fit for healthy sample, while the central and right panels to the scatter plots for the healthy and diseased samples.}
	\end{center} 
\end{figure}

\begin{figure}[H]
	\begin{center}
		\footnotesize
\renewcommand{\arraystretch}{0.1}
\setlength{\tabcolsep}{-2pt}
		\begin{tabular}{ccc}
		\multicolumn{3}{c}{(a)}\\
		\multicolumn{3}{c}{$\quad$}\\
		\multicolumn{3}{c}{$\quad$}\\
		$\widehat{\ROC}_{\bx, \cl}$ & $ \widehat{\ROC}_{\bx }$ & $ \widehat{\ROC}_{\bx, \cl}^{(-6)}$\\
			\includegraphics[scale=0.35]{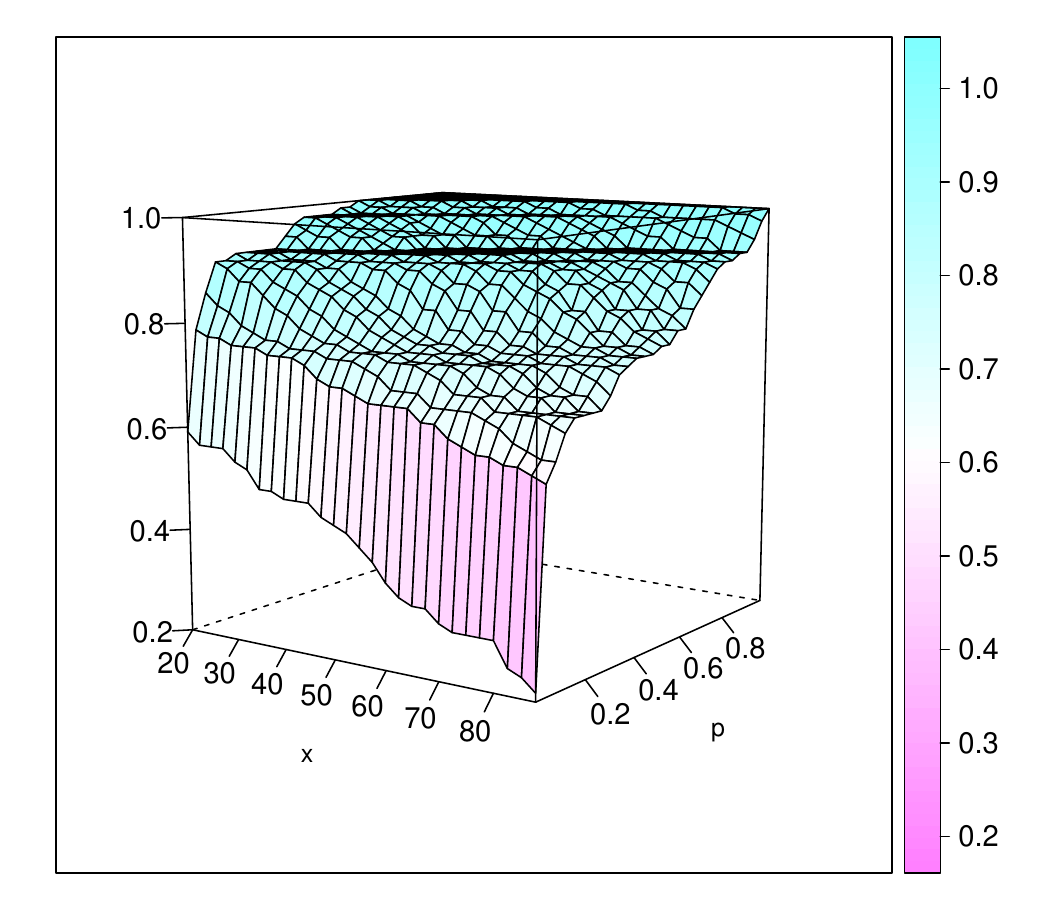} &  
			\includegraphics[scale=0.35]{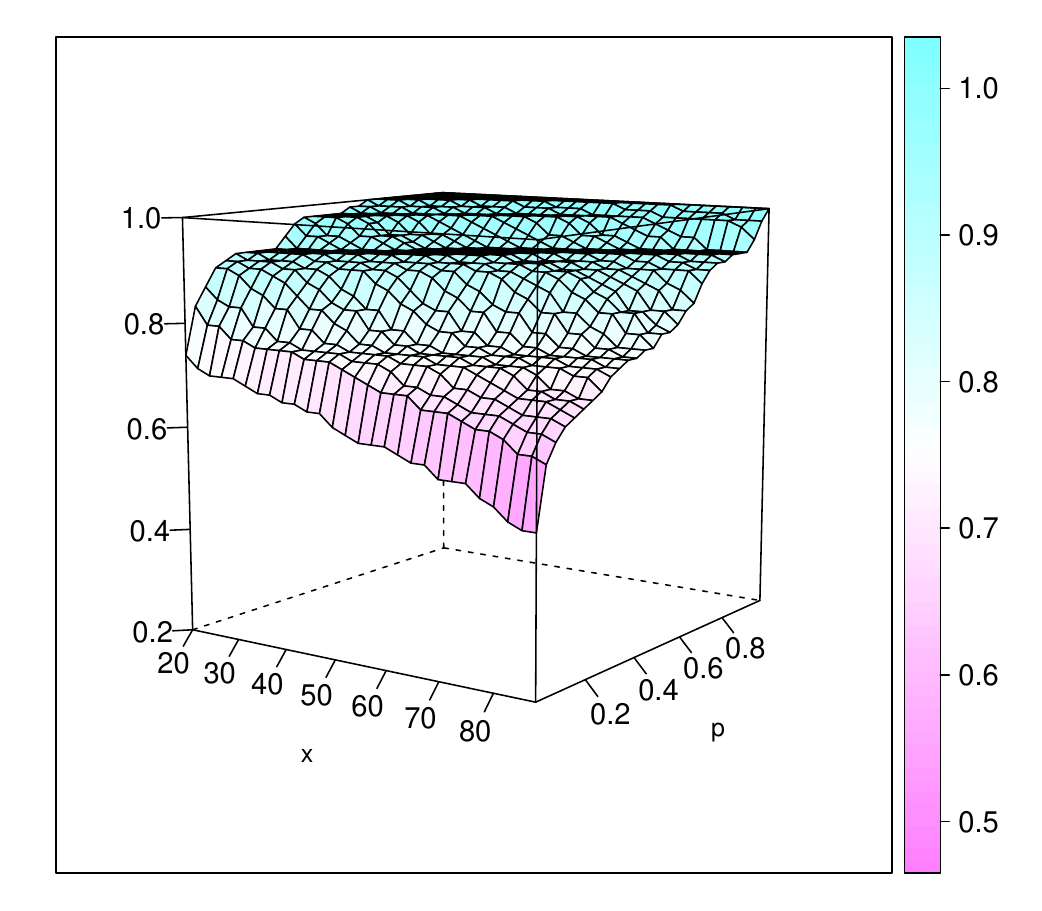} & 
			\includegraphics[scale=0.35]{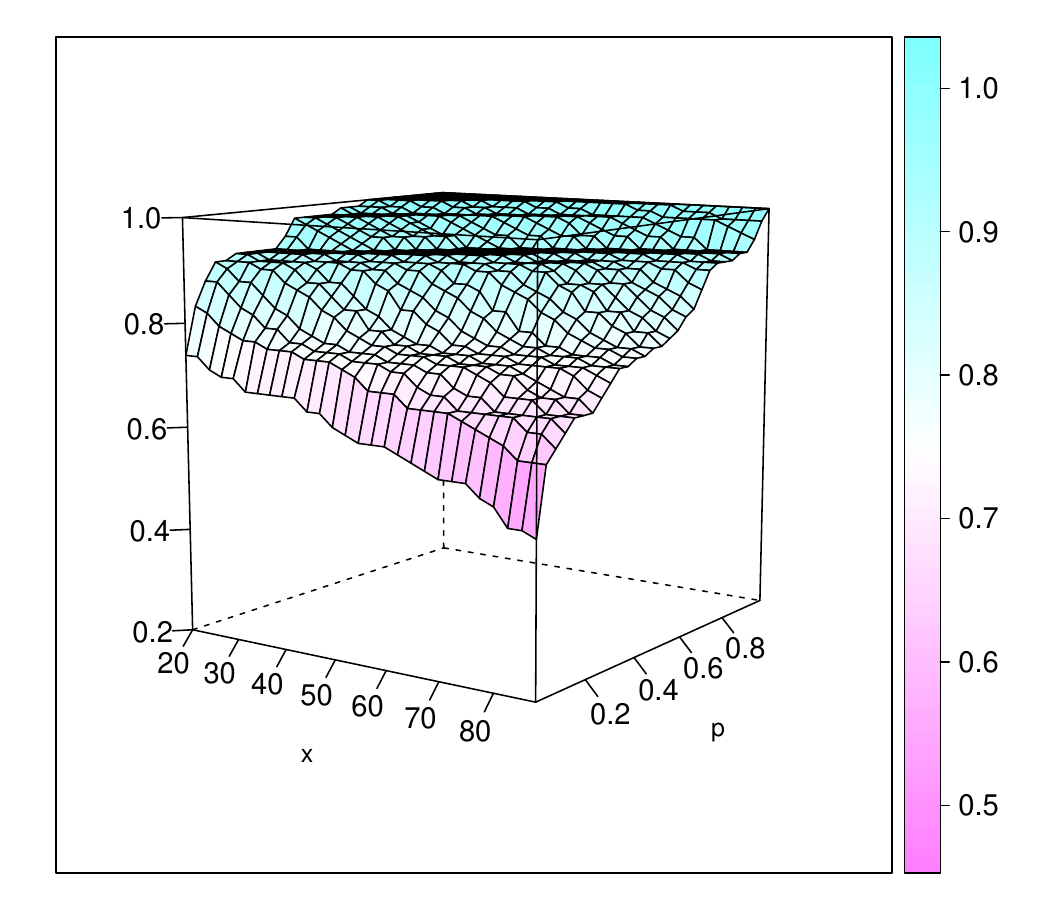}\\
			\multicolumn{3}{c}{(b)}\\
			\multicolumn{3}{c}{$\quad$}\\
			\multicolumn{3}{c}{$\quad$}\\
			$\widehat{\ROC}_{\bx }-\widehat{\ROC}_{\bx, \cl}$ & $\widehat{\ROC}_{\bx, \cl}^{(-6)}-\widehat{\ROC}_{\bx, \cl}$ & $\widehat{\ROC}_{\bx}-\widehat{\ROC}_{\bx, \cl}^{(-6)}$\\
			\includegraphics[scale=0.35]{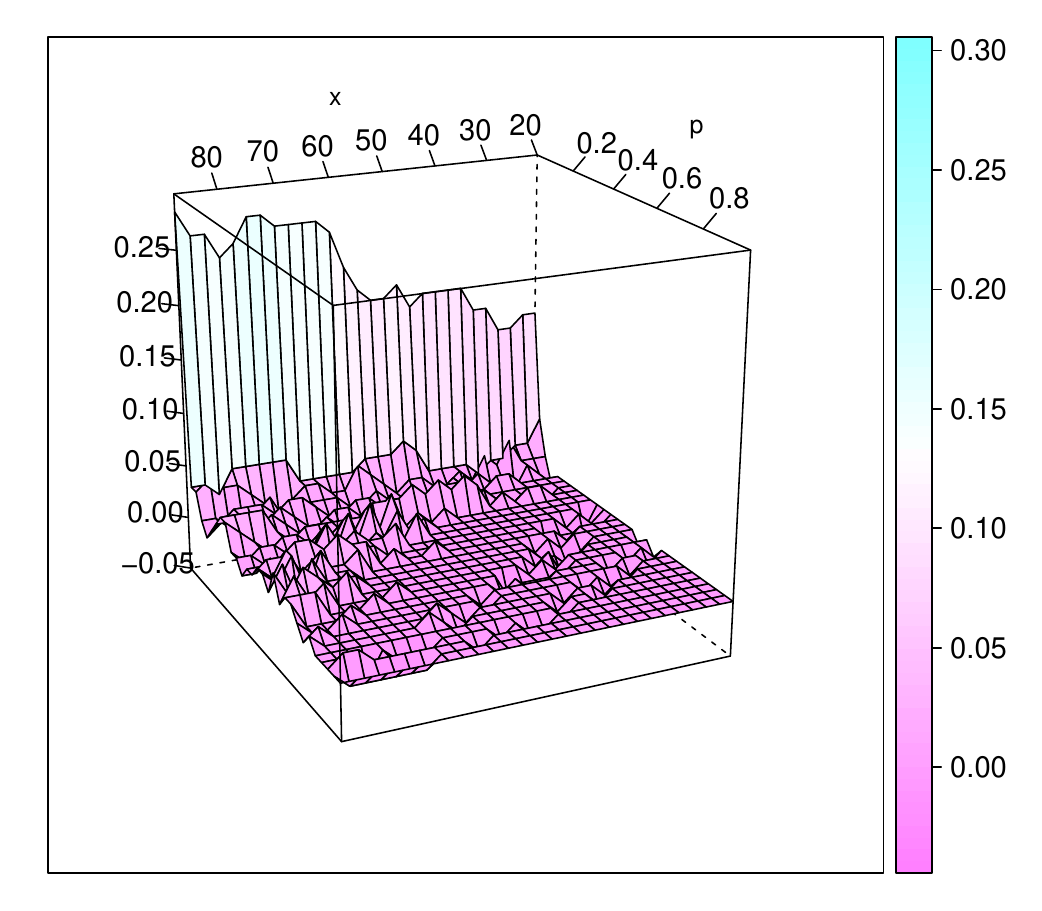} & 	
			\includegraphics[scale=0.35]{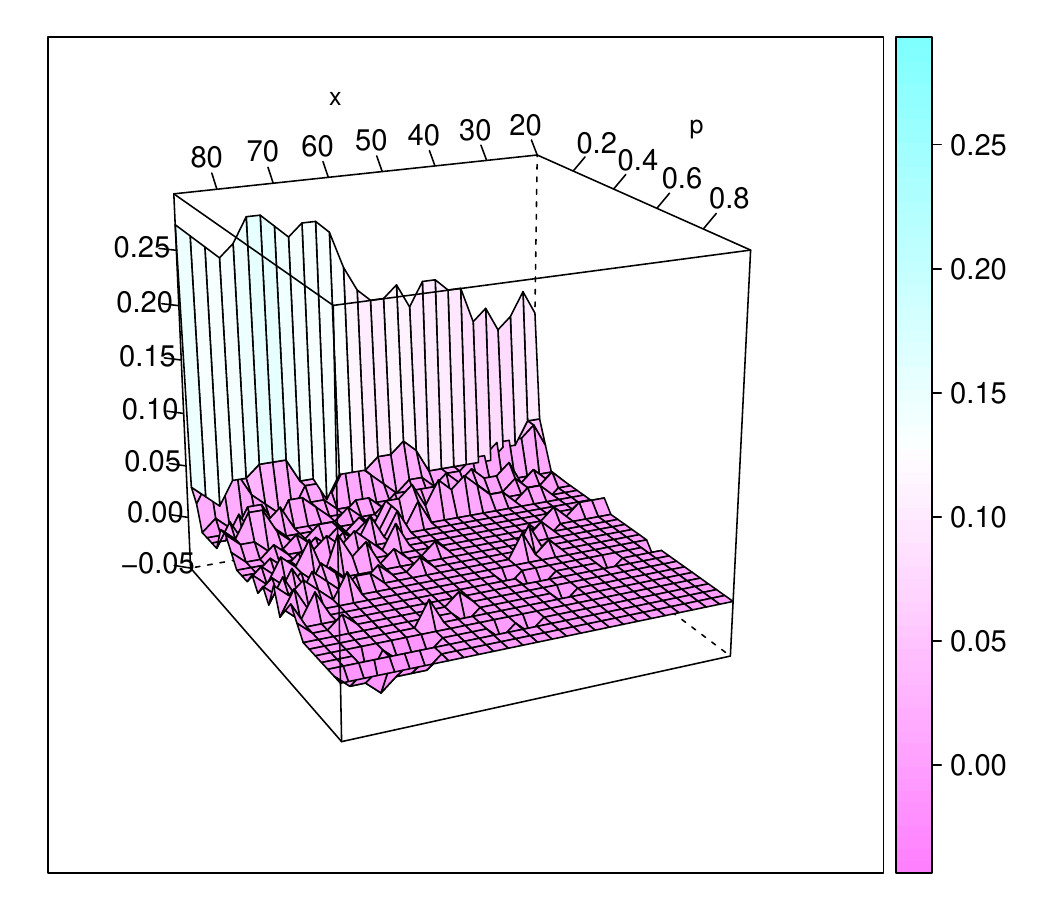} & 
			\includegraphics[scale=0.35]{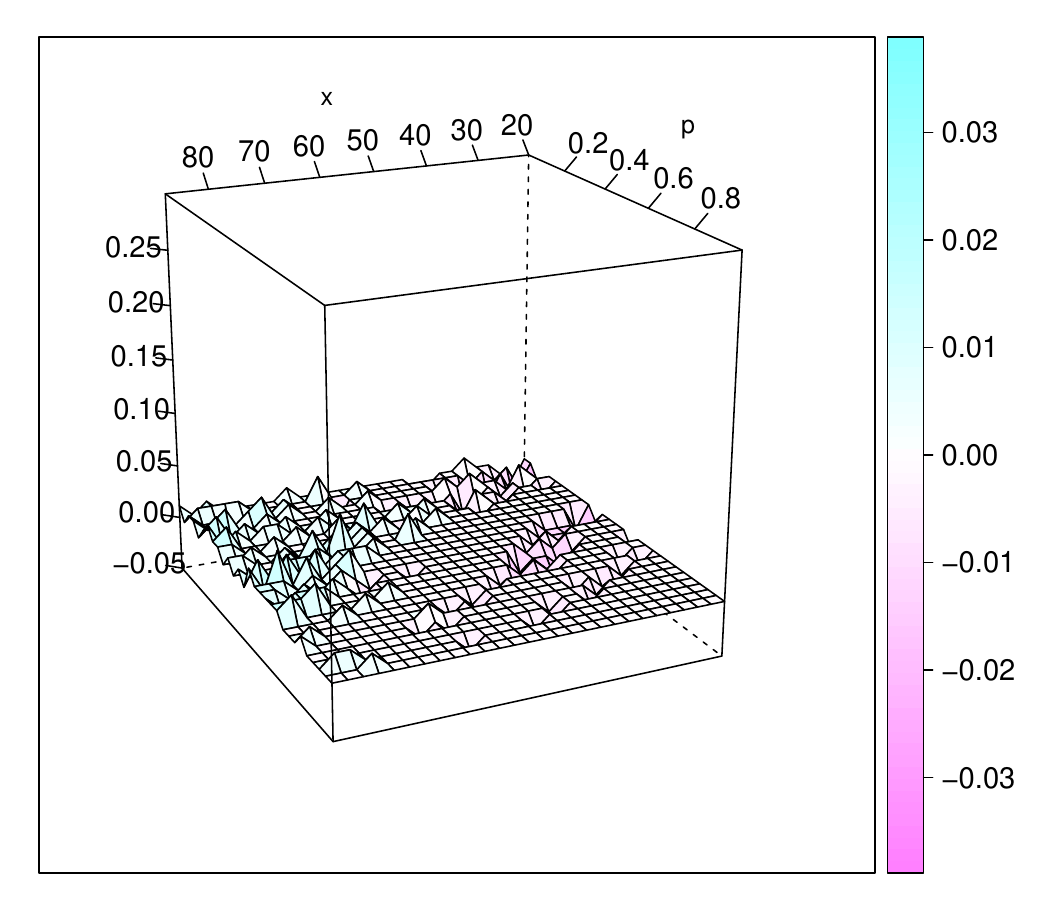}\\		
			\multicolumn{3}{c}{(c)}\\
			\multicolumn{3}{c}{$\quad$}\\
			\multicolumn{3}{c}{$\quad$}\\ 
			$\widehat{\ROC}_{\bx }-\widehat{\ROC}_{\bx, \cl}$ & $\widehat{\ROC}_{\bx, \cl}^{(-6)}-\widehat{\ROC}_{\bx, \cl}$ &
			\\
			\includegraphics[scale=0.35]{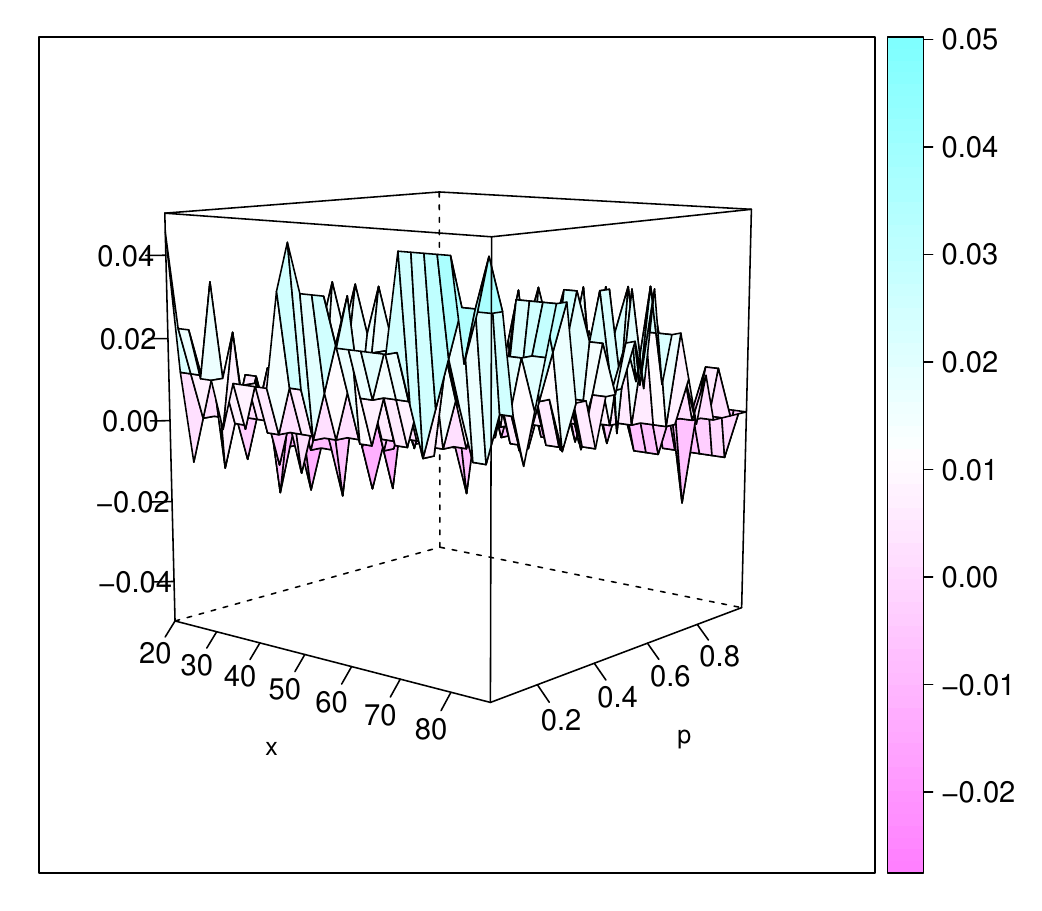} & 	
			\includegraphics[scale=0.35]{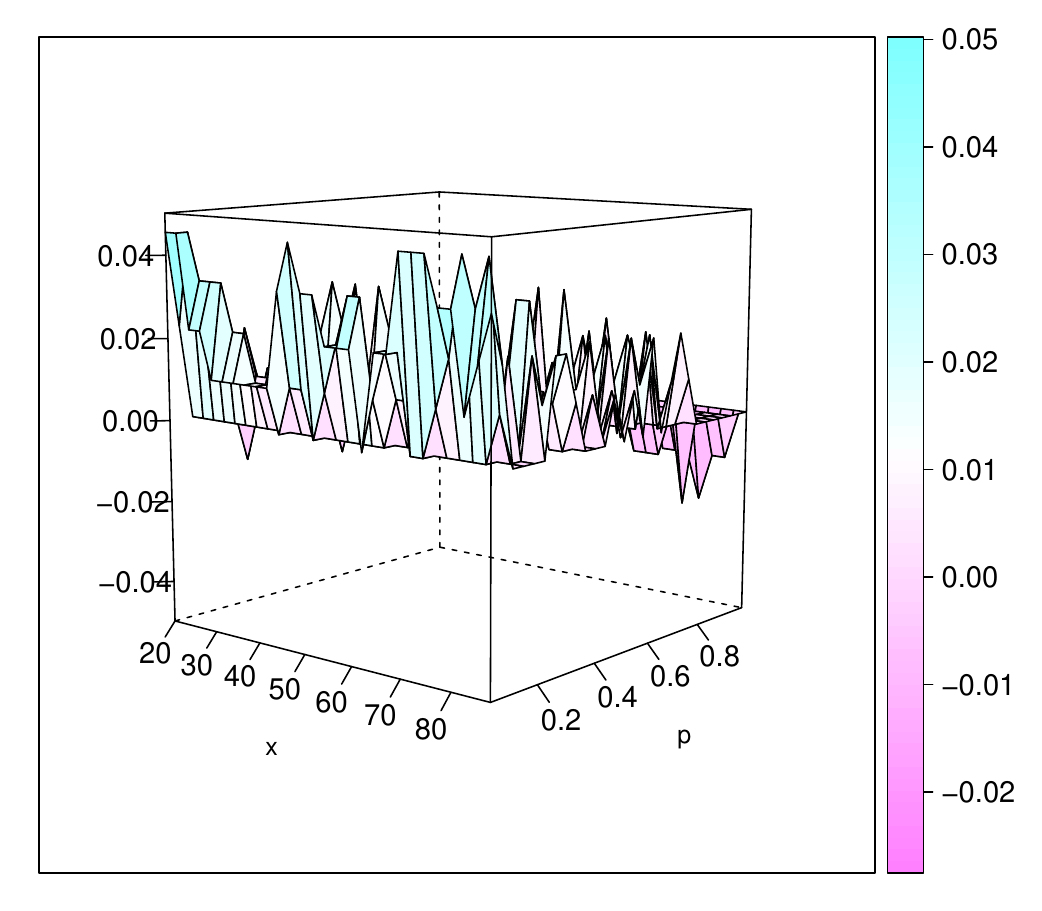} & 
		\end{tabular}
		\vskip-0.1in \caption{\label{fig:roc_ejemplo} Diabetes Data: (a) Estimated ROC surfaces and (b) Difference between the estimated ROC surfaces (c) Difference between the estimated ROC surfaces between 0.045 and 0.99.}
	\end{center} 
\end{figure}

Following the analysis given in Faraggi (2003), we transform the marker from both populations using power function $f(t)= -t^{-1/2}$. After this, we assume a  linear regression model in each population for the transformed marker $y$, i.e.,
\begin{eqnarray*}
y_{D,i} &=&  \beta_{D,1} +  \beta_{D,2}\, x_{D,i}  + \epsilon_{D,i} \,, 1\le i \le 88\,,\label{diseased}\\
y_{H,i} &=&   \beta_{H,1} + \beta_{H,2} \, x_{H,i}  +  \epsilon_{H,i}\; , 1\le i\le 198\,,\label{healthy}
\end{eqnarray*}
and we compute the classical and robust  estimators of the conditional ROC curves, denoted $\widehat{\ROC}_{\bx, \cl}$  and $ \widehat{\ROC}_{\bx }$, respectively.
Based on the residuals boxplots of a robust fit, 6 outliers were detected in the healthy sample,  labelled as 37, 78, 125, 137, 141 and 150, see the left panel of Figure \ref{fig:boxplotejemplo}. The filled red points on the central panel of Figure \ref{fig:boxplotejemplo} represent the   atypical observations encountered in the healthy sample which correspond to vertical outliers. After removing them,  the classical estimator of the conditional ROC curves is recomputed with the remaining points, namely $ \widehat{\ROC}_{\bx, \cl}^{(-6)}$. The upper panel of Figure \ref{fig:roc_ejemplo} displays the estimated surfaces with these three procedures using equidistant grids of points of size 29 and 28 in $p$ and $x$, respectively, between $p=0.01$ and $0.99$ and $x=20$ and $87.5$. In order to facilitate the differences between the estimated surfaces, the middle and lower panel in Figure \ref{fig:roc_ejemplo}  show  the differences between these estimators, making evident that the robust and classical estimator computed without the outliers are very similar all along the studied range, while the classical estimator computed from the whole sample shows  a different pattern,  clear in the   left panel of Figure \ref{fig:roc_ejemplo} especially for large values of age.



\normalsize

\section{Final Remarks}{\label{sec:conclusion}}
The ROC curve is a useful graphical tool that measures the discriminating power of a biomarker to distinguish between two conditions or classes. When   the practitioner may   measure   covariates related to the diagnostic variable which can increase the discriminating power, it is sensible to incorporate them in the analysis.  To have a deeper comprehension of the effect of the covariates, it would be advisable to incorporate the  covariates information to the ROC analysis instead of considering the marginal ROC curve. Conditional ROC curves may be easily estimated using a plug--in procedure. However, the use of classical regression estimators and empirical distribution and quantile functions may lead to estimates which breakdown in the presence of a small amount of atypical data.

In this piece of work, we introduce a   procedure to robustly estimate the conditional ROC curve. The methodology combines robust regression estimators with a weighted empirical distribution function which downweights the effect of large residuals. 
We prove that the estimators are uniformly strongly consistent under standard regularity
conditions.  A simulation study shows that our proposed estimators have good robustness and finite-sample statistical properties. Even though our numerical studies focus on a parametric regression approach, it should be mentioned that our proposal could also be implemented when considering  nonparametric or partly parametric regression models, using a robust fit. 

\noi {\small\textbf{Acknowledgment.} This work was partially developed while  Ana M. Bianco and Graciela Boente were visiting the Departamento de Estat\'{\i}stica, An\'alise Matem\'atica e Optimizaci\'on de la Universidad de Santiago de Compostela, Spain with the travel support of the program UBAINT DOCENTES 2019 from the University of Buenos Aires.  This research was partially supported by Grants   \textsc{pict} 2018-00740 from \textsc{anpcyt} and   20020170100022BA from the Universidad de Buenos Aires, Argentina  and also by the Spanish Project {MTM2016-76969P}  from the Ministry of Economy and Competitiveness   (MINECO/AEI/FEDER, UE), Spain.}  

\normalsize

\normalsize 

\setcounter{equation}{0}
\renewcommand{\theequation}{A.\arabic{equation}}

\setcounter{section}{0}
\renewcommand{\thesection}{\Alph{section}}

\section{Appendix A:  Proof of Theorem  \ref{theo:consist.2}.}\label{sec.appendix}

We begin by proving (i). Using assumption \ref{ass:A3} for $j=H$ and the continuity of the quantile functionals when \ref{ass:A1} holds, we get that, for the healthy subjects,  $\wG_{H}^{-1}(p) \convpp G_H^{-1}(p)$, for each $0<p<1$. To avoid burden notation denote as
\begin{eqnarray*}
\wDelta(\bx,p)&=& \frac{\wmu_{H}(\bx) -\wmu_{D}(\bx)}{\wsigma_{D} } + \frac{\wsigma_{H} }{\wsigma_{D}  } \, \wG_{H}^{-1}(1-p)\,,\\
\Delta(\bx,p)&=& \frac{\mu_{0,H}(\bx) -\mu_{0,D}(\bx )}{\sigma_{0,D}} + \frac{\sigma_{0,H}}{\sigma_{0,D}} \, G_{H}^{-1}(1-p)\,.
\end{eqnarray*}
Note that the consistency of   $\wsigma_j$ and \ref{ass:A61} together with the fact that $\wG_{H}^{-1}(p) \convpp G_H^{-1}(p)$, entail that for each fixed $p$ and $\bx$, $\wDelta(\bx,p)\convpp \Delta(\bx,p)$.
Therefore, we have that,
\begin{eqnarray*}
|\widehat{\ROC}_{\bx}(p) - {\ROC}_{\bx}(p)| &=& \left|\wG_{D}\left(\wDelta(\bx,p)\right)- G_{D}\left(\Delta(\bx,p)\right)\right|\\
&\le & \left|\wG_{D}\left(\wDelta(\bx,p)\right)- G_{D}\left(\wDelta(\bx,p)\right)\right| + \left| G_{D}\left(\wDelta(\bx,p)\right)- G_{D}\left(\Delta(\bx,p)\right)\right|\\
&\le & \left\|\wG_{D}- G_D\right\|_{\infty} + \left| G_{D}\left(\wDelta(\bx,p)\right)- G_{D}\left(\Delta(\bx,p)\right)\right|
\end{eqnarray*}
which together with the continuity of $G_D$ lead to $\widehat{\ROC}_{\bx}(p) \convpp {\ROC}_{\bx}(p)$, for each fixed $\bx$ and $0<p<1$. Note that for each fixed $\bx$, ${\ROC}_{\bx}(p)$ satisfies the conditions in Lemma S.1.1, so $\sup_{0<p<1} |\widehat{\ROC}_{\bx}(p)- {\ROC}_{\bx}(p)|\convpp 0$.

\noi (ii) Using that
\begin{eqnarray*}
\left|\wDelta(\bx,p)- \Delta(\bx,p)\right|&\le & \frac{1}{\wsigma_{D} } \left\{\left|\wmu_{H}(\bx) -\mu_{0,H}(\bx)\right|+ \left|\wmu_{D}(\bx) -\mu_{0,D}(\bx)\right|\right\} \\
& &+\left| \frac{1}{\wsigma_{D}} - \frac{1}{\sigma_{0,D}} \right|\; \left|\mu_{0,H}(\bx)) -\mu_{0,D}(\bx)\right|\\
&& + \frac{\wsigma_{H} }{\wsigma_{D}  } \,\left| \wG_{H}^{-1}(1-p)- G_{H}^{-1}(1-p)\right|+ | G_{H}^{-1}(1-p)|\, \left|\frac{\wsigma_{H} }{\wsigma_{D}}-\frac{\sigma_{0,H}}{\sigma_{0,D}} \right|\,,
\end{eqnarray*}
assumption \ref{ass:A7}, the consistency of  $\wsigma_j$ and the uniform consistency of $\wmu_j$, we get easily that $\sup_{\bx \in \itK} \left|\wDelta(\bx,p)- \Delta(\bx,p)\right|\convpp 0$.
Hence, 
\begin{eqnarray*}
\sup_{\bx \in \itK} |\widehat{\ROC}_{\bx}(p) - {\ROC}_{\bx}(p)| &\le & \left\|\wG_{D}- G_D\right\|_{\infty} + \| g_{D}\|_{\infty} \sup_{\bx \in \itK}\left| \wDelta(\bx,p)- \Delta(\bx,p) \right|
\end{eqnarray*}
leads to $\sup_{\bx \in \itK} |\widehat{\ROC}_{\bx}(p) - {\ROC}_{\bx}(p)| \convpp 0$. 

Denote as $\wB=\sup_{\delta<p<1-\delta}\sup_{\bx \in \itK} |\widehat{\ROC}_{\bx}(p) - {\ROC}_{\bx}(p)|$, then $\wB \le \sum_{\ell=1}^5 \wB_\ell$
where
\begin{eqnarray*}
\wB_1 &=& \left\|\wG_{D}- G_D\right\|_{\infty} \;,\\
\wB_2&=& \| g_{D}\|_{\infty} \left| \frac{1}{\wsigma_{D} } - \frac{1}{\sigma_{0,D} } \right|\; \left(A_H+A_D\right) \;,\\
\wB_3 &=&  \| g_{D}\|_{\infty} \frac{1}{\wsigma_{D} } \sup_{\bx \in \itK} \left\{\left|\wmu_{H}(\bx) -\mu_{0,H}(\bx)\right|+ \left|\wmu_{D}(\bx) -\mu_{0,D}(\bx)\right|\right\} \;,\\
\wB_4 &=&   \frac{\wsigma_{H} }{\wsigma_{D}  } \sup_{\delta<p<1-\delta} \,\left| \wG_{H}^{-1}(1-p)- G_{H}^{-1}(1-p)\right| \;,\\
\wB_5 &=& \sup_{\delta<p<1-\delta}| G_{H}^{-1}(1-p)| \,\left|\frac{\wsigma_{H} }{\wsigma_{D}  }-\frac{\sigma_{0,H}}{\sigma_{0,D}} \right| \;.
\end{eqnarray*}
Assumptions \ref{ass:A1} to \ref{ass:A3} together with \ref{ass:A6} and the consistency  of $\wsigma_j$  entail that $\wB_\ell\convpp 0$, for $\ell=1,2,3,5$. Note that  $\wG_H^{-1}$ is  a   non--decreasing function. Besides, using that \ref{ass:A1} entails that $G_H$ is    continuous and strictly increasing, we immediately obtain that  the quantile function $G_H^{-1}$, which in this case equals the inverse of $G_H$, is also strictly increasing and uniformly continuous over the compact interval $[\delta, 1-\delta]$. Hence,  taking into account that  $\wG_{H}^{-1}(p) \convpp G_H^{-1}(p)$, for each $0<p<1$, analogous arguments to those considered in the proof of Lemma \ref{lema:appendix1.1} below, when bounding $\sup_{a\le t\le b}|F_n(t)-F(t)|$ allow to derive that 
$\sup_{\delta\le p\le 1-\delta} \,\left| \wG_{H}^{-1}(1-p)- G_{H}^{-1}(1-p)\right| \convpp 0$,
concluding the proof of (ii).  

We now proceed to derive (iii). Let $\epsilon>0$ be fixed and choose $0<\eta<1$ such that,    $\sup_{\bx\in \itK}{\ROC}_{\bx}(\eta)<\epsilon/6$ and $\sup_{\bx\in \itK} (1-{\ROC}_{\bx}(1-\eta))<\epsilon/6$. Denote as  
\begin{align*}
\wB(\eta) & = \sup_{\eta<p<1-\eta}\sup_{\bx \in \itK} |\widehat{\ROC}_{\bx}(p) - {\ROC}_{\bx}(p)|\,,\\
\wB_1(\eta)& = \sup_{ p\le  \eta}\sup_{\bx \in \itK} |\widehat{\ROC}_{\bx}(p) - {\ROC}_{\bx}(p)|\,,\\
\wB_2(\eta)& = \sup_{1-\eta\le p}\sup_{\bx \in \itK} |\widehat{\ROC}_{\bx}(p) - {\ROC}_{\bx}(p)|\;.
\end{align*}
  Hence, 
 $\sup_{0<p<1 } \sup_{\bx\in \itK} |\widehat{\ROC}_{\bx}(p)- {\ROC}_{\bx}(p)|\le \wB(\eta)+\wB_1(\eta)+\wB_2(\eta) $.
From (ii), \linebreak$\wB(\eta)\convpp 0$. Besides,  using that ${\ROC}_{\bx}(p)$ is a distribution function and $\widehat{\ROC}_{\bx}(p)$ is non-decreasing in $p$, we get that for any $p\le \eta$, $\bx \in \itK$, 
$$|\widehat{\ROC}_{\bx}(p) - {\ROC}_{\bx}(p)| \le \max\left\{\widehat{\ROC}_{\bx}(\eta), {\ROC}_{\bx}(\eta)\right\}\,,$$
so 
$\wB_1(\eta)\le \sup_{\bx\in \itK} \max\left\{\widehat{\ROC}_{\bx}(\eta), {\ROC}_{\bx}(\eta)\right\}=\wC_1(\eta)$ .
Similarly, we obtain that \linebreak
$\wB_2(\eta)\le \sup_{\bx\in \itK} \max\left\{1-\widehat{\ROC}_{\bx}(1-\eta), 1-{\ROC}_{\bx}(1-\eta)\right\}=\wC_2(\eta)$.

Using that  $\sup_{\bx\in \itK}{\ROC}_{\bx}(\eta)<\epsilon$ and $\sup_{\bx\in \itK} (1-{\ROC}_{\bx}(1-\eta))<\epsilon$ and that for any fixed $0<p<1$,  $\sup_{\bx \in \itK} |\widehat{\ROC}_{\bx}(p) - {\ROC}_{\bx}(p)| \convpp 0$, we conclude that there exists $\itN$ such that $\prob(\itN)=0$ and for $\omega\notin \itN$, 
$ \wB(\eta)  \to 0$, $\wC_1(\eta) \to \sup_{\bx\in \itK}  {\ROC}_{\bx}(\eta)< \epsilon/{6}$ and $\wC_2(\eta)  \to \sup_{\bx\in \itK}  1- {\ROC}_{\bx}(1-\eta)< {\epsilon}/{6}$.
 Hence, for $n_H$ and $n_D$ large enough, we obtain that $ \wB(\eta)<\epsilon/3$, $\wC_\ell(\eta)<\epsilon/3$, for $\ell=1,2$ which leads to $\sup_{0<p<1 } \sup_{\bx\in \itK} |\widehat{\ROC}_{\bx}(p)- {\ROC}_{\bx}(p)|\le \epsilon$, concluding the proof.  \qed

\normalsize
\setcounter{equation}{0}
\renewcommand{\theequation}{B.\arabic{equation}}

\setcounter{section}{1}
\renewcommand{\thesection}{\Alph{section}}

\section{Appendix B}\label{sec:consistG}
In this section, we investigate the validity  of assumption  \ref{ass:A3}. For that purpose, we will derive the uniform strong consistency of $\wG_n(t)$ defined in  \eqref{empiricalw} in two situations, under a linear model or a non--linear one, since for the former we can also include a hard rejection weight function to define the weights $w_i$. It is worth noticing that our results generalize those given in Gervini and Yohai (2002) in two directions: we extend their results beyond the linear model to a non--linear one and we obtain almost surely uniform consistency  instead of   results in probability.  

 From now on, for any measure $Q$,  we denote as   $N(\epsilon, \itF, L_s(Q))$ and $N_{[\;]}(\epsilon, \itF, L_s(Q))$  the covering  and bracketing  numbers  of the class $\itF$ with respect to the distance in $ L_s(Q)$, as defined, for instance, in van der Vaart and Wellner (1996).

\subsection{Linear Model}{\label{sec:appendix1}}
Throughout this section, we will assume that we have a random sample $(y_{1},\bx_{1}),\dots,(y_{ n},\bx_{n})$, where $\bx_i$ is a vector of  $p$ explanatory variables and $y_i$ is a response variable that satisfy the linear regression model
$$ y_i= \bx_i \trasp \bbe_0 + u_i=   \bx_i \trasp \bbe_0 +\sigma_0 \epsilon_i, i=1 \dots n \, ,$$
with $\bbe_0 \in \real^p$ and the errors $\epsilon_i$ are i.i.d. and independent of $\bx_i$ with unknown distribution $G_0(\cdot)$ and $\sigma_0$ is the scale parameter. 

From now on, $\wbbe$ and $\wsigma$ stand for robust consistent estimators of $\bbe_0$ and $\sigma_0$, so the standardized residuals are given by
$$r_{i}= \dfrac{y_{i}-\bx_{i}\trasp \wbbe }{\wsigma} \, .$$
Based on the residuals the   adaptive weighted empirical distribution given in \eqref{empiricalw} is defined using the weights  $w_i=w(r_i/t_n)$ defined in \eqref{eq:pesos} with $t_n$ given in  \eqref{eq:tn}.
 
To derive uniform consistency results, we will need the following set of assumptions:
 \begin{enumerate}[label=\textbf{C\arabic*}]
 \item \label{ass:peso} The weight function $w: \real \to [0,1]$ is   even, non-increasing on $[0, +\infty)$,  continuous,
 $w(0)=1$, $w(u)>0$ for $0 < u <1$ and $w(u)=0$ for $   |u| \ge 1$.

\item\label{ass:G0} $G_0$ is a continuous distribution function. 

\item\label{ass:consis} The estimators  $\wbbe$ and $\wsigma$ are such that  $\wbbe\convpp \bbe_0$ and $\wsigma\convpp \sigma_0$.
 \end{enumerate}
 
 Define the values
 \begin{eqnarray*}
 d_0 &=&  \sup_{t \ge \eta} \{G^{+}(t)-G^{+}_0(t)\}^+=\sup_{t \ge 0} \{\max\left(G^{+}(t)-G^{+}_0(t),0\right)\} 
 \label{eq:d0}\\
 t_0 &=& (G_0^{+})^{-1}(1-d_0)=G_0^{-1}\left(1-\frac{d_0}2\right)
 \label{eq:t0}
 \end{eqnarray*}
 As mentioned in Gervini and Yohai (2002), when $G^{+}$ is stochastically larger or equal than $G_0^{+}$, we have that $t_0=\infty$, so $\wG_n$ defined in \eqref{empiricalw} will converge to $G_0$. Furthermore, consider the functions
 \begin{eqnarray}
  h_{\infty}(t) &=&  \esp_{G_0} w\left(\frac{\epsilon_1}{ \, t}\right) 
  \label{eq:hinfty}\\
 h_0(t,s) &=&  \esp_{G_0} w\left(\frac{\epsilon_1}{  \,  t}\right)\indica_{\epsilon_1\le s  }  
 \label{eq:h0}
 \end{eqnarray}
 
 The following lemma is a well known result regarding continuous distributions, whose proof we include for completeness.
 
 \vskip0.1in
 
\begin{lemma} \label{lema:appendix1.1}
Let $F_n:\real \to [0,1]$ and $F:\real \to [0,1]$ be non--decreasing functions such that $F$ is continuous, $\lim_{t\to +\infty}F(t)=1$ and $\lim_{t\to -\infty}F(t)=0$. Then, if $F_n(t)\convpp F(t)$, for any $t\in \real$, we also have that $\|F_n-F\|_{\infty} \convpp 0 $.
\end{lemma}
 
 \noi \textsc{Proof.} Given $\epsilon>0$, let $a$ and $b$ be such that  $F(a)<\epsilon$ and $F(b)>1-\epsilon$. Furthermore, using that $F$ is uniformly continuous on $[a,b]$, we get that there exists $\delta$ such that 
 $$ |t-s|<\delta, t,s \in [a,b] \Rightarrow |F(t)-F(s)|<\epsilon$$
 Let $a=a_0<a_2<\dots<a_k=b$, be a grid such that $a_j-a_{j-1}<\delta$, $1\le j\le k$. Then, we have that for any $t<a$,
 $F_n(t)-F(t)\le F_n(a)\le F_n(a)-F(a)+ F(a)\le |F_n(a)-F(a)|+\epsilon$, while $F(t)-F_n(t)\le F(a)<\epsilon$, so
 \begin{equation}
 \label{eq:Fna}
 \sup_{t<a}|F_n(t)-F(t)|\le |F_n(a)-F(a)|+\epsilon\,.
 \end{equation}
 Similarly, 
 \begin{equation}
 \label{eq:Fnb}
 \sup_{t>b}|F_n(t)-F(t)|\le |F_n(b)-F(b)|+\epsilon\,.
 \end{equation}
 Finally, given $t\in [a,b]$, there exists $1\le j\le k$ such that $t\in [a_{j-1}, a_j]$, so that
  \begin{eqnarray*}
  F_n(t)-F(t) & \le & F_n(a_j)-F(a_{j-1})\le F_n(a_j)-F(a_{j})+ F(a_{j})-F(a_{j-1})\\
   & \le & \epsilon+\max_{1\le j\le k}|F_n(a_j)-F(a_{j})| \,.
  \end{eqnarray*}
  Similarly, 
   \begin{eqnarray*}
   F (t)-F_n(t)  & \le & F(a_j)-F_n(a_{j-1})\le F(a_j)-F(a_{j-1})+ F(a_{j-1})-F_n(a_{j-1})\\
    & \le & \epsilon+\max_{1\le j\le k}|F_n(a_j)-F(a_{j})|\,,
      \end{eqnarray*}
   so 
   \begin{equation}
 \label{eq:Fnab}
 \sup_{a\le t\le b}|F_n(t)-F(t)|\le \epsilon+\max_{1\le j\le k}|F_n(a_j)-F(a_{j})|\,.
 \end{equation}
Let $\itN$ be such that, for $\omega\notin \itN$, $F_n(a_j)\to F(a_j)$ and $\prob(\itN)=0$. Then, using \eqref{eq:Fna}, \eqref{eq:Fnb} and \eqref{eq:Fnab} we get that
$$\prob(\limsup \|F_n-F\|_{\infty}<\epsilon)=1\,,$$
for any $\epsilon>0$, concluding the proof. \qed

\vskip0.1in 
 
\begin{lemma} \label{lema:appendix1.2} 
Under \ref{ass:G0} and \ref{ass:consis}, we have that 
 \begin{enumerate}[label=\alph*)]
 \item $\|G_n^{+}-G_0^{+}\|_{\infty}\convpp 0$.
 \item $d_n\convpp d_0$. 
 \item $t_n\convpp t_0$,  if, in addition, $ G_0$ is strictly increasing on its support. 
  \end{enumerate}
  \end{lemma}
 
 \noi \textsc{Proof.} a) Let us consider the family of functions
 $$\itF=\{f_{\bthech,\; \kappa}(u,\bx)=\indica_{|u-\bx\trasp\bthech|\le  \kappa\; t} \mbox{ for } (\bthe, t,\kappa)\in \real^p\times \real_{\ge 0}\times \real_{> 0}\} \,.$$
 First, note that  
 $$f_{\bthech,\; \kappa}(u,\bx)=\indica_{|u-\bx\trasp\bthech|\le  \kappa \; t}=\indica_{C_{(\kappa^{-1}, \bthech\; \kappa^{-1}, \; t)}} \,,$$
 where the set $C_{(s, \bthech,\; t)}= A_{(s, \bthech,\; t)}\cap B_{(s, \bthech,\; t)}$, with $ A_{(s, \bthech,\; t)}=\{(u,\bx)\in \real^{p+1}: su-\bx\trasp\bthe  -t \le 0\}$ and $ B_{(s, \bthech,\; t)}= \{(u,\bx)\in \real^{p+1}: 0\le su-\bx\trasp\bthe +t  \} $. 
   Define the classes of sets
 \begin{eqnarray*}
 \itA &=& \{  A_{(s, \bthech,\; t))}: (\bthe, s,\; t))\in \real^p\times\real_{\ge 0}\times \real_{>0}\}\\
 \itB &=& \{  B_{(s, \bthech,\; t))}: (\bthe, s,\; t))\in \real^p\times\real_{\ge 0}\times \real_{>0}\}\,.
 \end{eqnarray*}
 Taking into account that $\{g(u,\bx)=su-\bx\trasp\bthe  -t ; (\bthe, s,\; t))\in \real^p\times\real_{\ge 0}\times \real_{>0}\}$ is a finite--dimensional space of functions with dimension $p+2$, from Lemmas 9.6, 9.8 and 9.9 in Kosorok (2008) we get that $\itA$ and $\itB$ are VC-classes with index at most $p+4$. Furthermore, $\itC=\itA\cap\itB$ is also a VC-class with index  smaller or equal than $2p+7$. Taking into account that $C_{(s, \bthech)}\in \itC$, applying again Lemma  9.8 in Kosorok (2008), we get that the class of functions $\itF$ is a VC-class with index $V(\itF)$ smaller or equal than  $2p+7$. Note that the envelope of $\itF$ equals $F\equiv 1$. Hence, Theorem 2.6.7 in van der Vaart and Wellner (1996) entails that, there exists a universal constant $K$ such that, for any measure $Q$
 $$N(\epsilon, \itF, L_1(Q)) \le K \;  V(\itF) \left(16 e\right)^{V(\itF)} \left(\frac{1}{\epsilon}\right)^{V(\itF)-1}\, ,$$
 which together with Theorem 2.4.3 in van der Vaart and Wellner (1996) or Theorem 2.4 in Kosorok (2008), leads to
 \begin{equation}
 \label{eq:glivenko}
 \sup_{f\in \itF} | P_n f- P f| \convpp 0\,,
 \end{equation}
 where we have used the standard notation in empirical processes, i.e., $P f=\esp f(u,\bx)$ and $P_n f=(1/n) \sum_{i=1}^n f(u_i, \bx_i)$.
 
 Note that $G_n^{+}$ can be written as
 $$G_n^{+}(t) =P_n f_{\wbthech, \;\wsigma\,t}(u_i,\bx_i)$$
 with $\wbthe=\wbbe-\bbe_0$. Denote as
$M(\bthe, \kappa)= P f_{\bthech,\; \kappa}(u,\bx)$. Then, using \eqref{eq:glivenko}, we conclude that
$$\sup_{t\ge 0}\left|G_n^{+}(t) - M( \wbthech, \;\wsigma\,t)\right|\le \sup_{f\in \itF} | P_n f- P f| \convpp 0\,.$$
It remains to show that
$$\sup_{t\ge 0}\left| M( \wbthe, \;\wsigma\,t)- G_0^{+}(t)\right| \convpp 0\,.$$
Note that
$$ M(\bthe, \kappa)= P f_{\bthech,\; \kappa}(u,\bx)= \prob(|u-\bx\trasp\bthe|\le  \kappa) \,,$$
hence
$$M(0, \sigma_0\, t)= \prob( |u|\le \sigma_0 \, t)=G_0^{+}(t)\,.$$
Therefore, we have to show that
$$\sup_{t\ge 0}\left| M( \wbthe, \;\wsigma\,t)- M(0, \sigma_0\, t)\right| \convpp 0\,.$$
First observe that
\begin{eqnarray*}
 M(\bthe, \kappa)&=& \prob \left( -\kappa + \bx\trasp\bthe\le u \le \kappa+\bx\trasp\bthe\right)\\
 &=& \esp \left\{G_0(\bx\trasp\bthe+\kappa)- G_0(\bx\trasp\bthe- \kappa)\right\}\, .
 \end{eqnarray*}
 The continuity of $G_0$ and the Dominated Convergence Theorem entail that $M(\bthe, \kappa)$ is a continuous function of its arguments, which together with 
\ref{ass:consis}, entails that $M( \wbthe, \;\wsigma\,t)- M(0, \sigma_0\, t) \convpp 0$, for each fixed $t$. Now 
$$\wM(t)=M( \wbthe, \;\wsigma\,t)=\prob\left(\frac{|u-\bx\trasp\wbthe|}{\wsigma}\le  t\right)$$
is a bounded monotone function of $t$, while $M(0, \sigma_0\, t)=G_0^{+}(t)$ is also  bounded, monotone and continuous, thus, from Lemma \ref{lema:appendix1.1} we obtain that the convergence is indeed uniform, that is, $\sup_{t\ge 0}\left| M( \wbthe, \;\wsigma\,t)- M(0, \sigma_0\, t)\right| \convpp 0 $, concluding the proof of a).  

b) As in Gervini and Yohai (2002), $|d_n- d_0|\le \|G_n^{+}-G_0^{+}\|_{\infty}$ and the result follows.

c) To show that $t_n\convpp t_0$, it is enough to show that $G_0^{+}(t_n)\convpp G_0^{+}(t_0)=1-d_0$ which follows from Lemma 3.1 in Gervini and Yohai (2002) distinguishing the cases $t_0<\infty$ and $t_0=\infty$. \qed

\vskip0.2in

\begin{lemma} \label{lema:appendix1.3}  
Assume that either $w(t)= \indica_{[-1,1]}(t)$ or $w$ satisfies \ref{ass:peso}. Then,  we have that $\sup_{f\in \itF} | P_n f- P f| \convpp 0$, where 
$$\itF=\{f_{\bthech,\; \kappa, \nu}(u,\bx)=w\left(\nu({u -\bx\trasp\bthe})\right)\, \indica_{u-\bx\trasp\bthech\le \kappa\,  s} \mbox{ for } (\bthe, \kappa, \nu)\in \real^p\times \real_{\ge 0}\times \real_{\ge 0}\} \,.$$
\end{lemma}

\noi \textsc{Proof.}  Assume   that $w$ satisfies \ref{ass:peso} and note that $\itF\subset \itF_1 \cdot \itF_2$ where 
  \begin{eqnarray*}
  \itF_1 &=& \{f_{\bthech,\, \nu}(u,\bx)=w\left(\nu({u -\bx\trasp\bthe})\right)\,  \mbox{ for } (\bthe,   \nu)\in \real^p\times  \real_{\ge 0}\}\\
  \itF_2 &=& \{f_{\bthech,\; \kappa }(u,\bx)=  \indica_{u-\bx\trasp\bthech\le \kappa\,  s} \mbox{ for } (\bthe, \kappa )\in \real^p\times \real_{\ge 0}\}\,.
   \end{eqnarray*}
The classes $\itF$, $\itF_1$ and $\itF_2$ have envelope 1,  hence we have easily that,  for any measure $Q$,
 $$N(2\,\epsilon, \itF, L_1(Q))\le N(\epsilon, \itF_1, L_1(Q)) N(\epsilon, \itF_2, L_1(Q))\,,$$
so that to show $\sup_{f\in \itF} | P_n f- P f| \convpp 0$, it will be enough to prove that, for $j=1,2$,
\begin{equation}
\label{eq:NPn}
 \frac{1}{n} \log N(\epsilon, \itF_j, L_1(P_n)) \convprob 0\,.
 \end{equation}
As in the proof of Lemma \ref{lema:appendix1.2}, it is easy to see that $\itF_2$ is a VC-class with index $V_2=p+3$, so 
$$N(\epsilon, \itF_2, L_1(Q)) \le K \;  V_2 \left(16 e\right)^{V_2} \left(\frac{1}{\epsilon}\right)^{V_2-1}\, ,$$
leading to \eqref{eq:NPn}, when $j=2$.

On the other hand, the family  
$$\itR = \left\{\nu({u -\bx\trasp\bthe})\,:\; \bthe \in
  \real^p,   \nu \in \real_{\ge 0}\right\}$$
is a subset of the vector space of all linear functions in $p+1$
variables. It follows from Lemma 2.6.15 of van der Vaart and Wellner  (1996) that $\itR$ has VC-index at most $p+3$. Note that  $w$ is an even function, non-increasing on $[0,+\infty)$, hence it can be written as
$w = w^{(1)}+w^{(2)}$, where $w^{(1)}(x)=w(x)\indica_{[0,+\infty)}(x)$ is
non--increasing and $w^{(2)}(x)=w(x)\indica_{(-\infty,0)}(x)$ is
non--decreasing. 
Using the permanence property for VC-classes, see   Lemma 9.9 in Kosorok (2008), we obtain that the classes of functions
  $\itR_{w^{(1)}}=w^{(1)} \circ \itR$ and $\itR_{w^{(2)}}=w^{(2)} \circ \itR$ are VC--classes with  VC--index at most $p+ 3$. 
  Furthermore, the classes  $\itR_{w^{(j)}}$, $j=1,2$, have envelope 1. Then, Theorem 2.6.7 of Van der Vaart and
Wellner (1996) implies that there exists a universal constant $K$ such that, for any probability measure  $Q$ on $\real^{p+1}$ and any $0<\epsilon<1$,   we have that  
$$N(\epsilon , \itR_{w^{(j)}}, L_1(Q)) \leq K (p+3)\; (16e)^{(p+3)}\left(\frac 1\epsilon\right)^{p+2}\,.$$
Note that $\itR_{w^{(1)}} + \itR_{w^{(2)}}$ has also constant envelope equal to $2 $. Therefore,
\begin{align*}
  N(2 \epsilon  , \itR_{w^{(1)}} + \itR_{w^{(2)}}, L_1(Q)) 
  &\leq N( \epsilon ,\itR_{w^{(1)}}  , L_1(Q)) \times N( \epsilon ,\itR_{w^{(2)}}  , L_1(Q))  \\
  &\leq \left[K (p+3)\; (16e)^{(p+3)}\left(\frac 1\epsilon\right)^{p+2}\right]^2\,.
\end{align*}
Finally noting that $\itF_1 $ has constant envelope equal to 1 and  
$\itF_1 \subset \itR_{w^{(1)}} + \itR_{w^{(2)}}$, we get that 
$$N(2 \epsilon  , \itF_1 , L_1(P_n)) \le  \left[K (p+3)\; (16e)^{(p+3)}\left(\frac 1\epsilon\right)^{p+2}\right]^2\,,$$
concluding the proof. 

When  $w(t)= \indica_{[-1,1]}(t)$ the result is straightforward using that 
\begin{eqnarray*}
\itF_1 &=& \{f_{\bthech,\, \nu}(u,\bx)=\indica_{\nu({u -\bx\trasp\bthe})\le 1}\indica_{\,-\,\nu({u -\bx\trasp\bthe})\le 1}\,  \mbox{ for } (\bthe,   \nu)\in \real^p\times  \real_{\ge 0}\} 
\end{eqnarray*}
and similar arguments to those consider above. \qed

\vskip0.2in
   
\begin{proposition} \label{prop:appendix1.1} 
Assume that   $w(t)= \indica_{[-1,1]}(t)$ or $w$ satisfies \ref{ass:peso}. Under  \ref{ass:G0} to \ref{ass:consis},  we have that
\begin{enumerate}[label=\alph*)]
\item if $t_0<\infty$, 
$$\sup_{s\in \real} \left|\wG_n(s)- \frac{h_0(t_0,s)}{h_{\infty}(t_0)}\right|\convpp 0\,,$$
with $h_\infty(t_0)$ and $h_0(t_0,s)$ defined in \eqref{eq:hinfty} and \eqref{eq:h0}, respectively.
\item  if   $t_0=\infty$,  $\|\wG_n-G_0\|_{\infty}\convpp 0$.
\end{enumerate}
\end{proposition}
  
  \noi \textsc{Proof.}   
  When $t_0=\infty$, using that $G_0$ is a  bounded, monotone and continuous function and that $\wG_n$ is monotone, it will be enough to show that
  for each $s\in \real$, $\wG_n(s)\convpp G(s)$. On the other hand, when $t_0<\infty$, standard arguments allow to show that $F(s)= {h_1(t_0,s)}/{h_{\infty}(t_0)}$ is a  bounded, monotone and continuous function of $s$ and the uniform convergence also follows from the pointwise one.

  Denote as $\wnu_n= 1/(t_n\, \wsigma_n )$, $\nu_0= 1/(t_0\, \sigma_0 )$, where we understand that if $t_0=+\infty$, $\nu_0=0$. Then $\wnu_n\convpp \nu_0$.
  
  We will begin by showing that
  \begin{equation}
  \label{eq:aprobar1}
  \frac{1}{n} \sum_{i=1}^n  w_i I(r_i \le s) \convpp  h_0(t_0,s) =\left\{\begin{array}{lr}
  \esp_{G_0} w\left(\dfrac{\epsilon_1}{  \,  t_0}\right)\indica_{\epsilon_1\le s  }  & \mbox{ if } t_0<\infty\\
   \esp_{G_0} \indica_{\epsilon_1\le s  }=G_0(s)  & \mbox{ if } t_0=\infty \, .
  \end{array}
  \right.
  \end{equation}
  For that purpose and noting that $r_i=(u_i-\bx_i\trasp \wbthe)/\wsigma$ with $\wbthe=\wbbe-\bbe_0$, define the class of functions
   $$\itF=\{f_{\bthech,\; \kappa, \nu}(u,\bx)=w\left(\nu({u -\bx\trasp\bthe})\right)\, \indica_{u-\bx\trasp\bthech\le \kappa\,  s} \mbox{ for } (\bthe, \kappa, \nu)\in \real^p\times \real_{\ge 0}\times \real_{\ge 0}\} \,.$$
 Lemma \ref{lema:appendix1.3}  entails that
 $$
 \sup_{f\in \itF} | P_n f- P f| \convpp 0\,,
 $$
 then, using that 
 $$\frac{1}{n} \sum_{i=1}^n  w_i I(r_i \le s)= P_n f_{\wbthech, \;\wsigma,\; \wnu_n}\, ,$$
 we obtain that
 $$ \frac{1}{n} \sum_{i=1}^n  w_i I(r_i \le s) - P f_{\wbthech , \;\wsigma,\; \wnu_n} \convpp 0\,.$$
 It remains to show that   
  $$P f_{\wbthech , \;\wsigma,\; \wnu_n }\convpp P f_{0,\, \sigma_0,\; \nu_0} = h_0(t_0,s)\,,$$
 which will follow if we derive that  
 \begin{eqnarray}
 A_n&=&P f_{\wbthech, \;\wsigma,\; \wnu_n }-\esp w\left(\nu_0  u   \right)\, \indica_{u-\bx\trasp\wbthech\le  \wsigma \,  s}   \convpp 0
 \label{eq:An}\\
  B_n&=& \esp w\left(\nu_0  u   \right)\, \indica_{u-\bx\trasp\wbthech\le  \wsigma \,  s}  -h_0(t_0,s) \convpp 0 \, .
 \label{eq:Bn}
  \end{eqnarray}
  We begin by considering the situation where $w$ satisfies \ref{ass:peso}. Noting that
 \begin{eqnarray*}
 |A_n| &=& \left|\esp \left\{ w\left(\nu_n(u-\bx\trasp\wbthe)\right) - w\left(\nu_0\, u  \right)\right\} \indica_{u-\bx\trasp\wbthech\le  \wsigma \,  s} \right|\\
 &\le &  \esp \left| w\left(\nu_n(u-\bx\trasp\wbthe)\right) - w\left(\nu_0\, u  \right)\right|\, ,
 \end{eqnarray*}
using the Dominated Convergence Theorem, the continuity of $w$   and the fact that $\wnu_n\convpp \nu_0$ and $\wbthe\convpp 0$,   we obtain that $A_n\convpp 0$, concluding the proof of \eqref{eq:An}.

When $w= \indica_{[-1,1]}$, we have that
\begin{eqnarray*}
w\left(\nu_n(u-\bx\trasp\wbthe)\right) - w\left(\nu_0\, u  \right)&=&\indica_{\nu_n(u-\bx\trasp\wbthech)\le 1}\;\indica_{-1\le \nu_n(u-\bx\trasp\wbthech)}- \indica_{\nu_0\, u \le 1}\; \indica_{-1\le \nu_0\, u }\\
&=& \indica_{\nu_n(u-\bx\trasp\wbthech)\le 1}\;\left\{\indica_{-1\le \nu_n(u-\bx\trasp\wbthech)}- \indica_{-1\le \nu_0\, u }\right\}\\
&&+ \left\{\indica_{\nu_0\, u \le 1}-\indica_{\nu_n(u-\bx\trasp\wbthech)\le 1}\right\}\; \indica_{-1\le \nu_0\, u }\;,
\end{eqnarray*}
so
 \begin{eqnarray*}
 |A_n|  &\le &  \esp \left| w\left(\nu_n(u-\bx\trasp\wbthe)\right) - w\left(\nu_0\, u  \right)\right| \\
 & \le & \esp \left| \indica_{-1\le \nu_n(u-\bx\trasp\wbthech)}- \indica_{-1\le \nu_0\, u }\right| +\esp \left| \indica_{\nu_0\, u \le 1}-\indica_{\nu_n(u-\bx\trasp\wbthech)\le 1}\right|\\
  & \le & \esp \left| \indica_{-\frac{1}{\nu_n} +  \bx\trasp\wbthech \le u}- \indica_{-\frac{1}{\nu_0}\le \, u }\right| +\esp \left| \indica_{ u \le \frac{1}{\nu_0}}-\indica_{ u\le \frac{1}{\nu_n}+ \bx\trasp\wbthech}\right|\\
   & \le & \esp \left| \indica_{ u < -\frac{1}{\nu_n} +  \bx\trasp\wbthech}- \indica_{ u < -\frac{1}{\nu_0}}\right| +\esp \left| \indica_{ u \le \frac{1}{\nu_0}}-\indica_{ u\le \frac{1}{\nu_n}+ \bx\trasp\wbthech}\right|\\
  &\le & \esp \left|G_0\left(\frac{1}{\sigma_0}\left[ -\frac{1}{\nu_n} +  \bx\trasp\wbthe \right] \right) - G_0\left(-\frac{1}{\sigma_0\;\nu_0}  \right)  \right| + \esp \left|G_0\left( \frac{1}{\sigma_0}\left[  \frac{1}{\nu_n} +  \bx\trasp\wbthe \right] \right) - G_0\left( \frac{1}{\sigma_0\;\nu_0}  \right)  \right| \, ,
 \end{eqnarray*}
where we understand that $\indica_{ u < -1/{\nu_0}}=0$, $G_0\left(- {1}/{\sigma_0\;\nu_0}\right) =0$, $\indica_{ u < 1/{\nu_0}}=1$ and $G_0\left( {1}/{\sigma_0\;\nu_0}\right) =1$ if $\nu_0=0$. Now the proof follows from the continuity of $G_0$ is $t_0<\infty$ and from the fact that $\lim_{u\to -\infty} G_0(u)=0$ while $\lim_{u\to +\infty} G_0(u)=1$.

To derive \eqref{eq:Bn}, note that
 \begin{eqnarray*}
 | B_n|&=& \left|\esp w\left(\nu_0  u   \right)\, \indica_{u-\bx\trasp\wbthech\le  \wsigma \,  s}  - \esp  w\left(\nu_0\, u\right)\indica_{u\le \sigma_0 \, s  }\right|\\
 &\le &  \esp  \left|\indica_{u\le \bx\trasp\wbthech+  \wsigma \,  s} -\indica_{u\le \sigma_0 \, s  }\right| \, .
 \end{eqnarray*}
If  $\bx\trasp\wbthe+  \wsigma \,  s\le \sigma_0 \, s $, then $\indica_{u\le \bx\trasp\wbthech+  \wsigma \,  s} =1$ implies that $\indica_{u\le   \sigma_0 \,  s} =1$, so that $\Delta(u)=\indica_{u\le \bx\trasp\wbthech+  \wsigma \,  s} -\indica_{u\le \sigma_0 \, s  }=0$. Similarly, if $\indica_{u\le   \sigma_0 \,  s} =0$, then  $\indica_{u\le \bx\trasp\wbthech+  \wsigma \,  s} =0$ and $\Delta(u)=0$. Therefore, when  $\bx\trasp\wbthe +  \wsigma \,  s\le \sigma_0 \, s $, $\Delta(u)=1$ if and only if $ \bx\trasp\wbthe +  \wsigma \,  s< u\le \sigma_0 \,  s$.

On the other hand, if $\bx\trasp\wbthe +  \wsigma \,  s\ge \sigma_0 \, s $, then  $\Delta(u)=1$ if and only if $ \sigma_0 \, s < u\le \bx\trasp\wbthe +  \wsigma \,  s$.

Note that the fact that $\wbthe\convpp 0$ and $\wsigma\convpp \sigma_0$ entails that $\bx\trasp\wbthe+  \wsigma \,  s\convpp \sigma_0 \, s $, for each $\bx$. Let $\itC=\{\bx: \bx\trasp\wbthe+  \wsigma \,  s\le \sigma_0 \, s\}$ and $\overline{\itC}$ its complement, then 
\begin{eqnarray*}
 | B_n|  &\le &  \esp \;\indica_{\itC} \;\indica_{ \bx\trasp\wbthech+  \wsigma \,  s<u\le \sigma_0 \, s  } + \esp \;\indica_{\overline{\itC}} \;\indica_{ \sigma_0 \, s <u\le \bx\trasp\wbthech+  \wsigma \,  s }\\
  &\le &  \esp \;\indica_{\itC} \;\left\{G_0( \, s )- G_0\left(\frac{\bx\trasp\wbthe +  \wsigma \,  s}{\sigma_0} \right)\right\} + \esp \;\indica_{\overline{\itC}} \;\left\{ G_0\left(\frac{\bx\trasp\wbthe +  \wsigma \,  s}{\sigma_0} \right)- G_0(  \, s ) \right\} \\
  &\le & \esp \left | G_0\left(\frac{\bx\trasp\wbthe +  \wsigma \,  s}{\sigma_0} \right)- G_0(  \, s ) \right| 
 \end{eqnarray*}
and \eqref{eq:Bn} follows immediately from the continuity of $G_0$ and \ref{ass:consis}, concluding the proof of \eqref{eq:aprobar1}.

 Similar arguments allow to show that  
 \begin{equation}
  \label{eq:aprobar2}
  \frac{1}{n} \sum_{i=1}^n  w_i  \convpp  h_\infty(t_0) =\left\{\begin{array}{lr}
  \esp_{G_0} w\left(\dfrac{\epsilon_1}{  \,  t_0}\right)   & \mbox{ if } t_0<\infty\\
   1  & \mbox{ if } t_0=\infty
  \end{array}
  \right.
  \end{equation}
  and the desired result follows now easily combining \eqref{eq:aprobar1} and \eqref{eq:aprobar2}. \qed

\subsection{Non--linear Model}{\label{sec:appendix2}}
In this section, we assume that we have a random sample $(y_{1},\bx_{1}),\dots,(y_{ n},\bx_{n})$, where $\bx_i$ is a vector of  $p$ explanatory variables and $y_i$ is a response variable that satisfy
$$ y_i= f(\bx_i , \bbe_0) + u_i=   f(\bx_i , \bbe_0) +\sigma_0 \epsilon_i, i=1 \dots n \, ,$$
with $\bbe_0 \in \real^q$ and the errors $\epsilon_i$ are i.i.d. and independent of $\bx_i$ with unknown distribution $G_0(\cdot)$ and $\sigma_0$ is the scale parameter.  As above, the residuals are defined using  robust strongly consistent estimators of $\bbe_0$ and $\sigma_0$, let us say  $\wbbe$ and $\wsigma$ as
$$r_{i}= \dfrac{y_{i}-f(\bx_{i}, \wbbe) }{\wsigma} =  \dfrac{u_{i}-\left[f(\bx_{i}, \wbbe)-f(\bx_i , \bbe_0)\right] }{\wsigma} \, .$$
We compute the adaptive weighted empirical distribution at point $t$ as in \eqref{empiricalw} with 
$$w_i= w\left(\dfrac{ r_i }{t_n}\right)\,,$$  
where as in Section \ref{sec:appendix1},   the adaptive cut--off values are defined through \eqref{eq:tn}.
 
The following additional assumptions are required to provide a general framework to deal with non--linear models.

 \begin{enumerate}[label=\textbf{C\arabic*}]
 \setcounter{enumi}{3}
 \item \label{ass:clasef} The class of functions
 $$\itF=\{  f(\bx , \bbe)\,, \|\bbe-\bbe_0\|\le 1\}$$
 with enveloppe $F\in L^1(P_{\bx})$ is such that $N_{[\;]}(\epsilon, \itF, L_1(P_{\bx}))<\infty$, where $P_{\bx}$ is the probability measure of $\bx$.

 \item\label{ass:denG0} $G_0$ has a bounded density $g_0$.
 \item\label{ass:fcont} $f(\bx,\bbe)$ is a continuous function of $\bbe$ for each $\bx$ and $F(\bx)=\sup_{\|\bbe-\bbe_0\|\le 1} f(\bx,\bbe)\in L^1(P_{\bx})$.
 \end{enumerate}
  
  It is worth noticing that Lemma 3.10 in van der Geer (2000) entails that \ref{ass:clasef} holds if \ref{ass:fcont} holds.
  
  Lemma \ref{lema:appendix2.1} below is an intermediate result needed to derive  Lemma \ref{lema:appendix2.2} which is the non--linear counterpart of Lemma \ref{lema:appendix1.2}.
  
 \begin{lemma} \label{lema:appendix2.1} 
 Assume that  \ref{ass:G0},  \ref{ass:clasef} and \ref{ass:denG0} hold.   Denote $\itV_0=\{\bbe: \|\bbe-\bbe_0\|\le 1\}$ and $\itI_0=[\sigma_0/2, 2\,\sigma_0]$ and for any fixed $t\ge 0$  consider the family of functions
 $$\itH=\{h_{\bbech,\; \sigma}(y,\bx)=\indica_{|y-f(\bx , \bbech)|\le  \sigma\, t} \mbox{ for } (\bbe, \sigma)\in  \itV_0\times \itI_0\}\,.$$
Then, $ \sup_{h\in \itH} | P_n h- P h| \convpp 0$.
\end{lemma}
  
   \noi \textsc{Proof.}  First, note that  
 $$h_{\bbech,\; \sigma}(y,\bx)=\indica_{|y-f(\bx , \bbech)|\le  \sigma \; t}=  h^{(1)}_{\bbech,\; \sigma}(y,\bx)\,h^{(2)}_{\bbech,\; \sigma}(y,\bx)\,,$$
 where 
 \begin{eqnarray*}
 h^{(1)}_{\bbech,\; \sigma}(y,\bx) &=& \indica_{ y-f(\bx , \bbech) -  \sigma \; t\le 0}\\
 h^{(2)}_{\bbech,\; \sigma}(y,\bx) &=& \indica_{ 0\le y-f(\bx , \bbech)  +\sigma \; t  }\, .
 \end{eqnarray*}
 Denote as $\itH^{(j)}=\{h^{(j)}_{\bbech,\; \sigma}(y,\bx)\,, (\bbe, \sigma)\in  \itV_0\times \itI_0\} $. Taking into account that $\itH\subset  \itH^{(1)}\cdot \itH^{(2)}$ and that the functions $h^{(j)}_{\bbech,\; \sigma}$ are non--negative and bounded by 1,  to show that  
 $$
 \sup_{h\in \itH} | P_n h- P h| \convpp 0\,,
 $$
it will be enough to show that  $N_{[\;]}(\epsilon, \itH^{(j)}, L_1(P ))<\infty$, for $j=1,2$, where $P$ is the probability measure of $(y,\bx)$. We will derive the result for $\itH^{(1)}$, the proof for $\itH^{(2)}$ been analogous.
 
 Let $\epsilon>0$ and denote $\delta=\sigma_0\epsilon/(2\,\|g_0\|_{\infty})$. Then, the fact that $\itI_0$ is compact entails that there exist $k\le 2\, t\, \sigma_0/\delta$ $\sigma_0/2=\sigma_1\le \dots\le\sigma_k=2\, \sigma_0$     such that  $\sigma_j-\sigma_{j-1}\le \delta/ t $.

Denote $M=N_{[\;]}(\delta, \itF, L_1(P_{\bx}))$, then there exists $\{(f_{j,L}, f_{j,U})\}_{1\le j\le M}$ such that,  for any $f\in \itF$ there exists $j$ such that $f_{j,L} \le f\le f_{j,U} $ and $\esp f_{j,U}-f_{j,L} \le \delta$.

Fix $\bbe\in \itV_0$ and $\sigma\in \itI_0$ and let $1\le j\le M$ and $1\le \ell \le k-1$, be such that $\sigma\in [\sigma_\ell, \sigma_{\ell+1}]$ and $f_{j,L}(\bx) \le f(\bx)\le f_{j,U}(\bx) $, for all $\bx$. Then, using that $t\ge 0$ we obtain that
$$g_{\ell, j,L}(y,\bx)=y- f_{j,U}(\bx) -  \sigma_{\ell+1} \; t \le y-f(\bx , \bbe) -  \sigma \; t\le y-f_{j,L}(\bx) -  \sigma_{\ell}  \; t=g_{\ell, j, U}(y,\bx)\, ,$$
so that
$$\indica_{g_{\ell, j,U}(y,\bx)\le 0}\le  h^{(1)}_{\bbech,\; \sigma}(y,\bx)\le \indica_{g_{\ell, j,L}(y,\bx)\le 0}\,.$$
 Denote $h_{\ell,j,L}=\indica_{g_{\ell, j,U}(y,\bx)\le 0}$ and $h_{\ell,j,U}=\indica_{g_{\ell, j,L}(y,\bx)\le 0}$. We will show that $\esp |h_{\ell,j,U}- h_{\ell,j,L}|<\epsilon$, that is, $\{(h_{\ell,j,L}, h_{\ell,j,U}\}_{1\le \ell \le M, 1\le j\le k}$ is an $\epsilon-$bracket for $\itH^{(1)}$, so 
   $N_{[\;]}(\epsilon, \itH^{(j)}, L_1(P ))\le k M<\infty$.    
   Using that $h_{\ell,j,L}\le h_{\ell,j,U}$, $g_{\ell, j,L}(y,\bx)\le g_{\ell, j,U}(y,\bx)$ and that 
   $$g_{\ell, j,L}(y,\bx)=u+  f(\bx, \bbe_0)- f_{j,U}(\bx) -  \sigma_{\ell+1} \; t\quad   g_{\ell, j,U}(y,\bx)=u+  f(\bx, \bbe_0)- f_{j,L}(\bx) -  \sigma_{\ell} \; t$$
    we get that
 \begin{eqnarray*}
 \esp |h_{\ell,j,U}- h_{\ell,j,L}| &=& \esp \indica_{g_{\ell, j,L}(y,\bx)\le 0}-  \indica_{g_{\ell, j,U}(y,\bx)\le 0}   = \prob \left(g_{\ell, j,L}(y,\bx)\le 0\right) -\prob\left( g_{\ell, j,U}(y,\bx)\le 0 \right)\\
 &=& \prob \left(u\le  f_{j,U}(\bx) + \sigma_{\ell+1} \; t-  f(\bx, \bbe_0)\right)- \prob\left(u\le  f_{j,L}(\bx) + \sigma_{\ell} \; t-  f(\bx, \bbe_0)\right)\\
 &=& \esp\left\{G_0\left(\frac{ f_{j,U}(\bx) + \sigma_{\ell+1} \; t-  f(\bx, \bbe_0)}{\sigma_0}\right)-G_0\left(\frac{f_{j,L}(\bx) + \sigma_{\ell} \; t-  f(\bx, \bbe_0)}{\sigma_0}\right)\right\}
 \end{eqnarray*}  
   Thus, using that $G_0$ has a bounded density $g_0$, we obtain that
   \begin{eqnarray*}
 \esp |h_{\ell,j,U}- h_{\ell,j,L}| &\le &  \|g_0\|_{\infty}\esp\left\{ \left|\frac{ f_{j,U}(\bx) + \sigma_{\ell+1} \; t-  f(\bx, \bbe_0)}{\sigma_0} -\frac{f_{j,L}(\bx) + \sigma_{\ell} \; t-  f(\bx, \bbe_0)}{\sigma_0}\right|\right\}\\
 &\le &  \frac{\|g_0\|_{\infty}}{\sigma_0}\left\{  \left(\sigma_{\ell+1}-   \sigma_{\ell} \right)\; t+ \esp  \left|  f_{j,U}(\bx)   -f_{j,L}(\bx)   \right|\right\}\le 2\delta \; \frac{\|g_0\|_{\infty}}{\sigma_0} =\epsilon\,,
 \end{eqnarray*} 
 concluding the proof. \qed

   \vskip0.2in
  
 \begin{lemma} \label{lema:appendix2.2} 
 Assume that  \ref{ass:G0},  \ref{ass:consis}, \ref{ass:denG0} and \ref{ass:fcont} hold.   Then,  we have that 
 \begin{enumerate}[label=\alph*)]
 \item $\|G_n^{+}-G_0^{+}\|_{\infty}\convpp 0$, where 
 $$G^{+}_n(t)= \frac 1n \sum_{i=1}^n  I(|r_i| \le t)  \quad \quad r_i=  \dfrac{y_{i}-f(\bx_{i}, \wbbe) }{\wsigma}$$
and   $G^{+}_0(t)$ is the distribution of the absolute errors when $\epsilon_i \sim G_0$
 \item $d_n\convpp d_0$. 
 \item $t_n\convpp t_0$,  if in addition, $ G_0$ is strictly increasing on its support. 
  \end{enumerate}
  \end{lemma}
 
 \noi \textsc{Proof.} a) Using Lemma \ref{lema:appendix1.1}, it will be enough to show that for each fixed $t$
 \begin{equation}
 \label{eq:Gn+toG+}
 G_n^{+}(t)-G_0^{+}(t)\convpp 0\,. 
 \end{equation}
Denote $\itV_0=\{\bbe: \|\bbe-\bbe_0\|\le 1\}$ and $\itI_0=[\sigma_0/2, 2\,\sigma_0]$. 
Let us consider the family of functions
 $$\itH=\{h_{\bbech,\; \sigma}(y,\bx)=\indica_{|y-f(\bx , \bbech)|\le  \sigma\, t} \mbox{ for } (\bbe, \sigma)\in  \itV_0\times \itI_0\}\,.$$
Using that \ref{ass:fcont} implies \ref{ass:clasef}, Lemma \ref{lema:appendix2.1} entails that
 \begin{equation}
 \label{eq:glivenko2}
 \sup_{h\in \itH} | P_n h- P h| \convpp 0\,.
 \end{equation}
 On the other hand,  $G_n^{+}$ can be written as
 $$G_n^{+}(t) =P_n h_{\wbbech, \;\wsigma }(y_i,\bx_i)\,.$$
Hence, if we denote as
$M(\bbe, \sigma)= P h_{\bbech,\; \sigma}$, using \eqref{eq:glivenko2} and the fact that \ref{ass:consis} entails that with probability 1, for $n$ large enough, $(\wbbe,\wsigma)\in \itV_0\times \itI_0$, we conclude that
$$ \left|G_n^{+}(t) - M( \wbthech, \;\wsigma )\right|  \convpp 0\,.$$
It remains to show that
$$   M( \wbthe, \;\wsigma )- G_0^{+}(t)  \convpp 0\,.$$
Note that
$$ M(\bbe, \sigma)= P h_{\bbech,\; \sigma}= \prob(|y-f(\bx , \bbech)|\le  \sigma\, t) \, ,$$
hence
$$M(\bbe_0, \sigma_0 )= \prob( |u|\le \sigma_0 \, t)=G_0^{+}(t)\,.$$
Therefore, we have to show that
$$   M( \wbthe, \;\wsigma\,t)\convpp  M(\bbe_0, \sigma_0)\,.$$
First observe that
\begin{eqnarray*}
 M(\bbe, \sigma)&=& \prob \left( -\sigma\, t  + f(\bx,\bbe)-   f(\bx,\bbe_0)  \le u \le \sigma\, t+ f(\bx,\bbe)-  f(\bx,\bbe_0)  \right)\\
 &=& \esp \left\{G_0(\sigma\, t+ f(\bx,\bbe)-  f(\bx,\bbe_0))- G_0(-\sigma\, t  + f(\bx,\bbe)-   f(\bx,\bbe_0))\right\}
 \end{eqnarray*}
 The continuity of $G_0$ and $f(\bx, \bbe)$ and the Dominated Convergence Theorem entail that $M(\bbe, \sigma)$ is a continuous function of its arguments, which together with 
\ref{ass:consis}, entails that $M( \wbthe, \;\wsigma )- M(\bbe_0, \sigma_0 ) \convpp 0$, for each fixed $t$, concluding the proof of a).  

b) and c) follow as in Lemma Lemma \ref{lema:appendix1.2}. \qed

\vskip0.2in
As in Section \ref{sec:appendix1}, denote  $\wnu_n= 1/(t_n\, \wsigma_n )$, $\nu_0= 1/(t_0\, \sigma_0 )$, where we understand that if $t_0=\infty$, $\nu_0=0$. Furthermore, let $\itJ_0$ be a compact interval with non--empty interior, such that $\nu_0\in \itJ_0$. 

Lemma \ref{lema:appendix2.3}  is the non--linear counterpart of Lemma \ref{lema:appendix1.3}. Note that a bounded density is needed when a general non--linear model is considered, as well as a continuous weight function.

\begin{lemma} \label{lema:appendix2.3}
Under \ref{ass:peso}, \ref{ass:G0}, \ref{ass:denG0} and \ref{ass:fcont},  we have that $\sup_{g\in \itG} | P_n g- P g| \convpp 0$, where 
$$\itG=\{g_{\bbech,\; \sigma,\; \nu}(y,\bx)=w\left(\nu({y -f(\bx,\bbe)})\right)\, \indica_{y-f(\bx , \bbech)\le\sigma\, t} \mbox{ for } (\bbe, \sigma, \nu)\in \itV_0\times \itI_0 \times \itJ_0\} \,.$$
\end{lemma}
 
  \noi \textsc{Proof.} Note that $\itG\subset \itG_1 \cdot \itG_2$ where 
  \begin{eqnarray*}
  \itG_1 &=& \{g_{\bbech,\, \nu}(y,\bx)=w\left(\nu(y -f(\bx,\bbe))\right)\,  \mbox{ for } (\bbe,   \nu)\in \itV_0\times  \itJ_0\}\\
  \itG_2 &=& \{g_{\bbech,\; \sigma }(y,\bx)=  \indica_{y-f(\bx , \bbech)\le\sigma\, t} \mbox{ for } (\bbe, \sigma )\in \itV_0\times \itI_0  \}\,.
   \end{eqnarray*}
The classes $\itG$, $\itG_1$ and $\itG_2$ have envelope 1 and are classes of non--negative functions,  hence we have easily that,  
 $$N_{[\;]}(2\,\epsilon, \itG, L_1(P))\le N_{[\;]}(\epsilon, \itG_1, L_1(P)) N_{[\;]}(\epsilon, \itG_2, L_1(P))\,,$$
so that to show $\sup_{g\in \itG} | P_n g- P g| \convpp 0$, it will be enough to prove that, for $j=1,2$,
\begin{equation}
\label{eq:NP2}
 N_{[\;]}(\epsilon, \itG_j, L_1(P))<\infty\,.
 \end{equation}
Note that, when $j=2$, \eqref{eq:NP2} follows from the proof of Lemma \ref{lema:appendix2.1}. On the other hand, the continuity of $w$ and \ref{ass:fcont} entail that $w\left(\nu(y -f(\bx,\bbe))\right)$ is a continuous function of $(\nu, \bbe)$ for each $(y,\bx)$. Then, Lemma 3.10 in van der Geer (2000) entails that   $ N_{[\;]}(\epsilon, \itG_1, L_1(P))<\infty$, concluding the proof. \qed

\vskip0.2in
   
\begin{proposition} \label{prop:appendix2.1} 
Under \ref{ass:peso} to \ref{ass:consis} and \ref{ass:denG0} and \ref{ass:fcont},  we have that
\begin{enumerate}[label=\alph*)]
\item if $t_0<\infty$, 
$$\sup_{s\in \real} \left|\wG_n(s)- \frac{h_0(t_0,s)}{h_{\infty}(t_0)}\right|\convpp 0\,,$$
with $h_\infty(t_0)$ and $h_0(t_0,s)$ defined in \eqref{eq:hinfty} and \eqref{eq:h0}, respectively.
\item  if   $t_0=\infty$,  $\|\wG_n-G_0\|_{\infty}\convpp 0$.
\end{enumerate}
\end{proposition}
  
  \noi \textsc{Proof.}  When $t_0=\infty$, using that $G_0$ is a  bounded, monotone and continuous function and that $\wG_n$ is monotone, from  Lemma \ref{lema:appendix1.1}, it will be enough to show that
  for each $s\in \real$, $\wG_n(s)\convpp G(s)$. On the other hand, when $t_0<\infty$, standard arguments allow to show that $F(s)= {h_1(t_0,s)}/{h_{\infty}(t_0)}$ is a  bounded, monotone and continuous function of $s$ and the uniform convergence also follows from the pointwise one.
  
  Taking into account that $\wnu_n\convpp \nu_0$, we have that with probability 1, for $n$ large enough $\wnu_n\in \itJ_0$.
  
As in the proof of Proposition \ref{prop:appendix1.1}, we will begin by showing that
  \begin{equation}
  \label{eq:aprobar12}
  \frac{1}{n} \sum_{i=1}^n  w_i I(r_i \le s) \convpp  h_0(t_0,s) =\left\{\begin{array}{lr}
  \esp_{G_0} w\left(\dfrac{\epsilon_1}{  \,  t_0}\right)\indica_{\epsilon_1\le s  }  & \mbox{ if } t_0<\infty\;,\\
   \esp_{G_0} \indica_{\epsilon_1\le s  }=G_0(s)  & \mbox{ if } t_0=\infty \;.
  \end{array}
  \right. 
  \end{equation}
  For that purpose and noting that $r_i=(y_i-f(\bx, \wbbe))/\wsigma$, define the class of functions
  $$\itG=\{g_{\bbech,\; \sigma,\; \nu}(y,\bx)=w\left(\nu( y -f(\bx,\bbe))\right)\, \indica_{y-f(\bx , \bbech)\le\sigma\, s} \mbox{ for } (\bbe, \sigma, \nu)\in \itV_0\times \itI_0 \times \itJ_0\} \,.$$
 Lemma \ref{lema:appendix2.3}  entails that
 $$
 \sup_{g\in \itG} | P_n g- P g| \convpp 0\,,
 $$
 then, using that  
 $$\frac{1}{n} \sum_{i=1}^n  w_i I(r_i \le s)= P_n g_{\wbbech,  \;\wsigma, \; \wnu_n}\;,$$
 we obtain that
 $$ \frac{1}{n} \sum_{i=1}^n  w_i I(r_i \le s) - P g_{\wbbech, \;\wsigma, \; \wnu_n} \convpp 0\,.$$
 It remains to show that   
  $$P g_{\wbbech, \;\wsigma ,\;\wnu_n}\convpp P g_{\bbech_0,\, \sigma_0,\; \nu_0} = h_0(t_0,s)\,.$$
 which will follow if we derive that  
 \begin{eqnarray}
 A_n&=&P g_{\wbbech , \;\wsigma,\;\wnu_n }-\esp w\left(\nu_0  u   \right)\, \indica_{y-f(\bx,\wbbech)\le  \wsigma \, s}   \convpp 0
 \label{eq:An2}\\
  B_n&=& \esp w\left(\nu_0  u   \right)\, \indica_{y-f(\bx,\wbbech)\le  \wsigma \,  s}  -h_0(t_0,s) \convpp 0 \;.
 \label{eq:Bn2}
  \end{eqnarray}
  Noting that
 \begin{eqnarray*}
 |A_n| &=& \left|\esp \left\{ w\left(\nu_n\left[y-f(\bx,\wbbe)\right]\right) - w\left(\nu_0\, u  \right)\right\} \indica_{y-f(\bx,\wbbech)\le  \wsigma \,  s} \right|\\
 &\le &  \esp \left| w\left(\nu_n\left[y-f(\bx,\wbbe)\right]\right) - w\left(\nu_0\, u  \right)\right| \;,
 \end{eqnarray*}
using the Dominated Convergence Theorem, the continuity of $w$   and the fact that $\wnu_n\convpp \nu_0$ and $\wbthe\convpp 0$, we obtain that $A_n\convpp 0$, concluding the proof of \eqref{eq:An2}.  

To derive \eqref{eq:Bn2}, using that $0\le w(x)\le 1$, we get  that
 \begin{eqnarray*}
 | B_n|&=& \left|\esp w\left(\nu_0  u   \right)\, \indica_{y-f(\bx,\wbbech)\le  \wsigma \,  s}  - \esp  w\left(\nu_0\, u\right)\indica_{u\le \sigma_0 \, s  }\right|\\
 &\le &  \esp  \left|\indica_{u\le  f(\bx,\wbbech)-f(\bx,\bbech_0)+  \wsigma \,  s} -\indica_{u\le \sigma_0 \, s  }\right|\;.
 \end{eqnarray*}
As in the proof of Proposition \ref{prop:appendix1.1}, we have that,  if  $f(\bx,\wbbe )-f(\bx,\bbe_0)+  \wsigma \,  s \le \sigma_0 \, s $, then   $\Delta(u)=\indica_{u\le \bx\trasp\wbthech+  \wsigma \,  s} -\indica_{u\le \sigma_0 \, s  }= 1$ if and only if $ f(\bx,\wbbe )-f(\bx,\bbe_0)+  \wsigma \,  s < u\le \sigma_0 \,  s$.

On the other hand, if $f(\bx,\wbbe )-f(\bx,\bbe_0)+  \wsigma \,  s \ge \sigma_0 \, s $, then  $\Delta(u)=1$ if and only if $ \sigma_0 \, s < u\le f(\bx,\wbbe )-f(\bx,\bbe_0)+  \wsigma \,  s$.

Note that the fact that $\wbbe\convpp 0$ and $\wsigma\convpp \sigma_0$ together with the continuity of $f(\bx, \bbe)$ entails that $f(\bx,\wbbe )-f(\bx,\bbe_0)+  \wsigma \,  s\convpp \sigma_0 \, s $, for each $\bx$. Let $\itC=\{\bx: f(\bx,\wbbe )-f(\bx,\bbe_0)+  \wsigma \,  s \le \sigma_0 \, s\}$ and $\overline{\itC}$ its complement, then 
\begin{eqnarray*}
 | B_n|  &\le &  \esp \;\indica_{\itC} \;\indica_{f(\bx,\wbbech)-f(\bx,\bbech_0)+  \wsigma \,  s <u\le \sigma_0 \, s  } + \esp \;\indica_{\overline{\itC}} \;\indica_{ \sigma_0 \, s <u\le f(\bx,\wbbech)-f(\bx,\bbech_0)+  \wsigma \,  s }\\
  &\le &   \esp \left | G_0\left(\frac{f(\bx,\wbbe )-f(\bx,\bbe_0)+  \wsigma \,  s}{\sigma_0} \right)- G_0(  \, s ) \right| 
 \end{eqnarray*}
and \eqref{eq:Bn2} follows immediately from the continuity of $G_0$ and \ref{ass:consis}, concluding the proof of \eqref{eq:aprobar12}.

 Similar arguments allow to show that  
 \begin{equation}
  \label{eq:aprobar22}
  \frac{1}{n} \sum_{i=1}^n  w_i  \convpp  h_\infty(t_0) =\left\{\begin{array}{lr}
  \esp_{G_0} w\left(\dfrac{\epsilon_1}{  \,  t_0}\right)   & \mbox{ if } t_0<\infty\\
   1  & \mbox{ if } t_0=\infty
  \end{array}
  \right.
  \end{equation}
  and the desired result follows now easily combining \eqref{eq:aprobar12} and \eqref{eq:aprobar22}. \qed

\small
\section*{References}
 
\begin{description}

\item Alonzo, T. A. and Pepe, M. S. (2002). Distribution-free ROC analysis using binary
regression techniques. \textsl{Biostatistics}, \textbf{3}, 421-432. 

\item Bianco, A. M. and Spano, P. (2019). Robust inference for nonlinear regression models. \textsl{Test}, \textbf{28}, 369-398. 

\item Cai, T. (2004). Semiparametric ROC regression analysis with placement values. \textsl{Bio
statistics}, \textbf{5}, 45-60. 

\item  In\'acio  de Carvalho, V.,  Jara, A.,  Hanson, T. E.  and de Carvalho, M. (2013). Bayesian nonparametric ROC regression modeling, \textsl{Bayesian Analysis}, \textbf{8}, 623-646. 

\item Faraggi, D. (2003). Adjusting receiver operating characteristic curves and related
indices for covariates. \textsl{Journal of the Royal Statistical Society}, Ser. D, \textbf{52}, 1152-1174.

\item Farcomeni, A. and Ventura, L. (2012). An overview of robust methods in medical research. \textsl{Statistical Methods in Medical Research}, \textbf{21}, 111-133.

\item Gervini, D. and Yohai, V. J. (2002). A class of robust and fully efficient regression estimators. \textsl{Annals of Statistics}, \textbf{30}, 583-616.


\item Goncalves, L., Subtil, A., Oliveira, M. R. and  Bermudez, P. (2014) Roc Curve Estimation: An Overview. \textsl{REVSTAT-Statistical Journal}, \textbf{12}, 1-20.

\item Gonz\'alez-Manteiga, W., Pardo-Fern\'andez, J. C., and Van Keilegom, I. (2011). ROC
curves in non-parametric location-scale regression models.\textsl{Scandinavian Journal of
Statistics}, \textbf{38}, 169-184.

\item Greco, L. and Ventura, L. (2011). Robust inference for the stress-strength reliability. \textsl{Statistical Papers}, \textbf{52}, 773-788.

\item Kosorok, M. (2008). \textsl{Introduction to Empirical Processes and Semiparametric Inference}.  Springer--Verlag, New
  York.

\item Krzanowski, W. J. and Hand, D. J. (2009). \textsl{ROC curves for continuous data}. Chapman and Hall/CRC, Boca Raton.

\item Pardo-Fern\'andez, J. C., Rodr\'{\i}guez-Alvarez, M. X. and Van Keilegom, I. (2014). A review on ROC curves in the presence of covariates. \textsl{REVSTAT Statistical Journal}, \textbf{12}, 21-41.

\item Pepe, M. S. (1997). A regression modelling framework for receiver operating characteristic curves in medical diagnostic testing. \textsl{Biometrika}, \textbf{84}, 595-608.

\item Pepe, M. S. (1998). Three approaches to regression analysis of receiver operating
characteristic curves for continuous test results. \textsl{Biometrics}, \textbf{54}, 124-135. 

\item Pepe, M. S. (2003). \textsl{The Statistical Evaluation of Medical Tests for Classification
and Prediction}, Oxford University Press, New York.

\item Rodr\'{\i}guez-Alvarez, M. X., Roca-Pardi\~nas, J., and Cadarso-Su\'arez, C. (2011a). ROC
curve and covariates: extending the induced methodology to the non-parametric
framework. \textsl{Statistics and Computing}, \textbf{21}, 483-495. 

\item Rodr\'{\i}guez-Alvarez, M. X., Tahoces, P. C., Cadarso-Su\'arez, C., and Lado, M. J. (2011b).
Comparative study of ROC regression techniques—applications for the computer-aided
diagnostic system in breast cancer detection.  \textsl{Computational Statistics and
Data Analysis}, \textbf{55}, 888-902.

 \item Sun, Y. and Genton, M. G. (2011). Functional boxplots. \textsl{Journal of Computational and Graphical Statistics}, \textbf{20}, 316-334.

  \item Van de Geer, S. (2000). \textsl{Empirical Processes in M--Estimation}, Cambridge University Press, United States of America.

\item van der Vaart, A.  and Wellner, J. (1996). \textsl{Weak Convergence and Empirical Processes. With Applications to Statistics}. Springer--Verlag, New York.

\item Walsh, S. J. (1997). Limitations to the robustness of binormal ROC curves:
effects of model misspecification and location of decision thresholds on bias,
precision, size and power, \textsl{Statistics in Medicine}, \textbf{16}, 669-679.

\item Yohai, V. J. (1987). High Breakdown-Point and High Efficiency Robust Estimates for Regression. \textsl{Annals of Statistics}, \textbf{15}, 642-656.  

\end{description}

\end{document}

%% file: definiciones.tex
\newcommand\bb {\mathbf b}

\newcommand\bx {\mathbf x}

\newcommand\bX {\mathbf X}

\newcommand\indica {\mathbb{I}}

\newcommand\wB {\widehat{B}}
\newcommand\wC {\widehat{C}}

\newcommand\wG {\widehat{G}}

\newcommand\wM {\widehat{M}}

\newcommand\itA {{\mathcal{A}}}
\newcommand\itB {{\mathcal{B}}}
\newcommand\itC {{\mathcal{C}}}

\newcommand\itF {{\mathcal{F}}}
\newcommand\itG {{\mathcal{G}}}
\newcommand\itH {{\mathcal{H}}}
\newcommand\itI {{\mathcal{I}}}
\newcommand\itJ {{\mathcal{J}}}
\newcommand\itK {{\mathcal{K}}}

\newcommand\itN {{\mathcal{N}}}

\newcommand\itR {{\mathcal{R}}}
\newcommand\itS {{\mathcal{S}}}

\newcommand\itV {{\mathcal{V}}}

\newcommand\bbe {\mbox{\boldmath $\beta$}}

\newcommand\bbech {\mbox{\scriptsize${\bbe}$}}

\newcommand\bthe {\mbox{\boldmath $\theta$}}
\newcommand\bthech {\mbox{\scriptsize\boldmath $\theta$}}

\newcommand\wbbe {\widehat{\bbe}}
\newcommand\wbbech {\mbox{\scriptsize$\wbbe$}}

\newcommand\wmu {\widehat{\mu}}

\newcommand\wnu {\widehat{\nu}}

\newcommand\wsigma {\widehat{\sigma}}

\newcommand\wbthe {\widehat{\bthe}}
\newcommand\wbthech {\widehat{\bthech}}

\newcommand\wDelta {\widehat{\Delta}}

\def\real{\mathbb{R}}

\newcommand{\esp}{\mathbb{E}}
\newcommand{\prob}{\mathbb{P}}

\newcommand{\convpp}{ \buildrel{a.s.}\over\longrightarrow}

\newcommand{\convprob  }{ \buildrel{p}\over\longrightarrow}

\newcommand{\trasp}{^{\mbox{\footnotesize \sc t}}}

\def\ROC {\mathop{\mbox{ROC}}} 
\def\AUC {\mathop{\mbox{AUC}}}

\newcommand\noi{\noindent}

\def\square{\ifmmode\sqr\else{$\sqr$}\fi}
\def\sqr{\vcenter{
         \hrule height.1mm
         \hbox{\vrule width.1mm height2.2mm\kern2.18mm
\vrule width.1mm}
         \hrule height.1mm}}

\newcommand{\cl}{\mbox{\footnotesize \sc cl}}

